\documentclass[twocolumn,epjc3]{svjour3mod}

\usepackage[colorlinks,citecolor=blue,urlcolor=blue,linkcolor=blue]{hyperref}
\usepackage{flushend}
\usepackage{graphicx,colordvi}
\usepackage{amsmath,amssymb,slashed,cite}
\usepackage{booktabs,tabulary}
\usepackage{dsfont}
\usepackage{color}
\usepackage{cite}

\newcommand{\diff}{\text{d}}
\newcommand{\MeV}{\text{MeV}}
\newcommand{\GeV}{\text{GeV}}
\renewcommand{\Im}{\text{Im}}
\renewcommand{\Re}{\text{Re}}
\newcommand{\disc}{\text{disc}\,}
\newcommand{\nl}{\notag\\}
\newcommand{\BR}{\mathcal{B}}
\newcommand{\beq}{\begin{equation}}
\newcommand{\eeq}{\end{equation}}
\newcommand{\bsp}{\begin{sloppypar}}
\newcommand{\esp}{\end{sloppypar}}

\smartqed  

\begin{document}

\title{A new parametrization for the scalar pion form factors}

\author{Stefan Ropertz\thanksref{addr1,e1}
	\and
	Christoph Hanhart\thanksref{addr2,e2}
	\and
	Bastian Kubis\thanksref{addr1,e3}
}

\thankstext{e1}{e-mail: ropertz@hiskp.uni-bonn.de}
\thankstext{e2}{e-mail: c.hanhart@fz-juelich.de}
\thankstext{e3}{e-mail: kubis@hiskp.uni-bonn.de}

\institute{Helmholtz-Institut f\"ur Strahlen- und Kernphysik (Theorie) and
         Bethe Center for Theoretical Physics,
         Universit\"at Bonn,
         53115 Bonn, Germany\label{addr1}
         \and
         Institut f\"ur Kernphysik, 
         Institute for Advanced Simulation, and 
         J\"ulich Center for Hadron Physics, 
         Forschungszentrum J\"ulich, 
         52425 J\"ulich, Germany\label{addr2}
}

\date{}

\maketitle

\begin{abstract}
We derive a new parametrization for the scalar pion form factors that allows us to
analyze data over a large energy range via the inclusion of resonances, and at the same time to ensure consistency
with the high-accuracy dispersive representations available at low energies.  
As an application the formalism is used to extract resonance properties of
excited scalar mesons from data for $\bar B^0_s\to J/\psi \pi\pi$.
In particular we find for the pole positions of $f_0(1500)$ and $f_0(2020)$
$1465\pm 18 - i (50\pm 9)\,\MeV$ and $1910\pm 50 - i(199\pm 40)\,\MeV$,
respectively. In addition, from their residues we calculate the respective branching ratios into $\pi\pi$ to be $(58\pm31)\%$ and $(1.3\pm1.8)\%$.
\end{abstract}

\section{Introduction}\label{sec:intro}

The scalar isoscalar sector of the QCD spectrum up to $2\,\GeV$ has been of high theoretical and experimental interest for
many years. One of the main motivations for these investigations is the hunt for glueballs: their lightest
representatives are  predicted to occur
in the mass range between $1600$ and $1700\,\MeV$ with quantum numbers $0^{++}$~\cite{Bali:1993fb,Morningstar:1997ff,Lee:1999kv,Chen:2005mg}. 
The most straightforward way to identify glueball candidates is to count states with and without
flavor quantum number and see if there are supernumerary isoscalar states; see, e.g., the minireview on non-$\bar qq$ states
provided by the Particle Data Group (PDG)~\cite{PDG} or the reviews Refs.~\cite{Klempt:2007cp,Ochs:2013gi}.
Unfortunately, regardless of the year-long efforts, the scalar isoscalar spectrum is still not fully resolved: e.g.\ there is
still an ongoing debate whether the $f_0(1370)$ exists or not~\cite{Klempt:2007cp}. One problem might be that most analyses of
experimental data performed so far are based on fitting sums of Breit--Wigner functions, which
can lead to reaction-dependent results. To make further progress, it therefore appears compulsory to
employ parametrizations that allow one to extract pole parameters, for those by definition 
do not depend on the production mechanism. 
This requires amplitudes that
are consistent with the general principles of analyticity and unitarity.
In this paper we present a new parametrization for the scalar pion form factors that has these
features built in, and in addition maps smoothly onto well constrained low-energy amplitudes.

The two-pion system at low energies is well understood from sophisticated investigations based
on dispersion theory---in particular the $\pi\pi$--$K\bar K$ phase shifts and inelasticities can be assumed 
as known from threshold up to an energy of about 
$s=(1.1\,\GeV)^2$~\cite{Ananthanarayan:2000ht,Colangelo:2001df,CCL,GarciaMartin:2011cn,Buettiker:2003pp,Pelaez:2018qny}. 
From this information, quantities
like the scalar non-strange and strange form factors for both pions and kaons can be constructed, again
employing dispersion theory~\cite{Donoghue:1990xh,Moussallam:1999aq,DescotesGenon:2000ct,Hoferichter:2012wf,Daub:2012mu,Celis:2013xja,Daub:2015,Winkler:2018qyg}. The resulting amplitudes, which capture the physics of the $f_0(500)$ (or $\sigma$)
and the $f_0(980)$, were already applied successfully to analyze
various meson decays, see, e.g., Ref.~\cite{Daub:2015}. 
In particular the non-Breit--Wigner shape of these low-lying resonances~\cite{Gardner:2001gc} is taken 
care of automatically.
However, to also include higher energies in the analysis, 
where additional inelastic channels become non-negligible and higher resonances need to be included, 
one is forced to leave the safe grounds of fully model-independent dispersion theory and to employ a
model. Ideally this is done in a way that the amplitudes match smoothly onto those constructed rigorously
from dispersion relations. Moreover, to allow for an extraction of resonance properties, 
the extension needs to be performed in a way consistent with analyticity.

A formalism that has all of these features was introduced for the pion vector form factor in Ref.~\cite{H_1}.
In that case, the low-energy $\pi \pi$ interaction can  safely be treated as a single-channel problem in the
full energy range where high-accuracy phase shifts are available, since the two-kaon contribution to the 
isovector $P$-wave inelasticity is very small~\cite{Buettiker:2003pp,Niecknig:2012sj}.\footnote{Note that a 
recent analysis of $N_f=2$ and $N_f=2+1$ lattice data
revealed indications for the necessity to include a $K\bar K$ component for the $\rho$ meson in the
formalism~\cite{Hu:2017wli}.}
However, this is not true for the isoscalar $S$-wave, clearly testified
by the presence of the $f_0(980)$ basically at the $K\bar K$ threshold with a large coupling to this 
channel~\cite{Baru:2004xg,Moussallam:2011zg}. 
Thus, in order to apply the formalism of Ref.~\cite{H_1} to the scalar isoscalar channel
it needs to be generalized. This is the main objective of the present article. As an application we test
the amplitudes on data for $\bar B^0_s\to J/\psi \pi\pi/K\bar K$ recently measured with high accuracy at LHCb~\cite{LHCb_PiPi,LHCb_KK}, which allows us to extract the strange scalar form factor of pions and kaons up to about $2\,\GeV$ and to constrain pole parameters and branching fractions of two of the heavier $f_0$ resonances in that energy range.

This paper is organized as follows. In Sect.~\ref{sec:formalism}, we derive the unitary and analytic scalar form factor para\-metrization to be used. In Sect.~\ref{sec::application} we illustrate its application in a coupled-channel analysis of the decays $\bar{B}_s^0\rightarrow J/\psi\pi\pi$ and $\bar{B}_s^0\rightarrow J/\psi K\bar K$. Specifically, we discuss the stability of our fits under changing assumptions for the parametrization concerning the number of resonances, the degree of certain polynomials, as well as the approximation in the description of the effective four-pion channel.
In addition, in Sect.~\ref{sec::poles} we extract pole parameters, in particular for both the $f_0(1500)$ and the $f_0(2020)$, 
via the method of Pad\'e approximants for the analytic continuation to the unphysical sheets. 
The paper ends with a summary and an outlook in Sect.~\ref{sec:summary}.

\section{Formalism}\label{sec:formalism}

\bsp
The derivation of the form factor parametrization is presented for the strange scalar isoscalar pion (kaon) form factor $\Gamma^s_\pi$ ($\Gamma^s_K$). These are related to the matrix elements
\begin{align}
	\left< \pi^+(p_1)\pi^-(p_2)\left|m_s\bar{s}s\right|0\right>&=\frac{2M_K^2-M_\pi^2}{2}\, \Gamma^s_\pi(s) \,,\nl
	\left<K^+(p_1) K^-(p_2)\left|m_s\bar{s}s\right|0\right>&=\frac{2M_K^2-M_\pi^2}{2}\,\Gamma^s_K(s)\,, \label{eq:scalarFF-def}
\end{align}
where $s=(p_1+p_2)^2$. 
The $\Gamma_i^s(s)$, $i=\pi,\,K$, defined this way are invariant under the QCD renormalization group.
Since the scalar isoscalar $\pi\pi$ system is strongly coupled to the $K\bar{K}$ channel via the $f_0(980)$ resonance, a coupled-channel description becomes inevitable even for energies around $s=1\,\GeV^2$. In this paper we present a parametrization for the scalar form factors
valid at even higher energies.
This becomes possible via the explicit inclusion of further inelasticities and resonances. 
Below $1\,\GeV$ the system is strongly constrained by dispersion theory using a coupled-channel treatment 
of $\pi\pi$ and $K\bar{K}$~\cite{Daub:2015}. At higher energies experimental data indicate that further 
inelasticities are usually accompanied by resonances. We thus derive a parametrization that allows for resonance exchange at higher energies. Those resonances also act as doorways for the coupling of the system to the additional channels.
At the same time we make sure that their presence does not distort the amplitude at lower energies.
To be concrete, here we consider in addition to $\pi\pi$ (channel 1) and $K\bar{K}$ (channel 2) an effective $4\pi$ channel (channel 3), modeled by either $\rho\rho$ or $\sigma\sigma$. 
Three-channel models with an effective $\sigma\sigma$ channel have been considered in the literature before~\cite{Kaminski:1997gc,Moussallam:1999aq}, while some of the $f_0$ states between $1.3$ and $2\,\GeV$ have even been hypothesized to be dynamically generated by attractive interactions between $\rho$ mesons~\cite{Molina:2008jw,Geng:2008gx,Gulmez:2016scm,Du:2018gyn}.
It should become clear from the derivation, however, that the formalism allows for the inclusion of additional channels in a straightforward 
manner.

The derivation starts from the scalar isoscalar scattering amplitude $T(s)_{if}$, where $i$ and $f$ denote the initial- and final-state channels. To implement unitarity and analyticity we use the Bethe--Salpeter equation, which reads
\begin{align}
	T_{if}=\left(V+V G T\right)_{if}=V_{if}+V_{im} G_{mm} T_{mf} \label{eq::Formalism::BetheSalpeter}
\end{align}
in operator form.
Here $V_{if}$ denotes the scattering kernel of the initial channel $i$ into the final channel $f$. The loop operator $G$ is diagonal in channel space
and provides the free propagation of the particles of channel $m$.
For example, at the one-loop level the above equation generates  an expression of the form
\begin{align}
	V_{i1} G_{11} V_{1f} &\propto \int\frac{\diff^4k}{(2\pi)^4} V_{i1}(k,\dots)\,\frac{i}{k^2-M_\pi^2+i\epsilon}\nl
	&\quad \times \frac{i}{(k-P)^2-M_\pi^2+i\epsilon}\,V_{1f}(k,\dots) 
\end{align}
for $\pi\pi$ rescattering, with $P$ being the total 4-momentum of the system such that $P^2=s$. 
For $m=1,\,2$, the discontinuity of the loop operator element $G_{mm}$ reads
\beq
\disc G_{mm}=2i\sigma_m \,,
\label{G1122def}
\eeq
where $\sigma_m(s) = \sqrt{1-4M_m^2/s}$ is the two-body phase space in the given channel,
and $M_m$ denotes the pion and kaon masses for channels 1 and 2, respectively.
For the third channel, we need to include the finite width of the two intermediate ($\rho$ and $\sigma$) mesons; we write
\begin{align}
\disc G_{33}^k &=2i \int_{4M_\pi^2}^\infty \diff m_1^2\,\diff m_2^2\,\rho_k(m_1^2)\,\rho_k(m_2^2) \nl
& \qquad\qquad \times \frac{\lambda^{1/2}(s,m_1^2,m_2^2)}{s} \, ,
\label{G33def}
\end{align}
where $\lambda(a,b,c)=a^2+b^2+c^2-2(ab+ac+bc)$ is the K\"all\'en function.
Here the spectral density for the state $k$, $\rho_k(m^2)$, is given as 
\begin{align}
	\rho_k(q^2)=\frac{1}{\pi}\,\frac{m_k \Gamma_k(q^2)}{(q^2-m_k^2)^2+m_k^2 \,\Gamma_k^2(q^2)} \,,
\end{align}
with the energy-dependent width 
\begin{align}
\Gamma_k(s) &=\frac{\Gamma_k \,m_k}{\sqrt{s}}\,\left(\frac{p_\pi(s)}{p_\pi(m_k^2)}\right)^{2L_k+1}\left(F_R^{(L_k)}(s)\right)^2\,, \nl
p_\pi(s)  &= \frac{\sqrt{s}}{2}\sigma_\pi(s) \,, \label{eq::Decay::width}
\end{align} 
where $\Gamma_k$ ($m_k$) denotes the nominal width (mass) of the resonance and $L_k$ the angular momentum
of the decay with $L_k=1$ and $0$ for the $\rho$ and the $\sigma$, respectively. 
The $F_R^{(L)}(s)$ denote barrier factors that prevent the width from growing continuously. 
We employ the parametrization of Refs.~\cite{Blatt:1952ije,LHCb:2012ae}, where their explicit forms for $L=0,\,1,\,2$ are given by
\begin{align}
	F_R^{(0)}=1\,,~ F_R^{(1)}=\sqrt{\frac{1+z_0}{1+z}}\,,~ F_R^{(2)}=\sqrt{\frac{9+3z_0+z_0^2}{9+3z+z^2}} \,,\label{eq::Formalism::BlattWeisskopf}
\end{align}
with $z=r_R^2 \,p_\pi^2(s)$, $z_0 = r_R^2 \,p_\pi^2(m_k^2)$, and
the hadronic scale $r_R=1.5\,\GeV^{-1}$. 
Note that as long as no exclusive data are employed for the $4\pi$ final state, the amplitudes
are not very sensitive to the details how, e.g., the spectral density of the $\sigma$ meson is parametrized, since it enters
only as the integrand in the self energies of the resonances.
 However, the analysis is somewhat sensitive to the differences between
a $\rho\rho$ and a $\sigma \sigma$ self energy, since the energy dependence of the two is quite different, given the different resonance parameters
and the different threshold behavior. We come back to this discussion later in this section.

\begin{figure*}
	\includegraphics[width=0.49\linewidth]{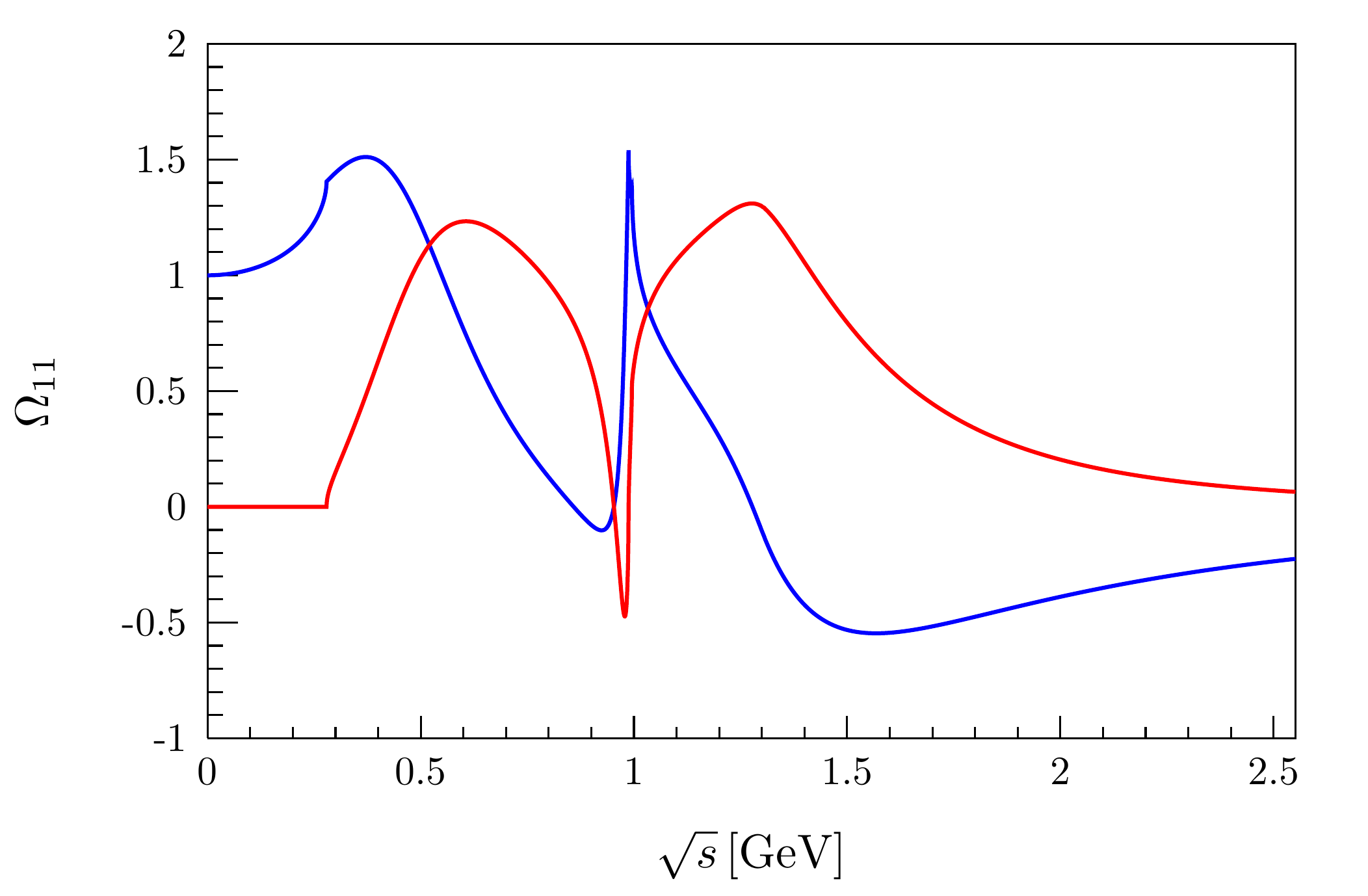}\hfill
	\includegraphics[width=0.49\linewidth]{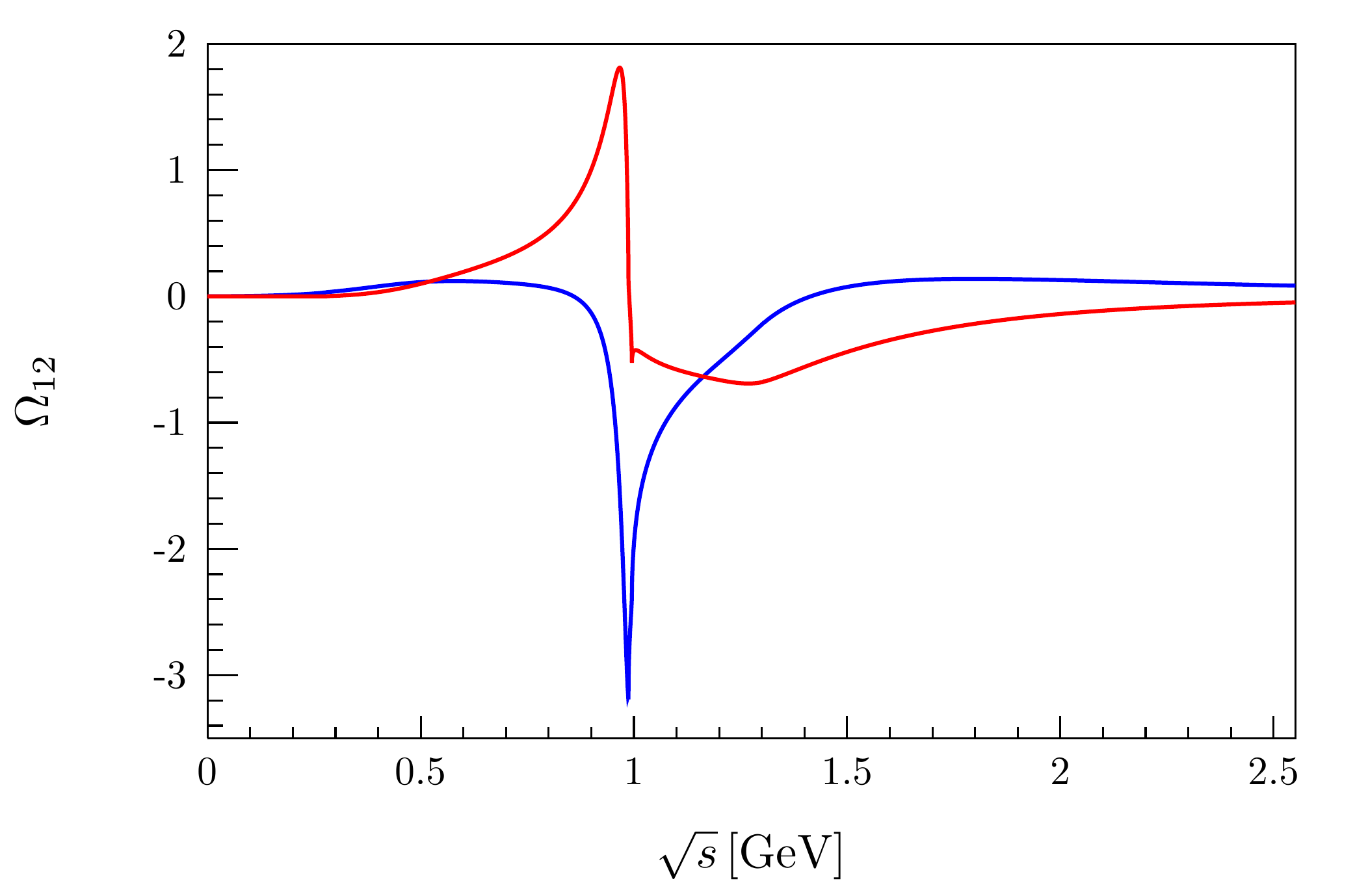}\\
	\includegraphics[width=0.49\linewidth]{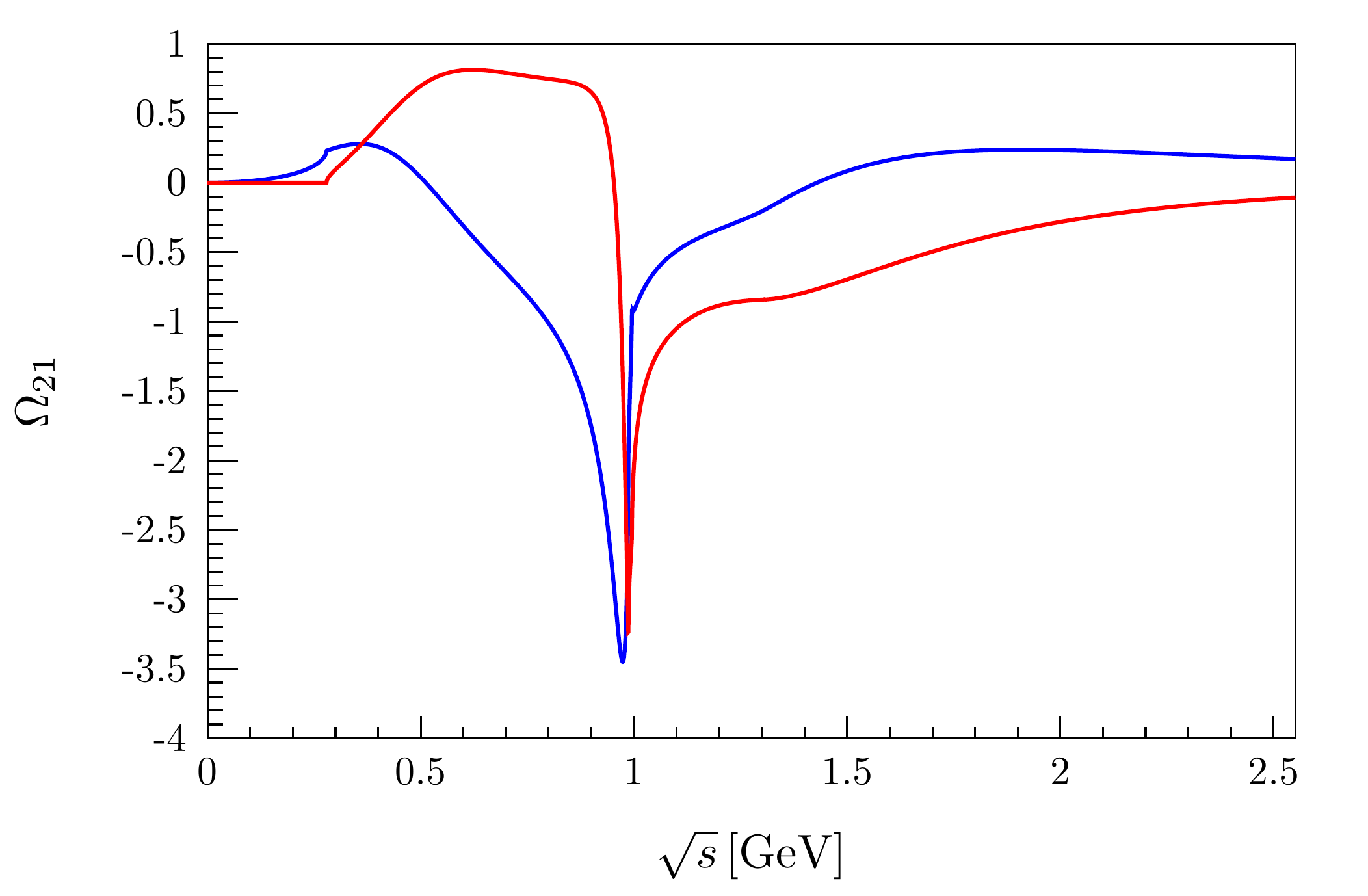}\hfill
	\includegraphics[width=0.49\linewidth]{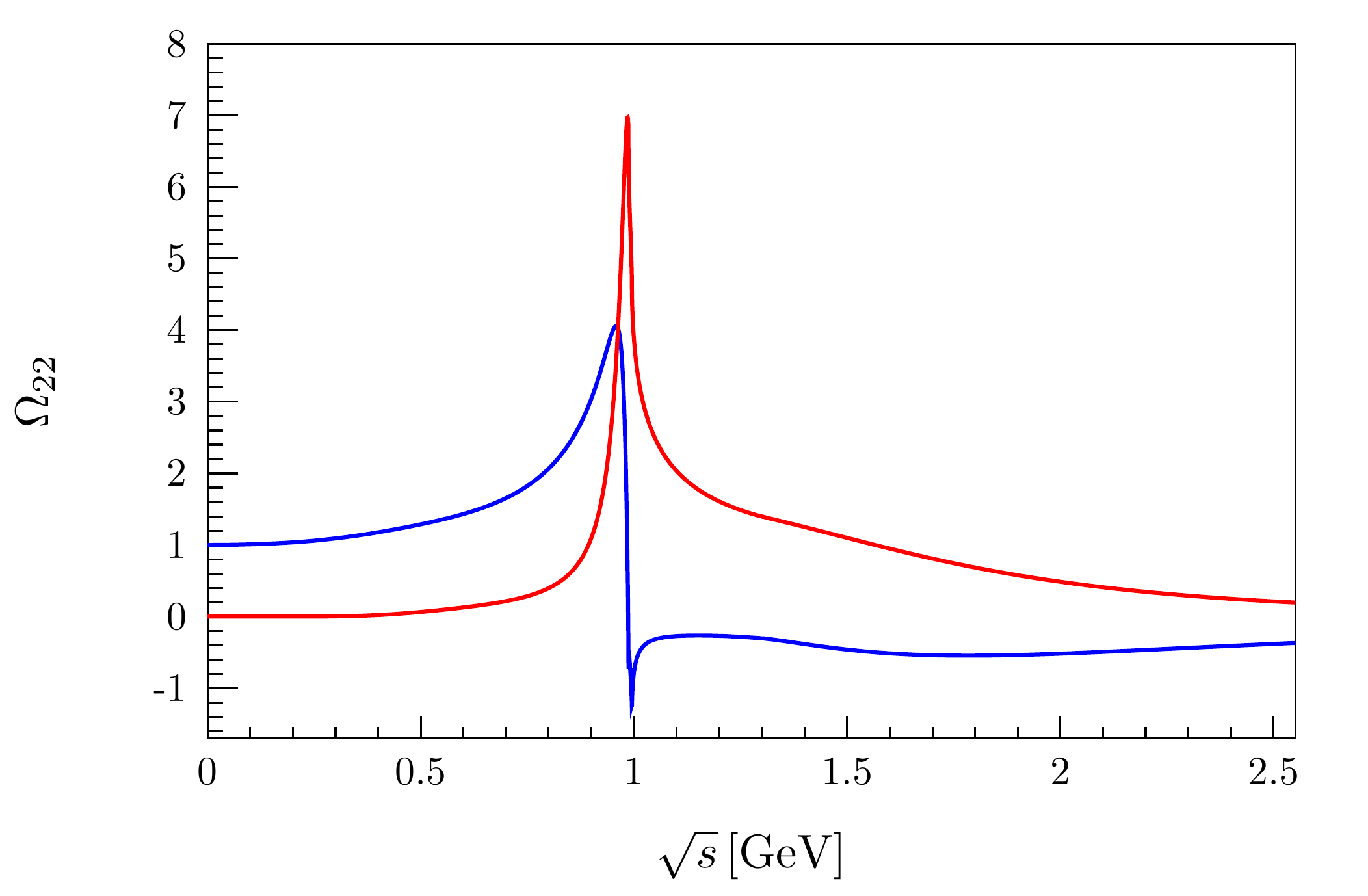}
\caption{Real (blue) and imaginary (red) parts of the Omn\`es matrix elements $\Omega_{11}$, $\Omega_{12}$, $\Omega_{21}$, and $\Omega_{22}$.}
\label{Fig::Parametrization::Omnes}
\end{figure*}

To proceed with the derivation we split the scattering kernel into two parts, $V=V_0+V_R$, conceptually following the so-called
two-potential formalism~\cite{Nakano:1982bc}. 
The effect of $V_0$ will eventually be absorbed into the dispersive piece fixed by the low-energy $\pi\pi$--$K\bar K$ $T$-matrix
input. Its explicit form is needed at no point; one may think of it as the 
driving term of a  Bethe--Salpeter equation 
\beq
T_0=V_0+V_0\,G\,T_0 \,. 
\label{T0def}
\eeq
Since $T_0$ it is the solution of a scattering equation,
$T_0$ is unitary. In particular, we may write
\begin{align}
	T_0=\begin{pmatrix}
	\frac{\eta_0 e^{2i\delta_0}-1}{2i\sigma_\pi} & g_0 e^{i\psi_0} & 0\\
	g_0 e^{i\psi_0} & \frac{\eta_0 e^{2i\left(\psi_0-\delta_0\right)}-1}{2i\sigma_K} & 0\\
	0 & 0& 0
	\end{pmatrix}\,,
	\label{eq:inputdelandin}
\end{align}
where $\delta_0$ is the scalar isoscalar $\pi\pi$ phase shift, $\psi_0$ the phase of the $\pi\pi \to K\bar{K}$ scattering amplitude, and $g_0$ its absolute value. The inelasticity $\eta_0$ is related to $g_0$ via
\begin{align}
	\eta_0=\sqrt{1-4 \left(g_0\right)^2 \sigma_\pi\,\sigma_K\,\Theta\left(s-4M_K^2\right)}\,.
\end{align}
The effects of resonances heavier than the $f_0(980)$ enter the amplitude via $V_R$. By means of $V_R$ we can construct 
the resonance $T$-matrix $T_R$, related to the full $T$-matrix via $T=T_R+T_0$.
Since $T_0$ is unitary by itself, $T_R$ cannot be independent of $T_0$ in order to respect the Bethe--Salpeter equation~\eqref{eq::Formalism::BetheSalpeter}. Solving for $T_R$ we obtain 
\beq
	\left(1-V_0G-V_RG\right)T_R=V_R\left(1+GT_0\right)\,.
\eeq
To proceed, we define the vertex function $\Omega$ via 
\beq
\Omega=1+T_0 G \,. 
\label{vertexdef}
\eeq
Its discontinuity is given by
\begin{align}
	\disc\Omega_{ij}=2i\,\left(T_0\right)^*_{im}\,\sigma_m\,\Omega_{mj}\,,
\end{align} 
which agrees with the discontinuity of the Omn\`{e}s matrix derived from the scattering
$T$-matrix $T_0$~\cite{Muskhelishvili,Omnes}. 
Therefore it can be constructed from dispersion theory:
\begin{align}
	\Omega=\begin{pmatrix}
	\Omega_{11} & \Omega_{12} & 0\\
	\Omega_{21} & \Omega_{22} & 0\\
	0 & 0 & 1
	\end{pmatrix} ,~
\Omega_{ij}(s)=\frac{1}{2\pi i}\int_{4M_\pi^2}^\infty \diff z \frac{\disc\Omega_{ij}(z)}{z-s-i\epsilon}\,.
\end{align}
Numerical results for $\Omega_{ij}(s)$ based on the $T$-matrix of Ref.~\cite{Dai:2014zta} 
are shown in Fig.~\ref{Fig::Parametrization::Omnes}. One observes in particular the signature
of the $f_0(500)$ or $\sigma$-meson, i.e.\ the broad bump in the imaginary part of $\Omega_{11}(s)$
below $1\,\GeV$, accompanied by a quick variation of the real part, which clearly cannot be parametrized
by a Breit--Wigner form. For an earlier discussion about this fact see Ref.~\cite{Gardner:2001gc}.

\begin{figure*}
	\includegraphics[width=0.49\linewidth]{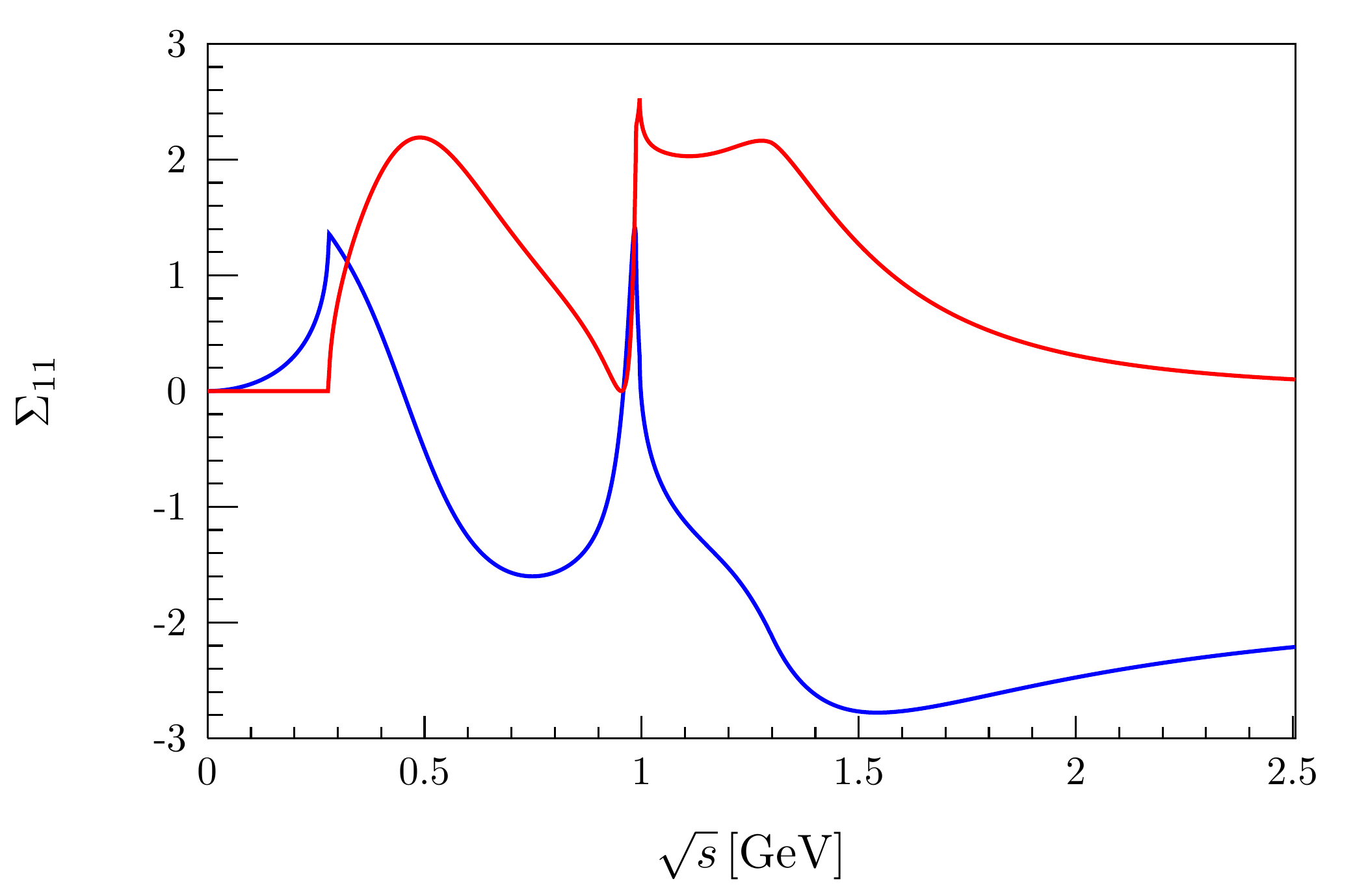} \hfill
	\includegraphics[width=0.49\linewidth]{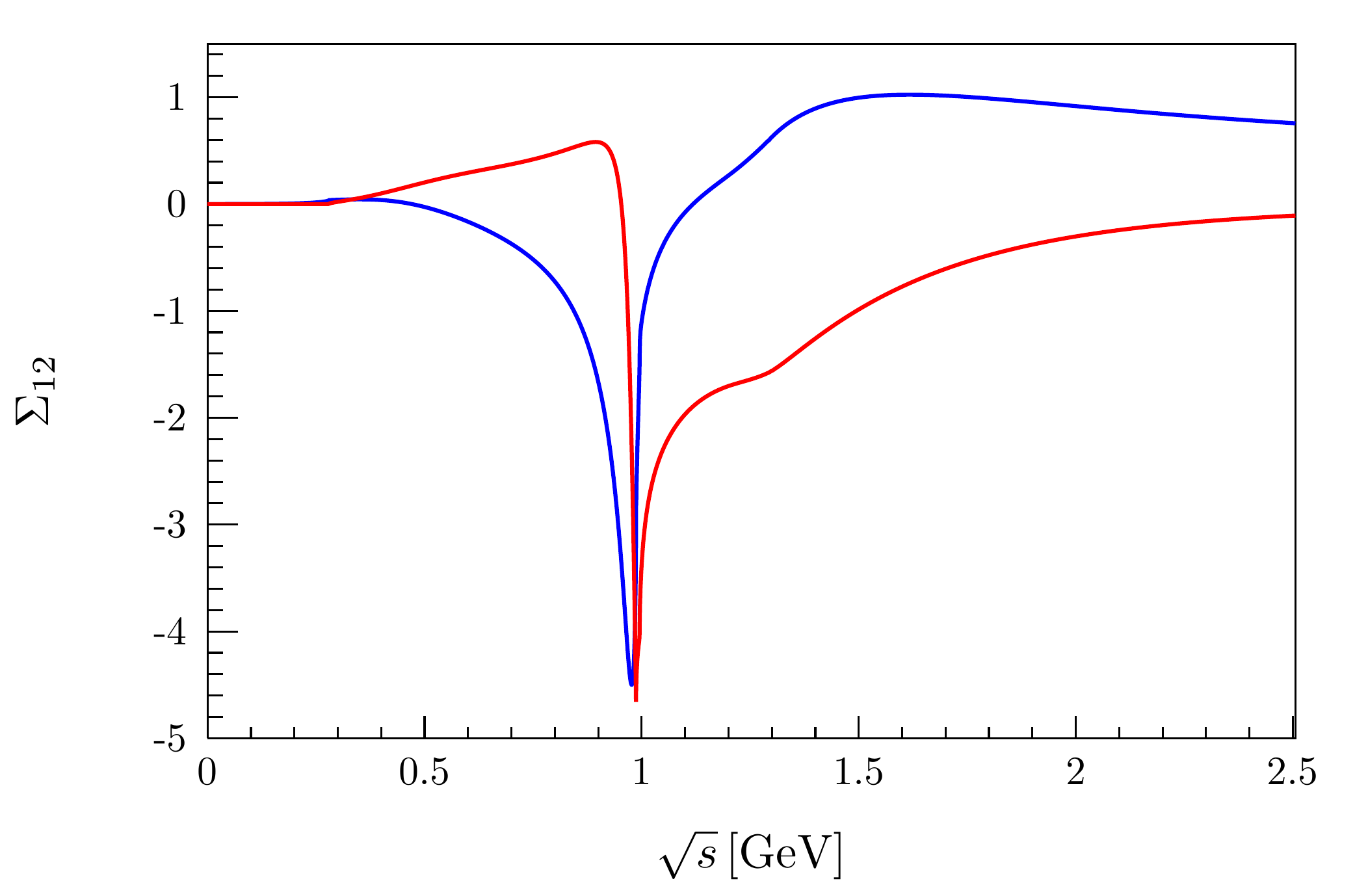} \\
	\includegraphics[width=0.49\linewidth]{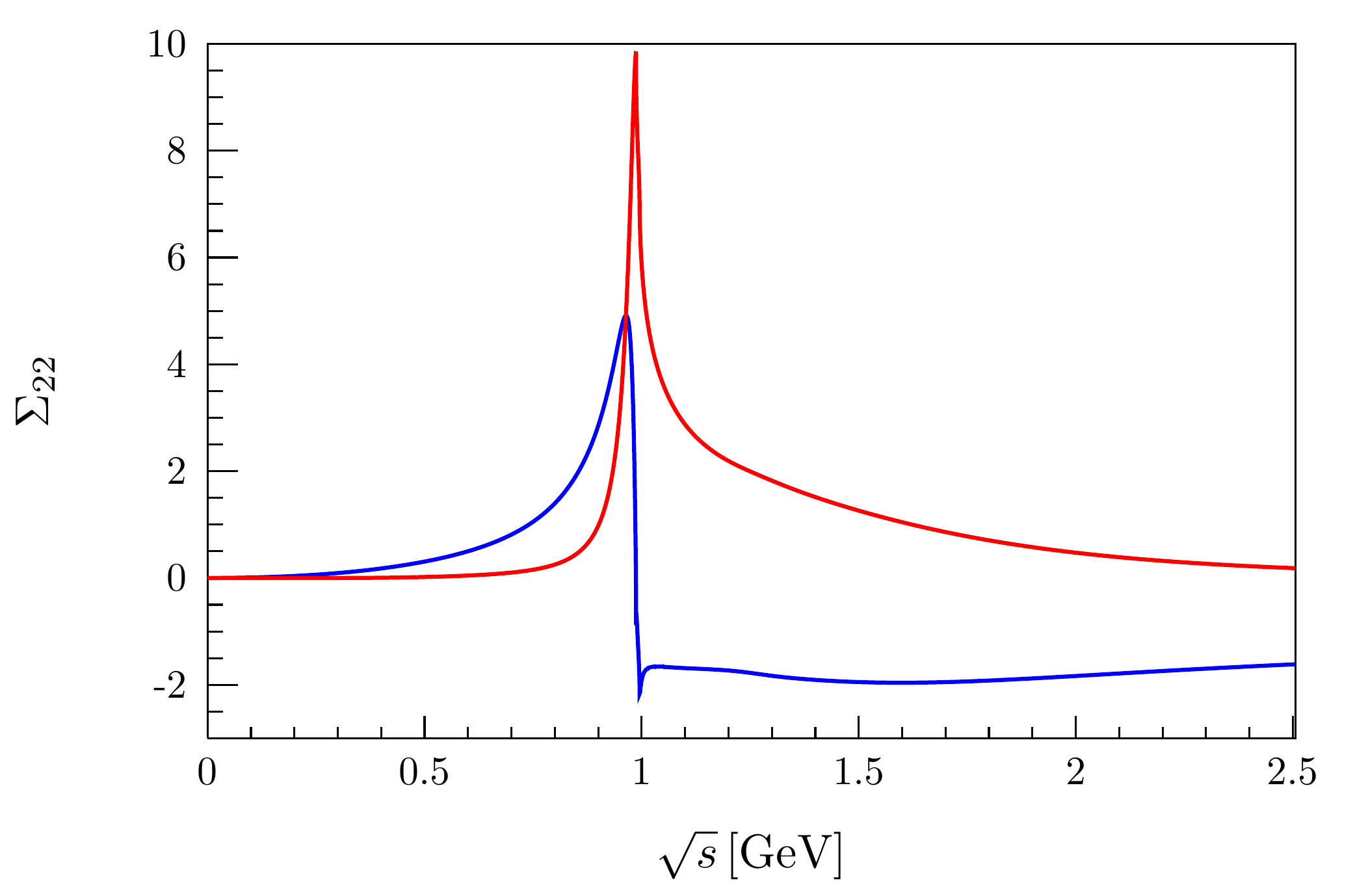} \hfill
	\includegraphics[width=0.49\linewidth]{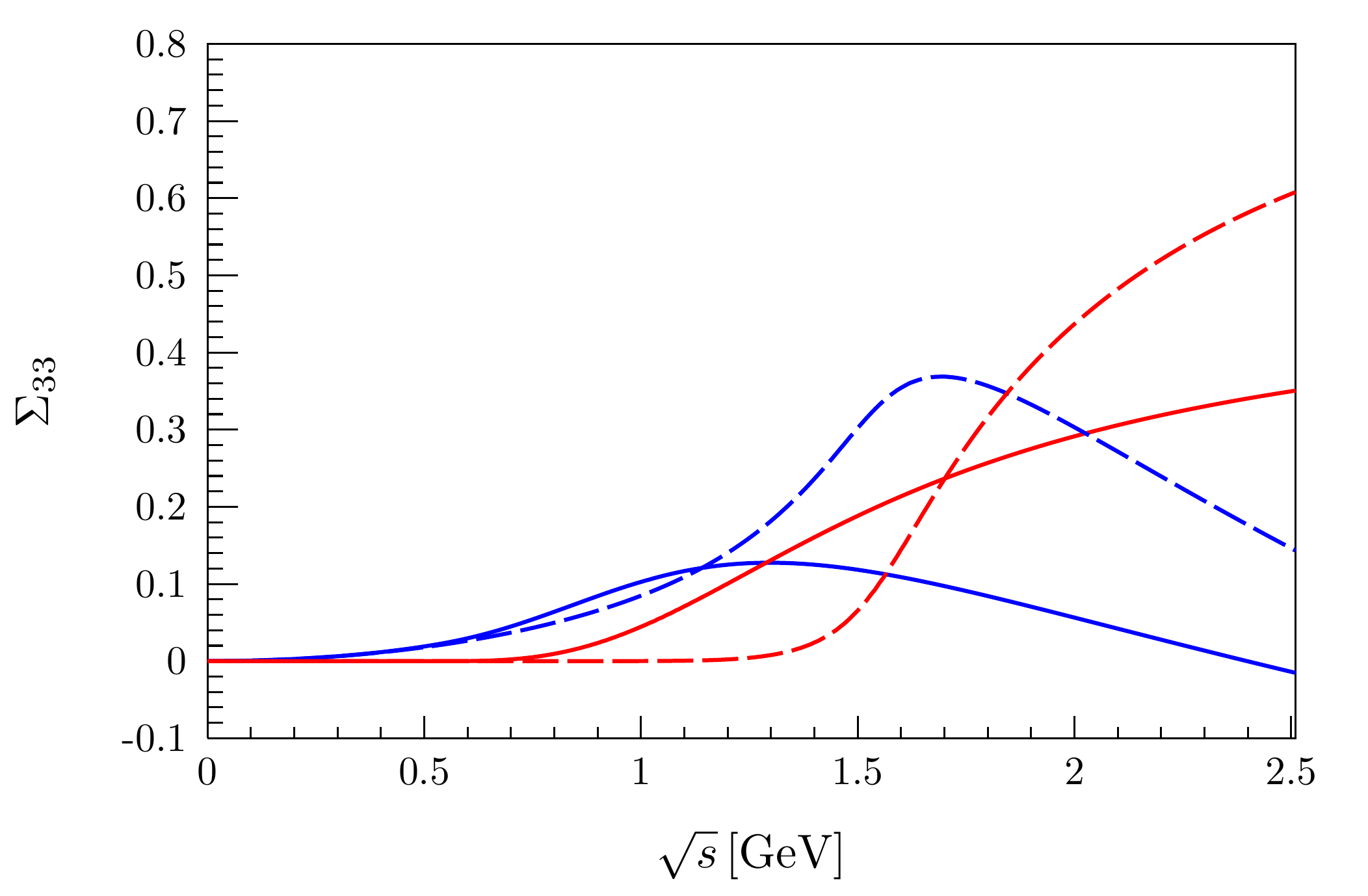}
\caption{Real (blue) and imaginary (red) parts of the self-energy functions $\Sigma_{11}$, $\Sigma_{12}=\Sigma_{21}$, $\Sigma_{22}$, and $\Sigma_{33}$, using the Omn\`es matrix displayed in Fig.~\ref{Fig::Parametrization::Omnes}. Note that $\Sigma_{33}$ is a once-subtracted dispersion integral over the four-particle phase space factor taken as a $\sigma\sigma$ (solid) or a $\rho\rho$ (dashed) state.}
\label{Fig::Parametrization::SelfEnergy}
\end{figure*}

Using $T_R=\Omega\,t_R\,\Omega^t$ and $V_0G\Omega=\Omega-1$, which follows 
from inserting Eq.~\eqref{T0def} into Eq.~\eqref{vertexdef}, one obtains  a Bethe--Salpeter equation for $t_R$,
\beq
t_R=V_R+V_R\, \left(G\Omega\right)\,t_R\,. \label{eq:tR}
\eeq
Note that Eq.~\eqref{eq:tR} does not depend on $V_0$ explicitly.
It appears only implicitly, since 
the loop operator $G$, describing the free propagation of the two-meson states, needs to be replaced 
by the dressed loop operator $(G\Omega)$, which describes the propagation of the two-meson state
in the presence of the interaction $T_0$, in order to preserve unitarity. 
The discontinuity of this self-energy matrix $\Sigma=G\Omega$ is given by
\begin{align}
	\disc \Sigma_{ij}(s)=\Omega_{im}^\dagger(s)\, \disc G_{mm}(s) \,\Omega_{mj}(s)\, .
	\label{discSigma}
\end{align}
The discontinuities of the loop functions for the two--body channels and the $4\pi$ channel were
given in Eqs.~\eqref{G1122def} and \eqref{G33def}, respectively.
Equation~\eqref{discSigma} allows us to write $\Sigma$ as a once-subtracted dispersion integral,
\beq
\Sigma_{ij}(s) = \Sigma_{ij}(0) +\frac{s}{\pi}\int \frac{\diff z}{z}\frac{\disc \Sigma_{ij}(z)}{z-s-i\epsilon} \,.
\label{sigmadef}
\eeq
The resulting self-energy functions $\Sigma_{ij}(s)$ are displayed in Fig.~\ref{Fig::Parametrization::SelfEnergy}. 
The subtraction constants can be absorbed in a redefinition of the yet undefined potential $V_R$.
Please observe that the component $\Sigma_{33}$ looks very different for the two different model
assumptions employed. For example,  the self energy from $\sigma\sigma$ intermediate states rises very
quickly right from the $4\pi$ threshold, while the one for $\rho\rho$ sets in significantly later.
This difference reflects that the $\sigma$ decays into two pions in an $S$-wave, while the $\rho$ decays in a $P$-wave. 
On the other hand, since the discontinuity of $G_{33}$ enters in the expression for $\Sigma_{33}$
only as the integrand, this component of the self energy is not very sensitive to the details of the concrete parametrizations
employed for the spectral functions.

The full solution for the scattering matrix is thus given by
\begin{align}
	T=T_0+\Omega\left[1-V_R\Sigma\right]^{-1} V_R \Omega^t\,.\label{eq::Formalism::TMatrix}
\end{align}
In order to obtain a parametrization for the form factor, we adapt the $P$-vector formalism~\cite{Aitchison:1972ay}
to the system at hand.
 The isoscalar scalar form factor $\Gamma^{s}_i$ is written as
\begin{align}
	\Gamma^{s}_i=M_i+T_{ij} G_{jj} M_j\,,
\end{align}
where $M_i$ is an analytic term describing the transition from the source to the channel $i$. Inserting the parametrization of Eq.~\eqref{eq::Formalism::TMatrix} we obtain, after some straightforward algebra,
\begin{align}
	\Gamma^{s}_i=\Omega_{im}\left[1-V_R\Sigma\right]^{-1}_{mn} M_{n}\,.\label{eq::Formalism::Formfactor}
\end{align}
As $T_0$ captures the physics in the $\pi\pi$ and $K\bar K$ channels at energies below $1\,\GeV$ including the $f_0(500)$, the $f_0(980)$, and the impact of the corresponding 
left-hand cuts (left-hand cuts in the other channel(s) are neglected by construction), 
the potential $V_R$ should predominantly describe the resonances above $1\,\GeV$. In order to reduce their impact at low energies, we subtract $V_R$ at $s=0$ and arrive at
\begin{align}
	\left(V_R\right)_{ij}= \sum_{r} g^r_{i} \,\frac{s}{m_r^2\left(m_r^2-s\right)}\, g_{j}^r\,.
\label{eq::Formalism::subtractedPotential}
\end{align}
The bare resonance masses, $m_r$, as well as the bare resonance--channel coupling constants, $g_i^r$, 
are free parameters that need to 
be determined by a fit to data. The subtraction constants are effectively absorbed into $T_0$ that
by construction captures all physics close to $s=0$.

The most general ansatz for $M$ reads
\begin{align}
	M_i=c_i+\gamma_i\,s+\cdots - \sum_r g_i^r\,\frac{s}{m_r^2-s}\,\alpha_r\,,\label{eq::Formalism::source}
\end{align}
where the parameters $c_i =\Gamma^{s}_i(0)$ provide the normalizations of the
different form factors. Here the isospin Clebsch--Gordan coefficients were absorbed into the definition of the 
form factors. This means explicitly
\begin{align}
	\Gamma_2^s\rightarrow\frac{2}{\sqrt{3}}\Gamma_2^s\quad\text{and}\quad M_2\rightarrow \frac{2}{\sqrt{3}}M_2 \,,
\end{align}
while for the third channel we absorb these factors into the coupling constants. 
The bare resonance masses and the corresponding couplings $g_i^r$ are the same
as before. The parameters $\alpha_r$, which quantify the resonance--source couplings, and the slope parameters $\gamma_i$ 
are additional free parameters.

This completely defines the formalism. Clearly, the number of inelastic channels can be extended in a straightforward way,
however, for the concrete application studied in the following section, three channels turn out to be sufficient as long as
no exclusive data for additional channels become available.
\esp

\section{Application: $\bar{B}_s^0\rightarrow J/\psi\,\pi^+\pi^-$ and $\bar{B}_s^0\rightarrow J/\psi\,K^+K^-$}\label{sec::application}
\subsection{Parametrization of the decay amplitudes}\label{sec::decay}

\begin{figure}
\centering
\includegraphics[width=0.7\linewidth]{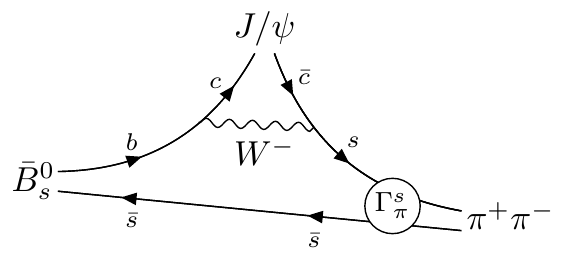}
\caption{Tree-level $W$-exchange diagram for the decay process $\bar{B}_s^0\rightarrow J/\psi\,\pi^+\pi^-$. The hadronization of the $\bar{s}s$ quark pair into $\pi^+\pi^-$ ($S$-wave dominated) is given by the scalar form factor $\Gamma^s_\pi$.}
\label{Fig::Decay_Process}
\end{figure}

\bsp
As an example, we now apply the formalism introduced in the previous section to
the decays $\bar{B}_s^0\rightarrow J/\psi\,\pi^+\pi^-(K^+K^-)$, analyzing data taken by the LHCb collaboration~\cite{LHCb_PiPi,LHCb_KK}. 
The dominant tree-level diagram for the corresponding weak transition on the quark level is displayed in Fig.~\ref{Fig::Decay_Process}.

It has been argued previously~\cite{Daub:2015,Albaladejo:2016mad} that the $S$-wave projection of the appropriate helicity-0 amplitude
for $\bar{B}_s^0\rightarrow J/\psi\,M_1M_2$ transitions are proportional to the corresponding strange scalar form factors of the 
light dimeson system $M_1M_2$; in particular, there are chiral symmetry relations between the different dimeson channels that fix the 
\textit{relative} strengths to be equal to those of the matrix elements in Eq.~\eqref{eq:scalarFF-def} at leading order 
in a chiral expansion~\cite{Albaladejo:2016mad}.
We conjecture here that the same will still hold true for the inclusion of the effective third ($4\pi$) channel.  In this sense, 
the $\bar{B}_s^0$ decays allow to test the pion and kaon strange scalar form factors, up to a common overall normalization.

A previous dispersive analysis~\cite{Daub:2015}, which considered the $\pi\pi$--$K\bar{K}$ coupled-channel system, 
worked well in the energy region up to $1.05\,\GeV$. However, due to higher resonances and the onset of additional inelasticities 
the framework could not be applied beyond this energy. Our new parametrization allows us to overcome this limitation,
while it guarantees at the same time a smooth matching onto the amplitudes employed in Ref.~\cite{Daub:2015}.
The data are provided in terms of angular moments $Y_L^0(\sqrt{s})$, which are given as angular averages of the differential decay rates
\begin{align}
	\left<Y_L^0\right>=\int \diff\cos\Theta\,\frac{\diff\Gamma}{\diff\sqrt{s}\,\diff\cos\Theta}\,Y_L^0(\cos\Theta)\,,\label{eq::Decay::AngularMoments}
\end{align}
where $\Theta$ is the scattering angle between the momentum of the dipion system in the $\bar{B}_s^0$ rest frame and the momentum of one of the pions. We express the decay amplitude in terms of the partial-wave-expanded helicity amplitudes $\mathcal{H}^L_\lambda$, where $L$ denotes the angular momentum of the pion or kaon pair, and $\lambda = 0,\parallel,\perp$ refers to the helicity of the $J/\psi$.  The angular moments are then given as
\beq
	\left<Y_0^0\right>=\frac{p_\psi p_\pi}{\sqrt{4\pi}} \bigg\{\left|\mathcal{H}_0^0\right|^2 + \sum_{\lambda=0,\parallel,\perp}\left( \left|\mathcal{H}_\lambda^1\right|^2 +  \left|\mathcal{H}_\lambda^2\right|^2\right)\bigg\}\label{eq::Y00}
\eeq
and
\begin{align}
\left<Y_2^0\right>&=\frac{p_\psi p_\pi}{\sqrt{4\pi}}\bigg\{2 \Re\left[\mathcal{H}_0^0 \left(\mathcal{H}_0^2\right)^*\right] \nl
& \quad +\frac{1}{\sqrt{5}}\left[2 \left|\mathcal{H}^1_0\right|^2 
-\left|H^1_\parallel\right|^2-\left|H^1_\perp\right|^2\right] \nl
& \quad +\frac{\sqrt{5}}{7}\left[2\left|\mathcal{H}_0^2\right|^2+\left|\mathcal{H}_\parallel^2\right|^2+\left|\mathcal{H}_\perp^2\right|^2 \right]\bigg\} \label{eq::Y02}
\end{align}
for the moments of relevance for this work; see Refs.~\cite{Daub:2015,Albaladejo:2016mad} for details.
In addition to the pion momentum in the dipion rest frame $p_\pi$ introduced earlier, we also use the $J/\psi$ momentum in the $\bar B_s^0$ rest frame, $p_\psi=\lambda^{1/2}(s,M_\psi^2,m_B^2)/(2m_B)$.

The scalar helicity amplitude $\mathcal{H}_0^0$ can be related to the scalar isoscalar form factor $\Gamma^{s}_{i}$ as
\begin{align}
	\mathcal{H}_0^0=\mathcal{N}p_\psi m_B \Gamma^{s}_i \,,\label{Eq::Decay::Helicity0} 
\end{align}
where the normalization factor $\mathcal{N}$ absorbs weak coupling constants and the pertinent Wilson coefficients,
as well as meson mass factors and decay constants~\cite{Daub:2015,Albaladejo:2016mad}. 
Here $i$ denotes the relevant channel. For the form factors we use the parametrization introduced in Sect.~\ref{sec:formalism}. Since the main focus of our analysis lies on the $S$-waves, we approximate the $P$- and $D$-waves as Breit--Wigner functions~\cite{Zhang:2012},
\begin{align}
\frac{\mathcal{H}^L_\lambda}{\sqrt{2L+1}}&=w_\lambda^L\sum_R h_\lambda^R\,e^{i\phi_\lambda^R}\mathcal{A}_{R}\nl
& \qquad\qquad \times F_B^{(J)} F_R^{(L)} \Big(\frac{p_\psi}{m_B}\Big)^{J} \Big(\frac{p_\pi}{\sqrt{s}}\Big)^L \,,
\end{align}
for $L\geq 1$.
The free parameters introduced here are the strength $h_\lambda^R$ of the resonance $R$ with helicity $\lambda$, its phase $\phi_\lambda^R$, and a total rescaling factor $w_\lambda^L$ for the helicity amplitude $\mathcal{H}_\lambda^L$. The factors $F_B^{(J)}$ and $F_R^{(L)}$ are the Blatt--Weisskopf factors of Eq.~\eqref{eq::Formalism::BlattWeisskopf}. Two different scales are employed therein: while $F_B^{(J)}$ depends on the argument $z=r_B^2 \, p_\psi^2$ with $r_B=5.0\,\GeV^{-1}$, for $F_R^{(L)}$ we use $z=r_R^2 \,p_\pi^2$ with $r_R=1.5\,\GeV^{-1}$ as in Eq.~\eqref{eq::Formalism::BlattWeisskopf}~\cite{LHCb:2012ae}. The position as well as width of the corresponding resonance is then included in the Breit--Wigner function
\begin{align}
	\mathcal{A}_R(s)=\frac{1}{m_R^2-s-i m_R\, \Gamma_R(s)}
\end{align}
with an energy-dependent width $\Gamma_R(s)$~\eqref{eq::Decay::width}.
Since the only interference term in the angular moments considered, Eqs.~\eqref{eq::Y00} and \eqref{eq::Y02},
is the $S$-$D$-wave interference in $\left<Y_2^0\right>$, our fits are only sensitive to the relative phase motion of $\mathcal{H}_0^0$ and $\mathcal{H}_0^2$. 
To reduce the total number of free parameters for all partial waves except the $S$-wave,
we fix the resonance masses $m_R$ as well as their respective widths to the central values found in Refs.~\cite{LHCb_PiPi,LHCb_KK}. 
Furthermore we fix both $h_\lambda^R$ as well as $\phi_\lambda^R$ with $\lambda=\parallel,\perp$ to the central values of the LHCb fits. 
However, since the phase motion of our $S$-wave will be different from the one 
of the LHCb parametrization~\cite{Daub:2015},
we allow $w_\lambda^R$ to vary. For the helicity amplitude $\mathcal{H}_0^2$ we keep both $h_0^R$ as well as $\phi_0^R$ flexible. To avoid unnecessary parameters we set $w_0^2=1$. The number of free parameters is discussed in more detail in Sect.~\ref{sec:fit}.
\esp

\subsection{Fits to the decay data}\label{sec:fit}
\bsp
In this section we discuss the fit using the form factor parametrization of Eq.~\eqref{eq::Formalism::Formfactor} to the data measured for $\bar{B}_s^0\rightarrow J/\psi \pi^+\pi^-$~\cite{LHCb_PiPi} and $\bar{B}_s^0\rightarrow J/\psi K^+ K^-$~\cite{LHCb_KK}, which are
 presented as angular moments related to the helicity amplitudes via Eqs.~\eqref{eq::Y00} and~\eqref{eq::Y02}.
Note that these angular moments have an arbitrary normalization and need to be rescaled to their physical values. The integrated partial width is given by
\begin{align}
\Gamma\left(\bar{B}_s^0\rightarrow J/\psi\, h^+h^-\right) &=\int\diff\sqrt{s}\,\diff\cos\Theta\,\frac{\diff\Gamma}{\diff\sqrt{s}\,\diff\cos\Theta} \nl
	&=\sqrt{4\pi}\int\diff\sqrt{s}\left<Y_0^0\right>\,.
\end{align}
The correctly normalized angular moments, $\left<Y_L^0\right>_\mathrm{norm}$, can be obtained from those published, 
$\left<Y_L^0\right>_\mathrm{LHCb}$, by
\begin{align}
	\left<Y_L^0\right>_{\mathrm{norm}}=\frac{\Gamma\left(\bar{B}_s^0\rightarrow J/\psi\,h^+h^-\right)}{\sqrt{4\pi}\int\diff\sqrt{s}\,\left<Y_0^0\right>_{\mathrm{LHCb}}}\,\left<Y_L^0\right>_{\mathrm{LHCb}}\,.
\end{align}
We determine the partial decay rates $\Gamma\left(\bar{B}_s^0 \rightarrow J/\psi\,h^+ h^-\right)$ via the total decay rate $\Gamma_{\bar{B}_s^0 } =\tau_{\bar{B}_s^0}^{-1}$ with~\cite{PDG} 
\beq
\tau_{\bar{B}_s^0}=\left(1.509\pm0.004\right)10^{-12}\,\mathrm{s} 
\eeq
and the branching ratios~\cite{PDG}
\begin{align}
\BR\left(\bar{B}_s^0\rightarrow J/\psi\,\pi^+\pi^-\right) &=\left(2.09\pm 0.23\right)\times 10^{-4} \,, \nl
\BR\left(\bar{B}_s^0\rightarrow J/\psi\,K^+K^-\right) &=\left(7.9\pm0.7\right)\times 10^{-4} \,.
\end{align}

The dispersive approach using the Omn\`es matrix already
captures the physics of the $f_0(500)$ and $f_0(980)$ resonances. 
In order to extend the description further, we use $N_R$ additional resonances. 
As outlined above, the $S$-wave contains in total up to $(N_c+N_s+1)N_R+2N_cN_s$ parameters,
where $N_c$ ($N_s$) denotes the number of channels (sources) included; in this study $N_s=1$, $N_c=3$, and
$N_R$ is either 2 or 3, depending on the fit. The last term in
the sum above comes from the non-resonant couplings of the system to the source.
The number of those parameters can be reduced from the observation
that the normalizations of the pion and the kaon form factors can be fixed
to $c_1=0$ and $c_2=1$~\cite{Daub:2015}. 
Since the four-pion channel is expected to couple similarly weakly to an $\bar{s}s$ source as the 
two-pion one (given OZI suppression at $s=0$), we also set $c_3=0$. Thus the only free parameter from the
constant terms in the sources $M_i$ can be absorbed into the overall
 normalization $\mathcal{N}$ introduced in Eq.~\eqref{Eq::Decay::Helicity0}.
 Below we present fits without
($\gamma_i=0$, resulting in $5N_R$ parameters) as well as with linear terms in the production vertex defined in Eq.~\eqref{eq::Formalism::source}
($\gamma_i\neq 0$, providing three more free constants). 
\esp

For the decay $\bar{B}_s^0\rightarrow J/\psi\,\pi\pi$ the dipion system is in an isoscalar configuration; 
due to Bose symmetry the pions can therefore only emerge in even partial waves. 
Since we restrict ourselves to a precision analysis of the $S$-wave, we adopt the $D$-waves of Ref.~\cite{LHCb_PiPi}
and accordingly include two resonances, namely $f_2(1270)$ and $f_2^\prime(1525)$. For the $0$ polarization we introduce four new parameters given by the amplitude $h_0^R$ and $\phi_0^R$, while we fix $w_0^0=1$. For the other two helicity amplitudes we constrain $h_\lambda^R$ and $\phi_\lambda^R$ while keeping $w_\lambda^0$ variable. This gives another two free parameters. In total we obtain six additional free parameters.

Since $K^+$ and $K^-$ do not belong to the same isospin multiplet, they do not follow the Bose symmetry restrictions. Thus the $P$-wave in the decay 
$\bar{B}_s^0\rightarrow J/\psi\,K^+K^-$ is non-negligible and, in fact, dominant. It shows large contributions of the $\phi(1020)$ as well as of the
 $\phi(1680)$. Since the $P$-wave does not interfere with $S$- or $D$-waves in the angular moments $\left<Y_0^0\right>$ 
 and $\left<Y_2^0\right>$, we adopt the parameters of LHCb~\cite{LHCb_KK}. In order to allow for some flexibility, we also fit $w_\lambda^1$, resulting in 
 three parameters. The $D$-wave includes the resonances $f_2(1270)$, $f_2^\prime(1525)$, $f_2(1750)$, and $f_2(1950)$. For 
 $\lambda=0$ we fit both $h_0^R$ as well as $\phi_0^R$ with fixed $w_0^2=1$, resulting in eight free parameters. For the other
  helicity amplitudes we stick to the LHCb parametrization and keep $w_\lambda^2$ free, which results in two additional fit parameters. 
  Therefore in total we have $13$ additional free parameters for this channel.

All in all we have $5N_R+20(+3)$ free parameters for $\gamma_i=0$ ($\gamma_i\neq 0$).
Clearly this number is larger than the number of parameters of two single-channel Breit--Wigner analyses, however,
the advantage of the approach advocated here is that it allows for a combined analysis of all channels in a way
that preserves unitarity, and for a straightforward inclusion of the $4\pi$ channel in the analysis. 
Note that the scalar resonances studied here are known to have prominent decays into four pions~\cite{PDG};
cf.\ also theoretical approaches modeling some of them as dynamically generated $\rho\rho$ 
resonances~\cite{Molina:2008jw,Geng:2008gx,Gulmez:2016scm,Du:2018gyn}.

\begin{table}
\centering
\renewcommand{\arraystretch}{1.4}
	\begin{tabular}{ccc}
\toprule
${\chi^2}/{\mathrm{ndf}}$		& $\sigma\sigma$ & $\rho\rho$\\
\midrule
Fit~1 & $\frac{429.9}{384-30-1}=1.22$ & $\frac{376.2}{384-30-1}=1.07$\\
Fit~2 & $\frac{413.3}{384-33-1}=1.18$ & $\frac{361.4}{384-33-1}=1.03$\\
Fit~3 & $\frac{366.9}{384-35-1}=1.05$ & $\frac{335.4}{384-35-1}=0.96$\\
\bottomrule
	\end{tabular}\renewcommand{\arraystretch}{1.4}
	\caption{Reduced $\chi^2$ for the best fits. See main text for details.}
	\label{Tab::Fit::ReducedChi2}
\renewcommand{\arraystretch}{1.0}
\end{table}

\begin{figure*}[t!]
\centering
	\includegraphics[width=0.49\linewidth]{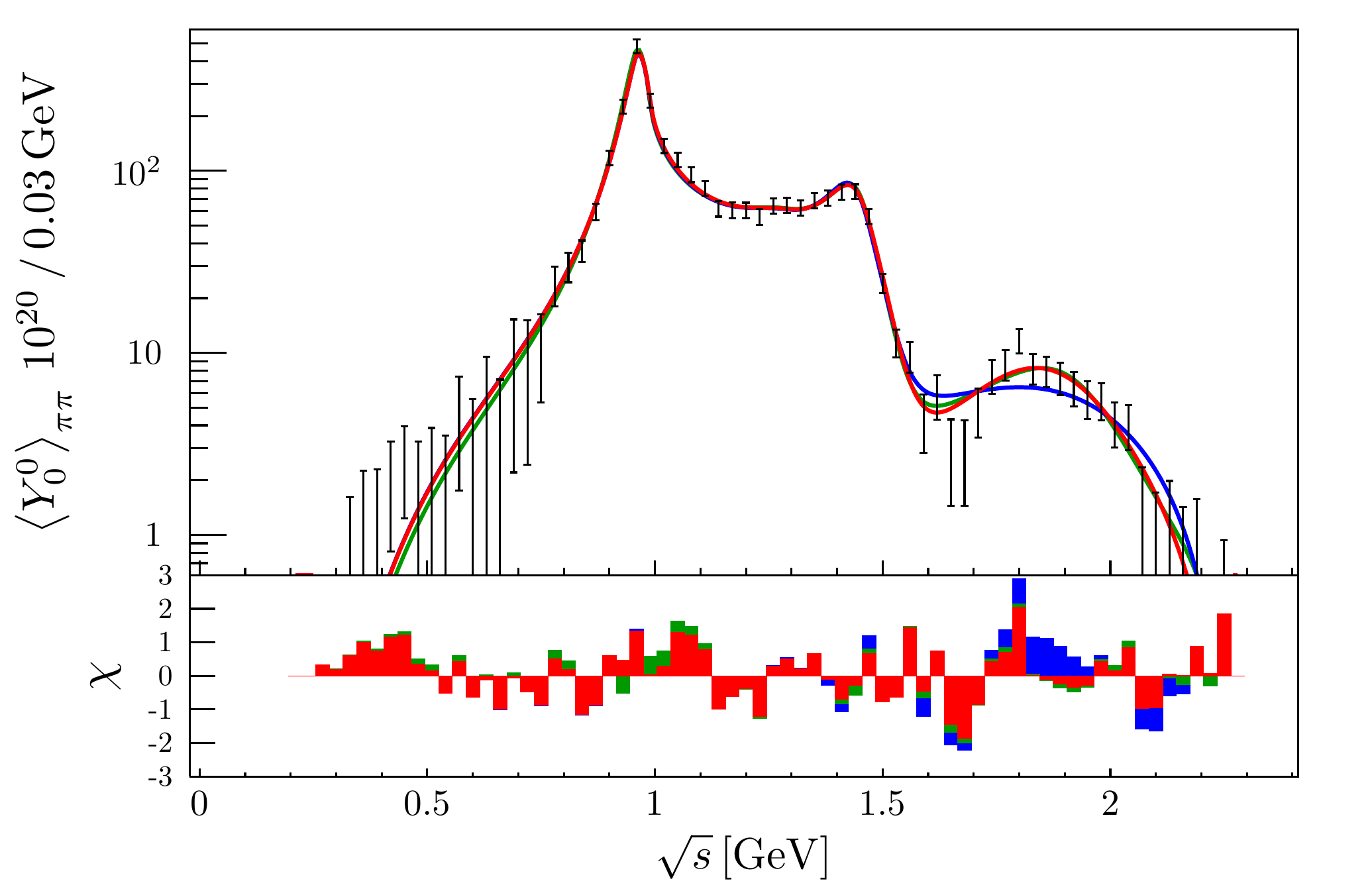}
	\includegraphics[width=0.49\linewidth]{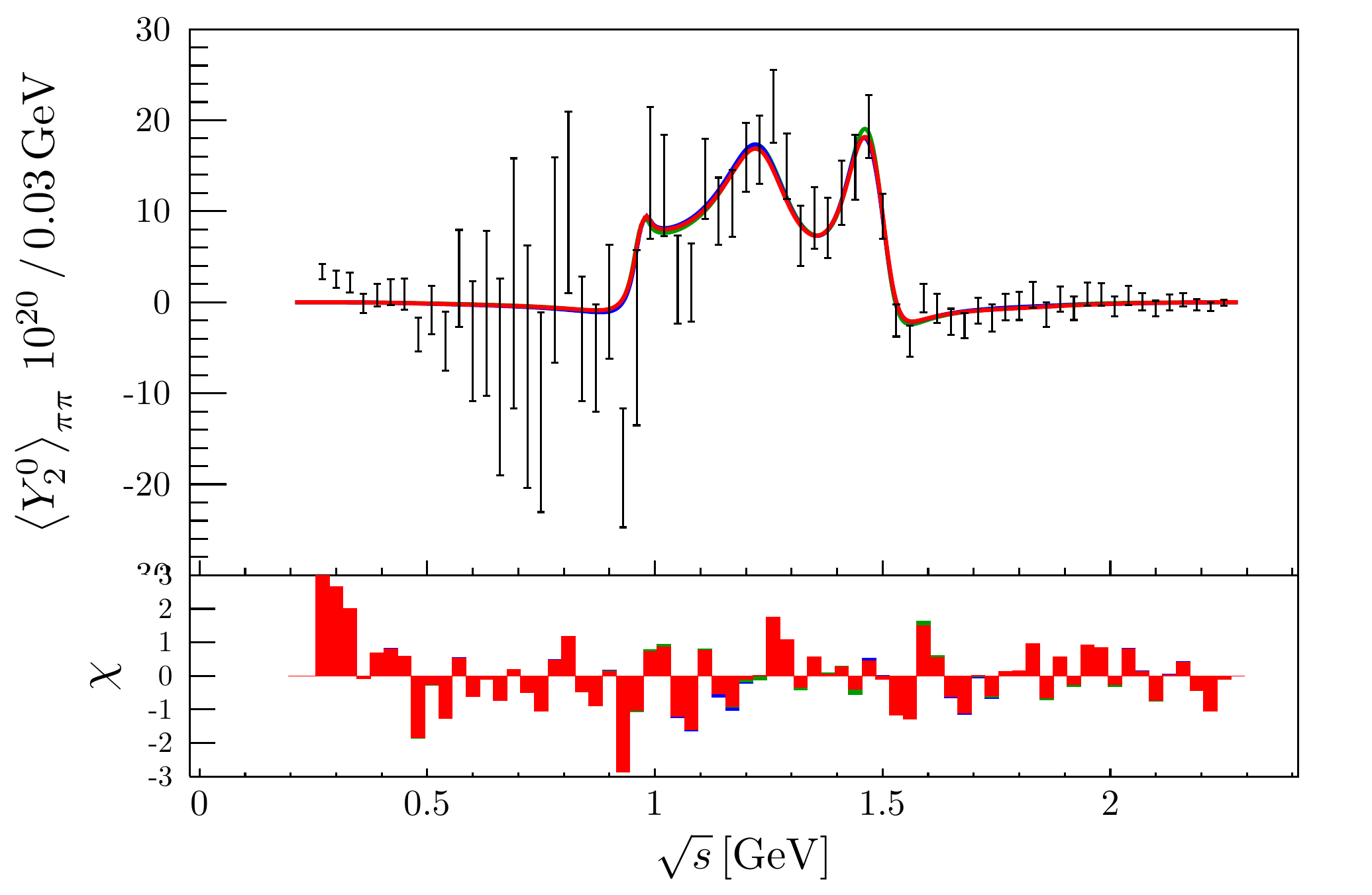}\\
	\includegraphics[width=0.49\linewidth]{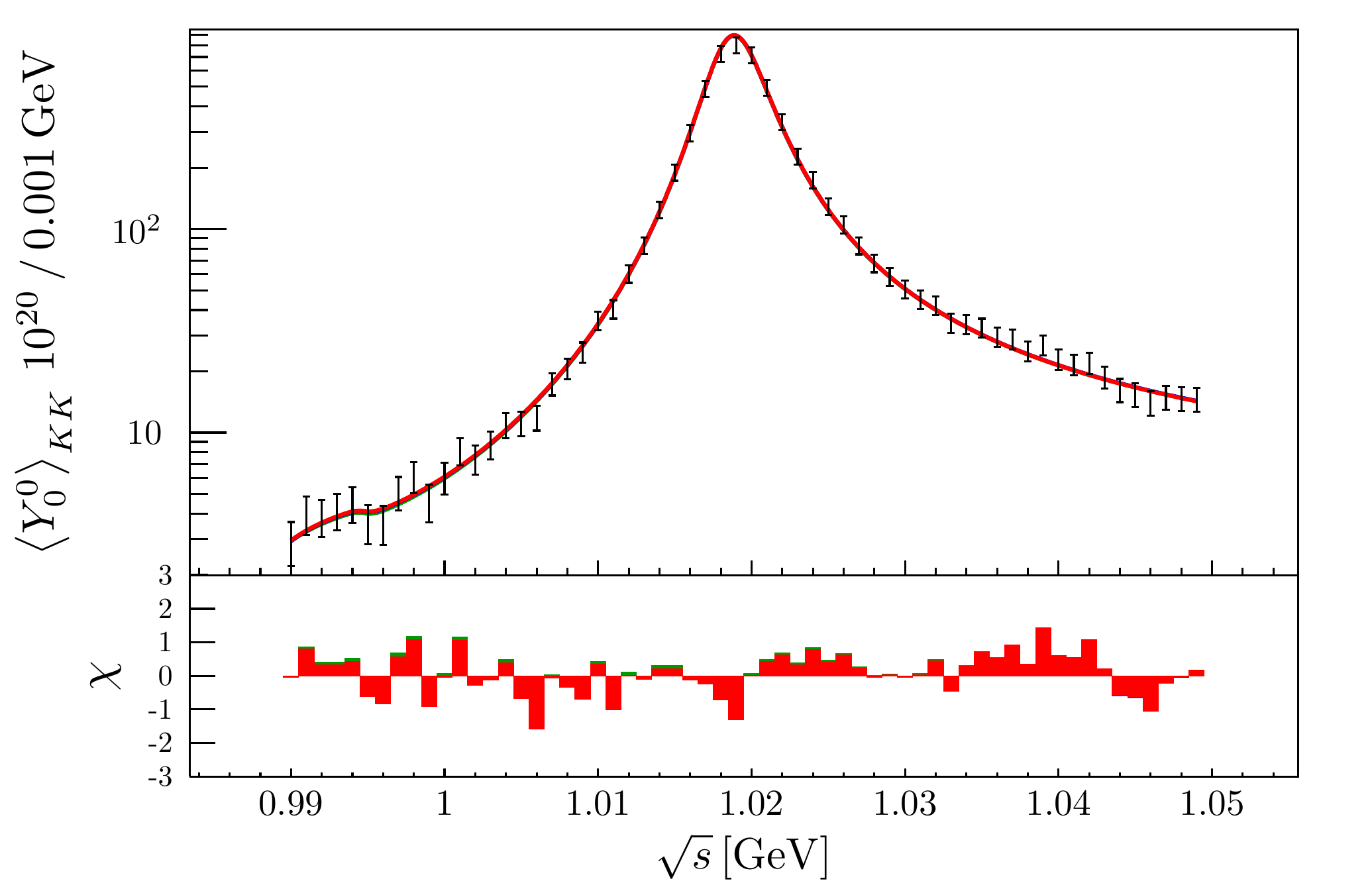}
	\includegraphics[width=0.49\linewidth]{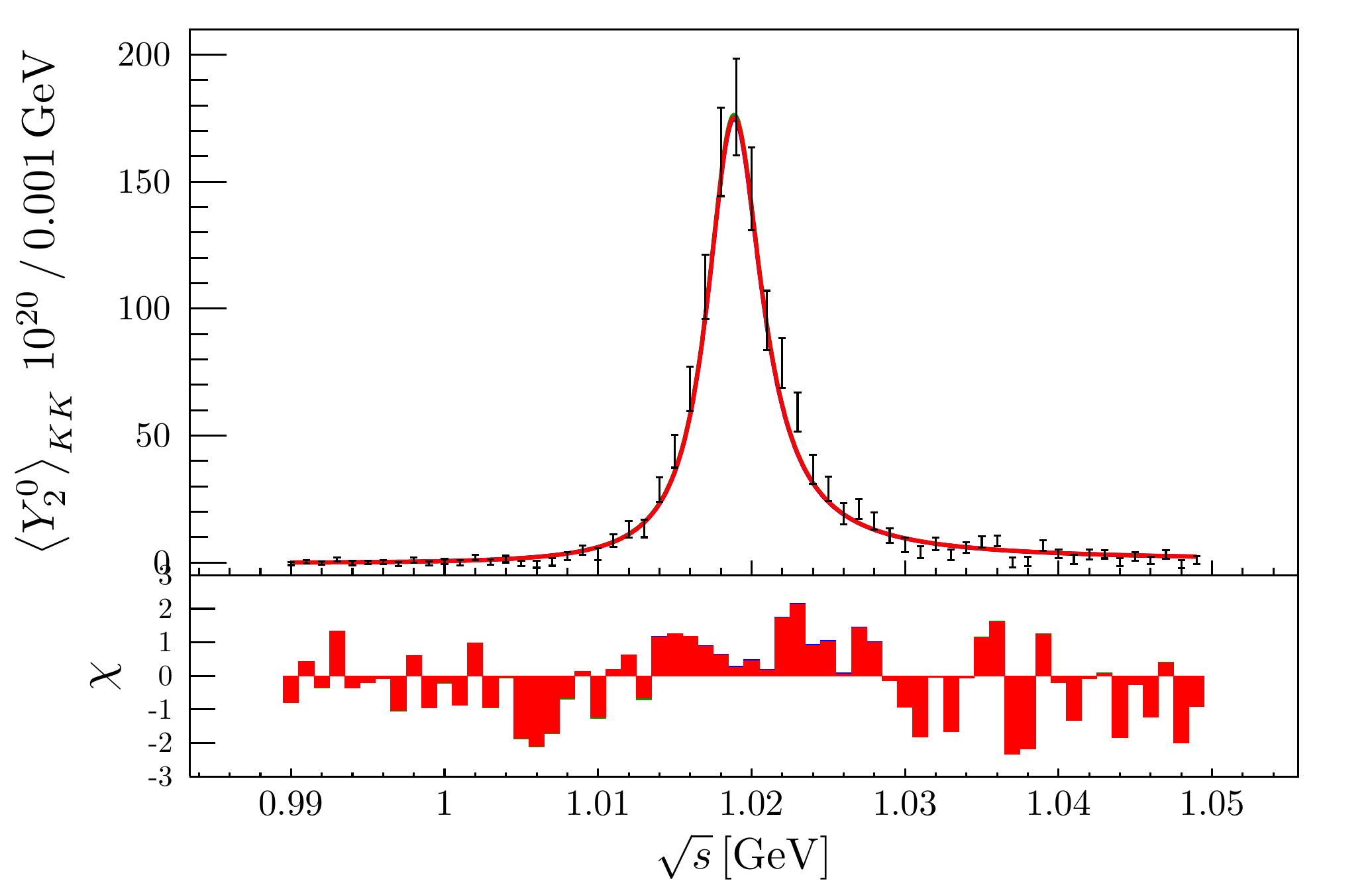}\\
	\includegraphics[width=0.49\linewidth]{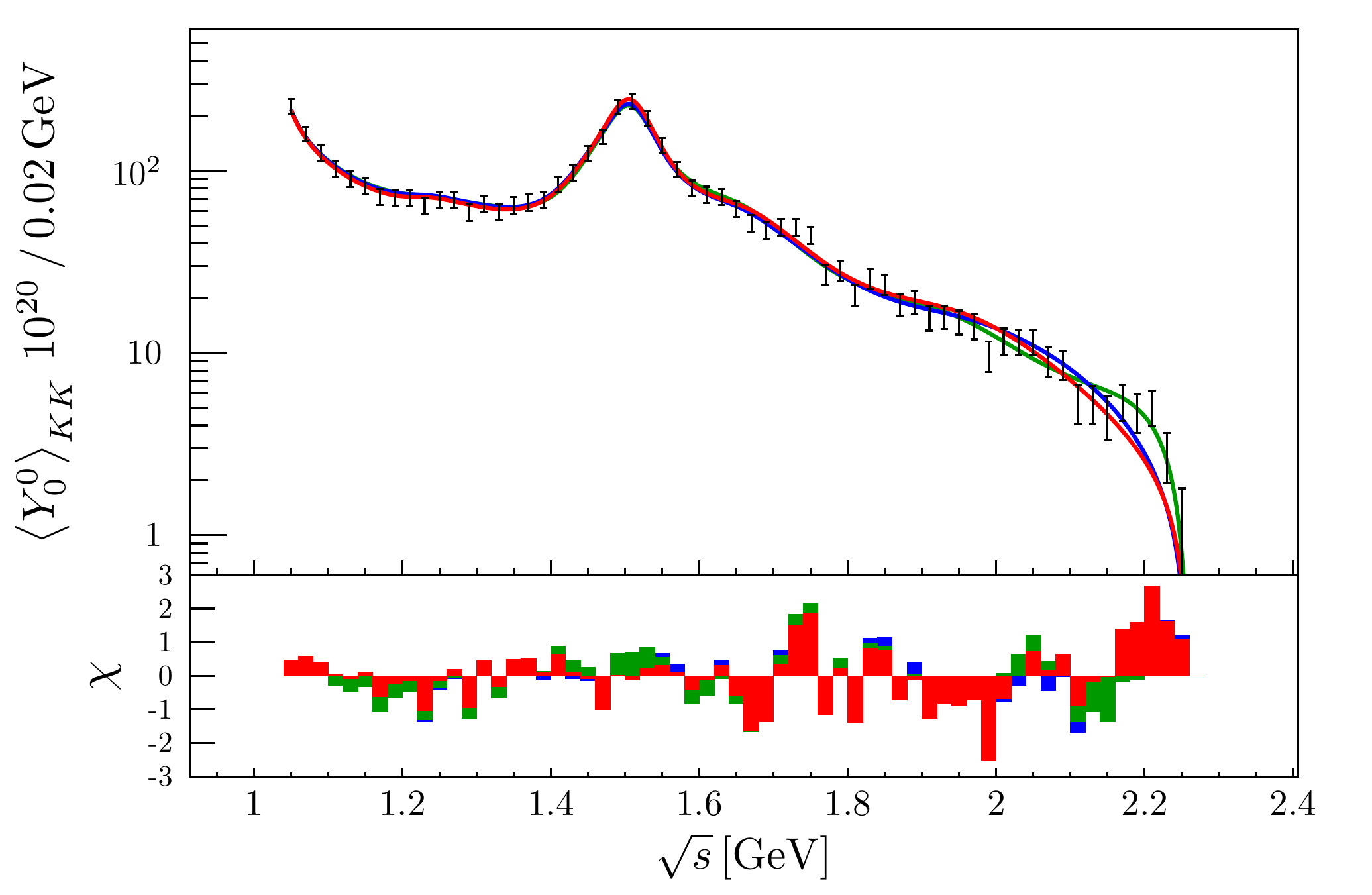}
	\includegraphics[width=0.49\linewidth]{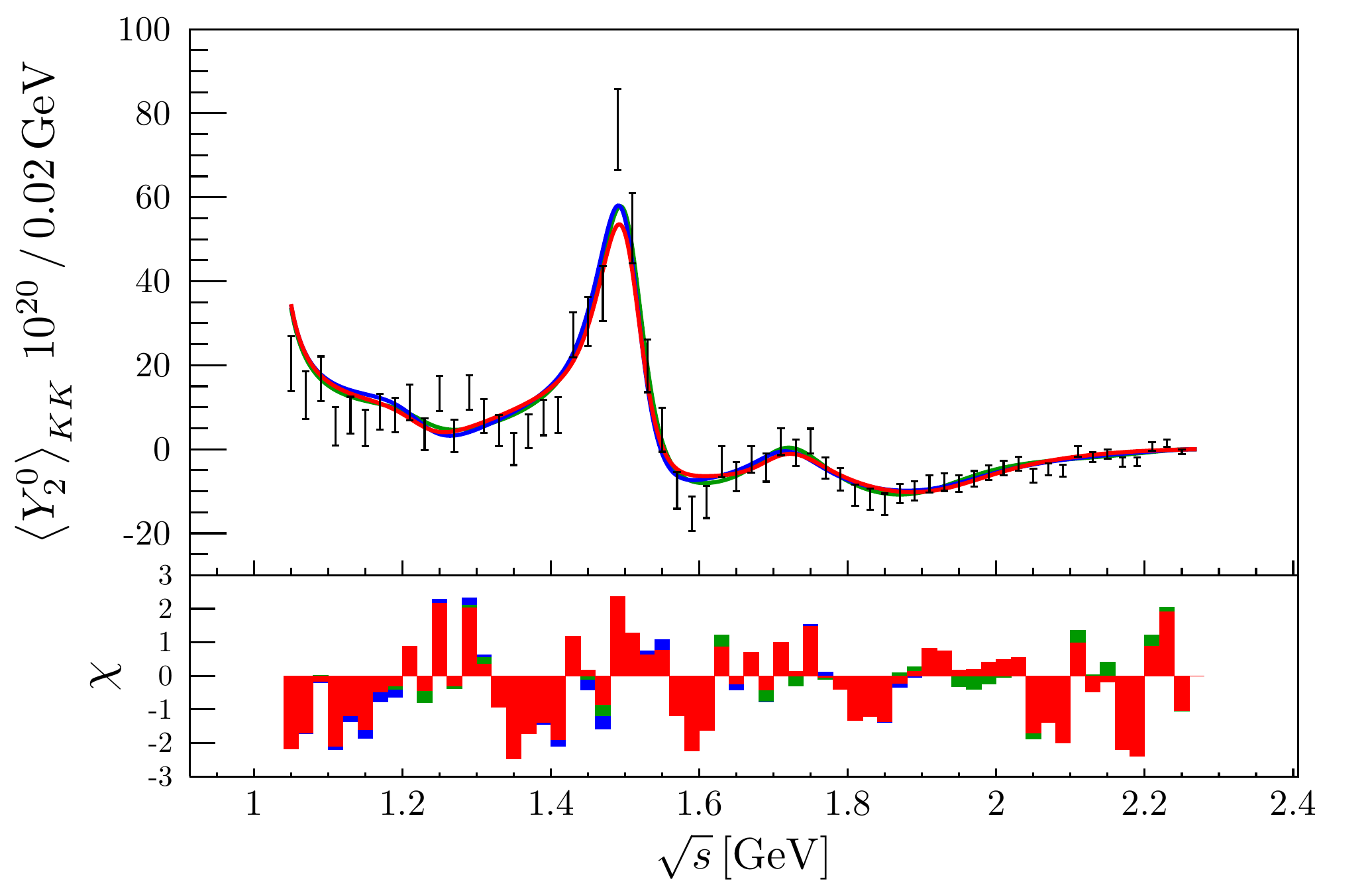}
\caption{Angular moments $\left<Y_0^0\right>$ and $\left<Y_2^0\right>$ for the decay $\bar{B}_s^0\rightarrow J/\psi\,\pi^+\pi^-$ (top two) and $\bar{B}_s^0\rightarrow J/\psi\,K^+K^-$ (bottom four) with an effective $\rho\rho$ channel. The picture shows Fit~1 in blue, Fit~2 in red, and Fit~3 in green. On the lower axis we show the fit residuals defined by $\chi=\left(\left<Y_L^0\right>_\mathrm{measured}-\left<Y_L^0\right>_\mathrm{fit}\right)/{\sigma_\mathrm{measured}}$. 
}
\label{Fig::Fit::AngularMoments_RhoRho}
\end{figure*}

\begin{figure*}[t!]
\centering
	\includegraphics[width=0.49\linewidth]{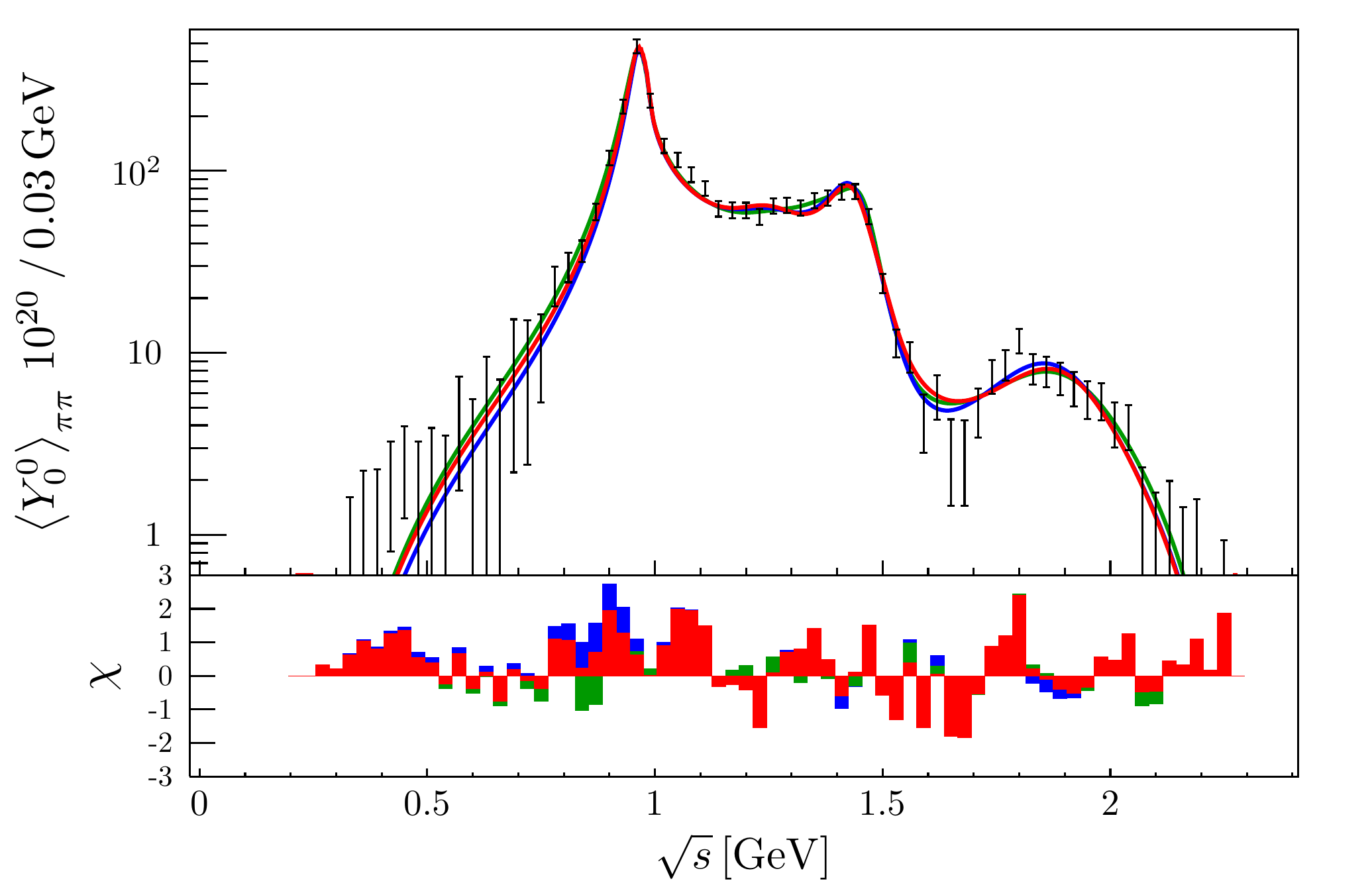}
	\includegraphics[width=0.49\linewidth]{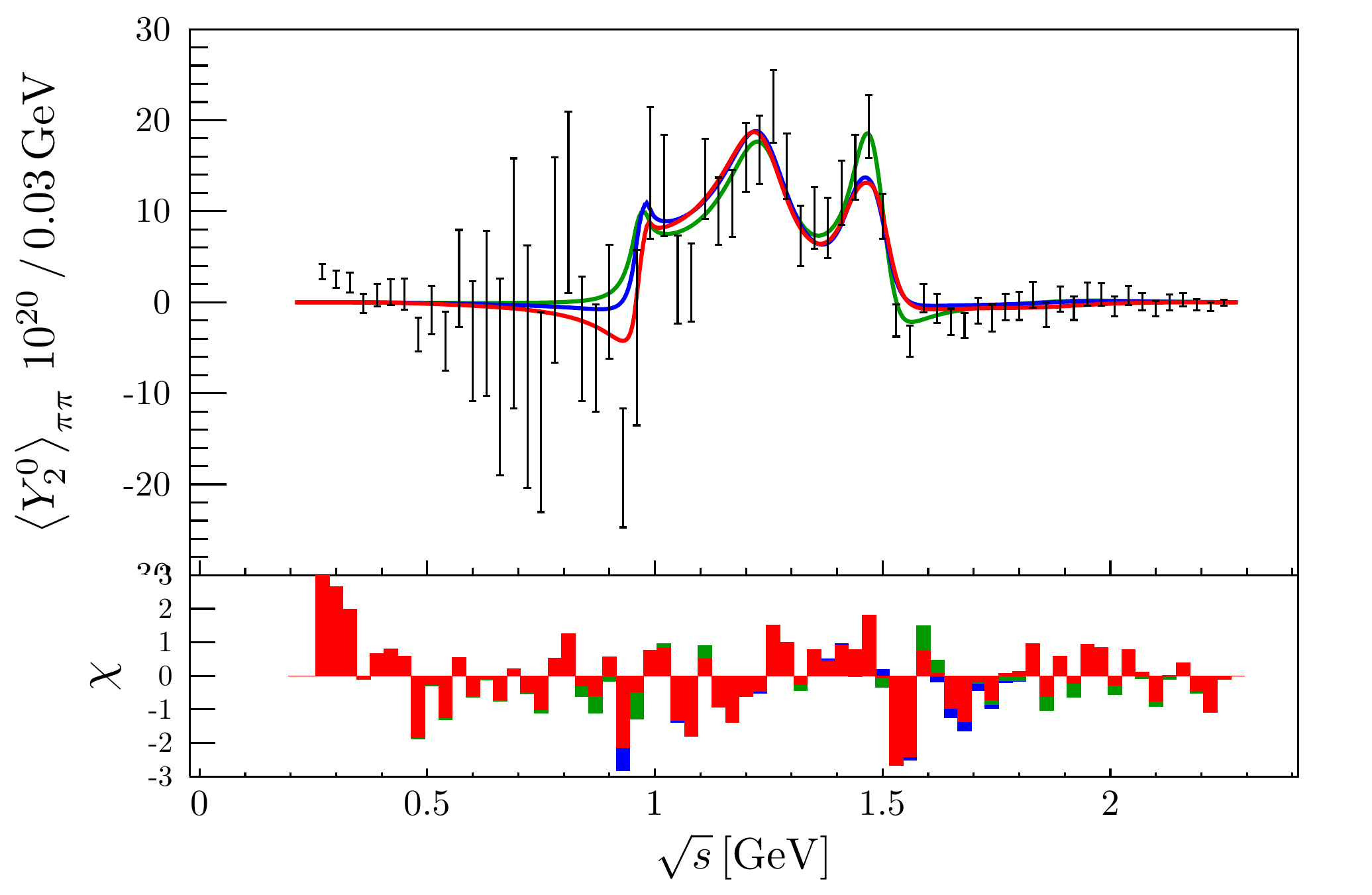}\\
	\includegraphics[width=0.49\linewidth]{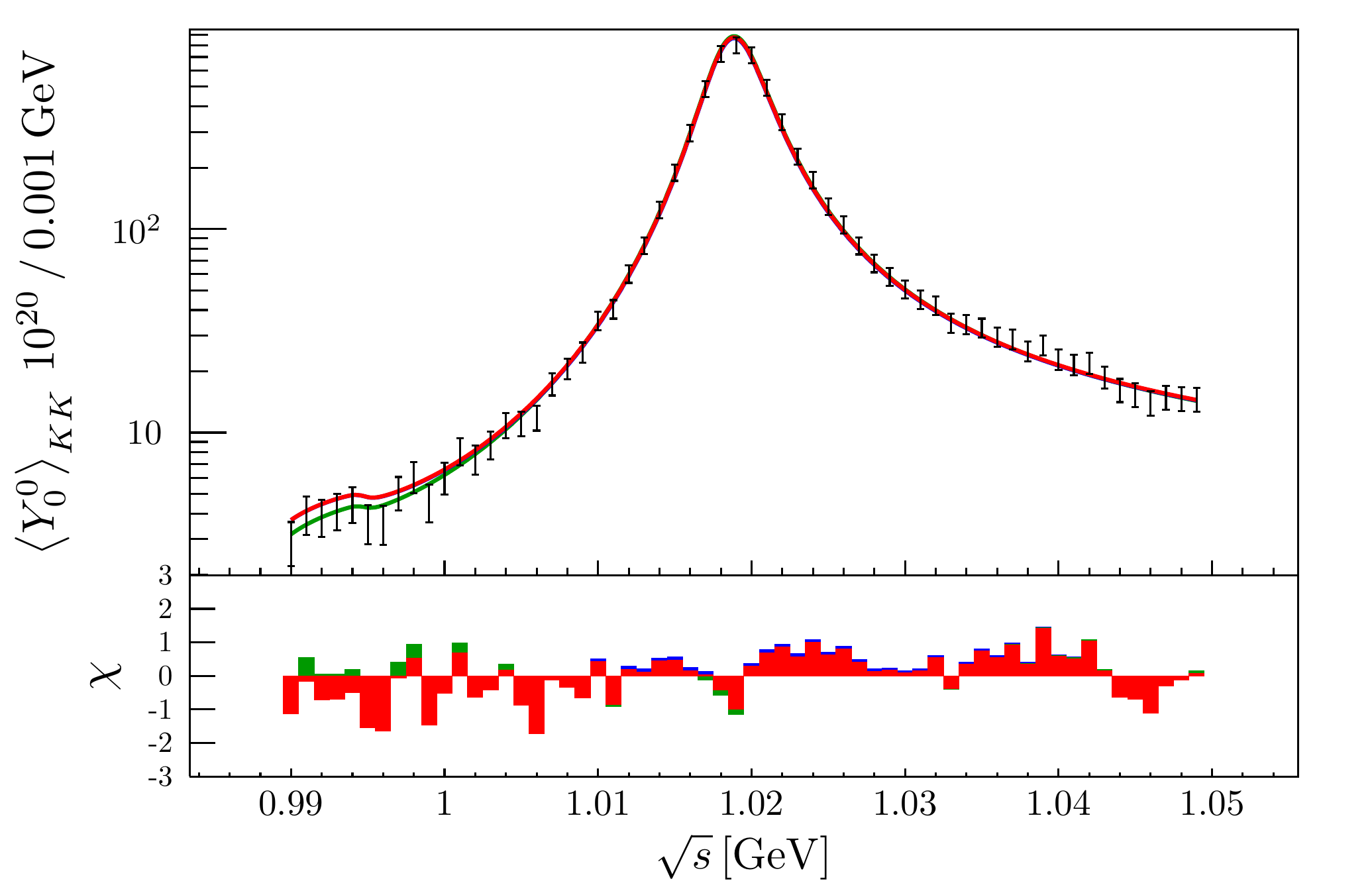}
	\includegraphics[width=0.49\linewidth]{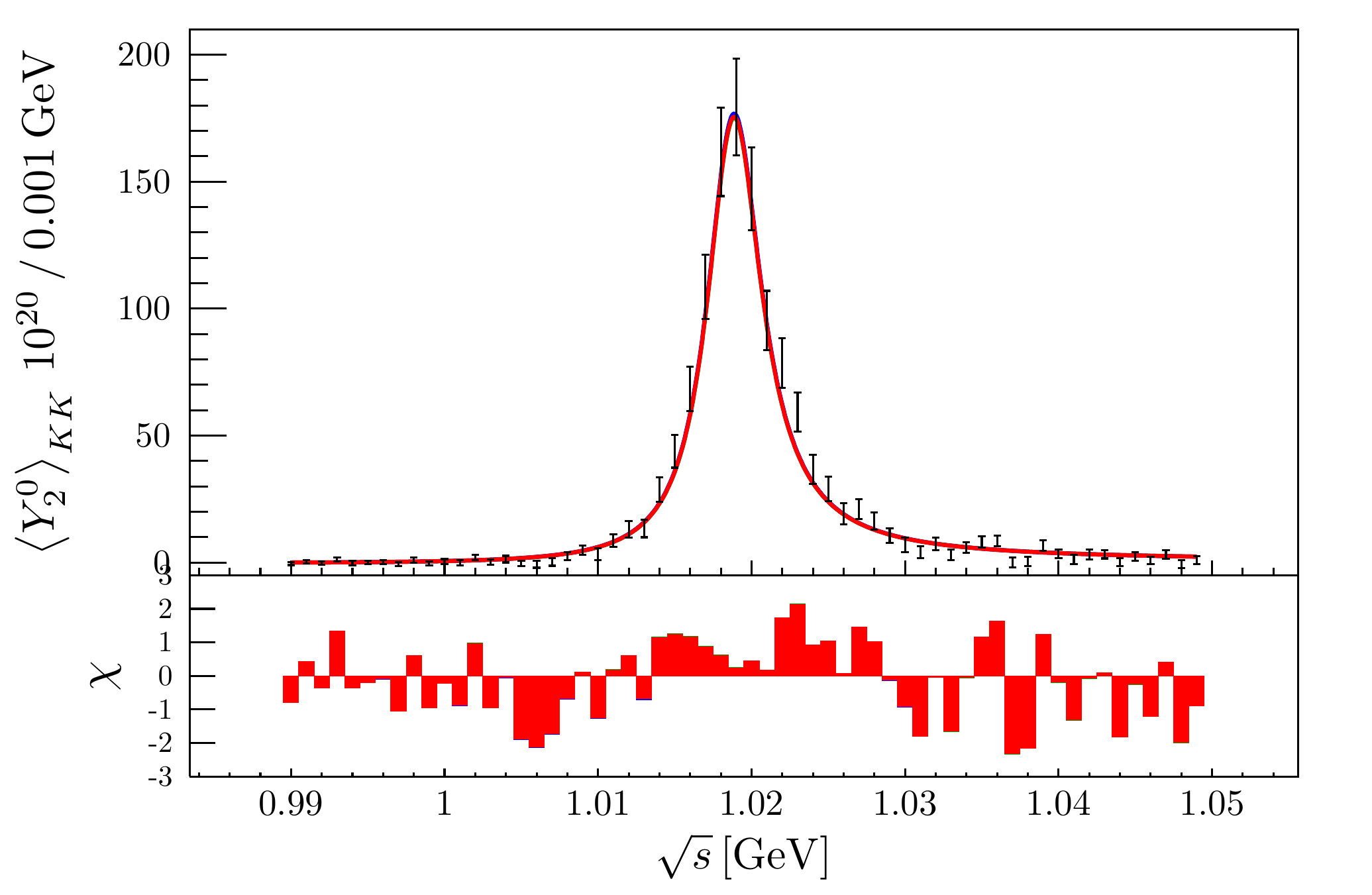}\\
	\includegraphics[width=0.49\linewidth]{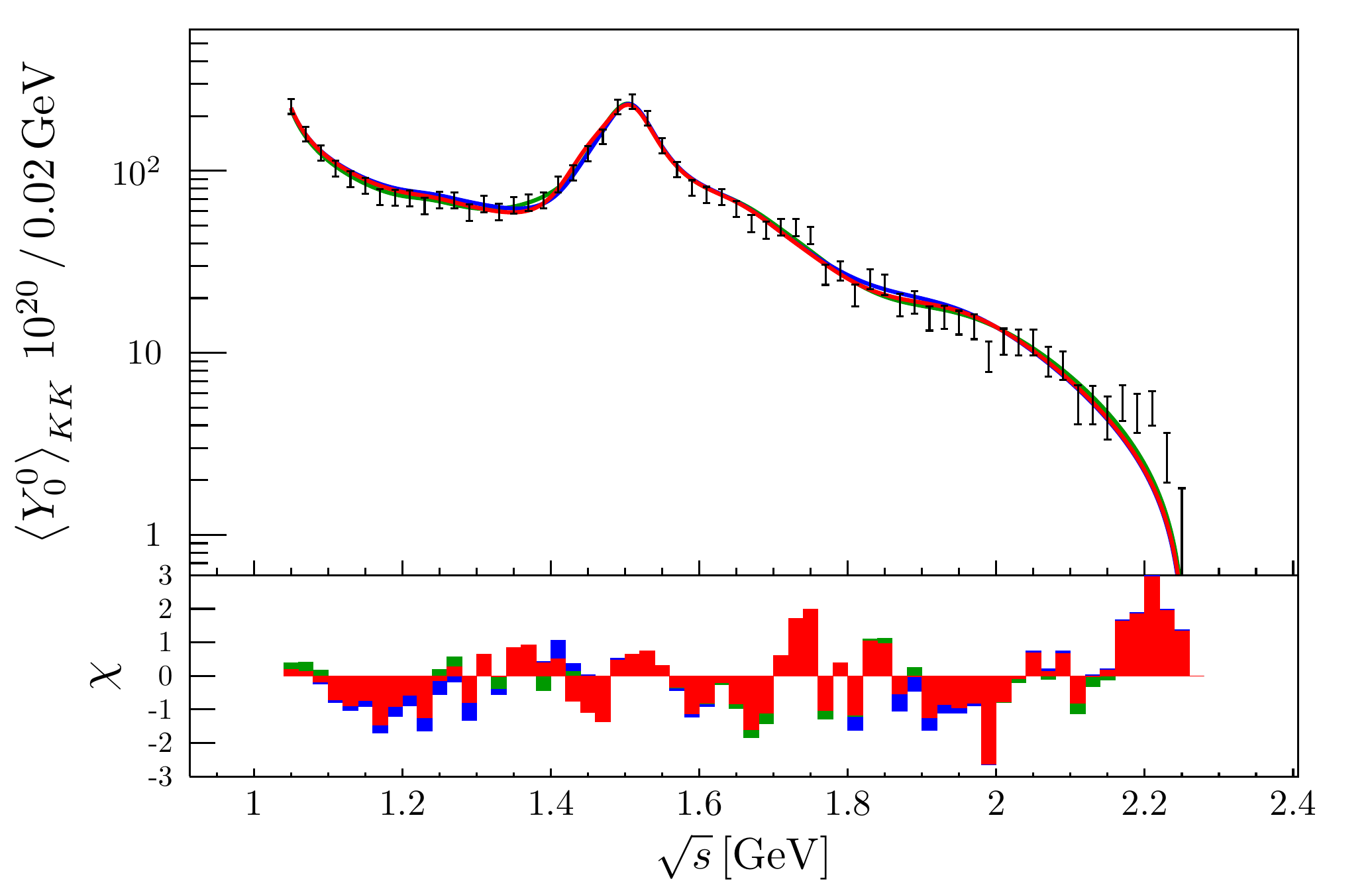}
	\includegraphics[width=0.49\linewidth]{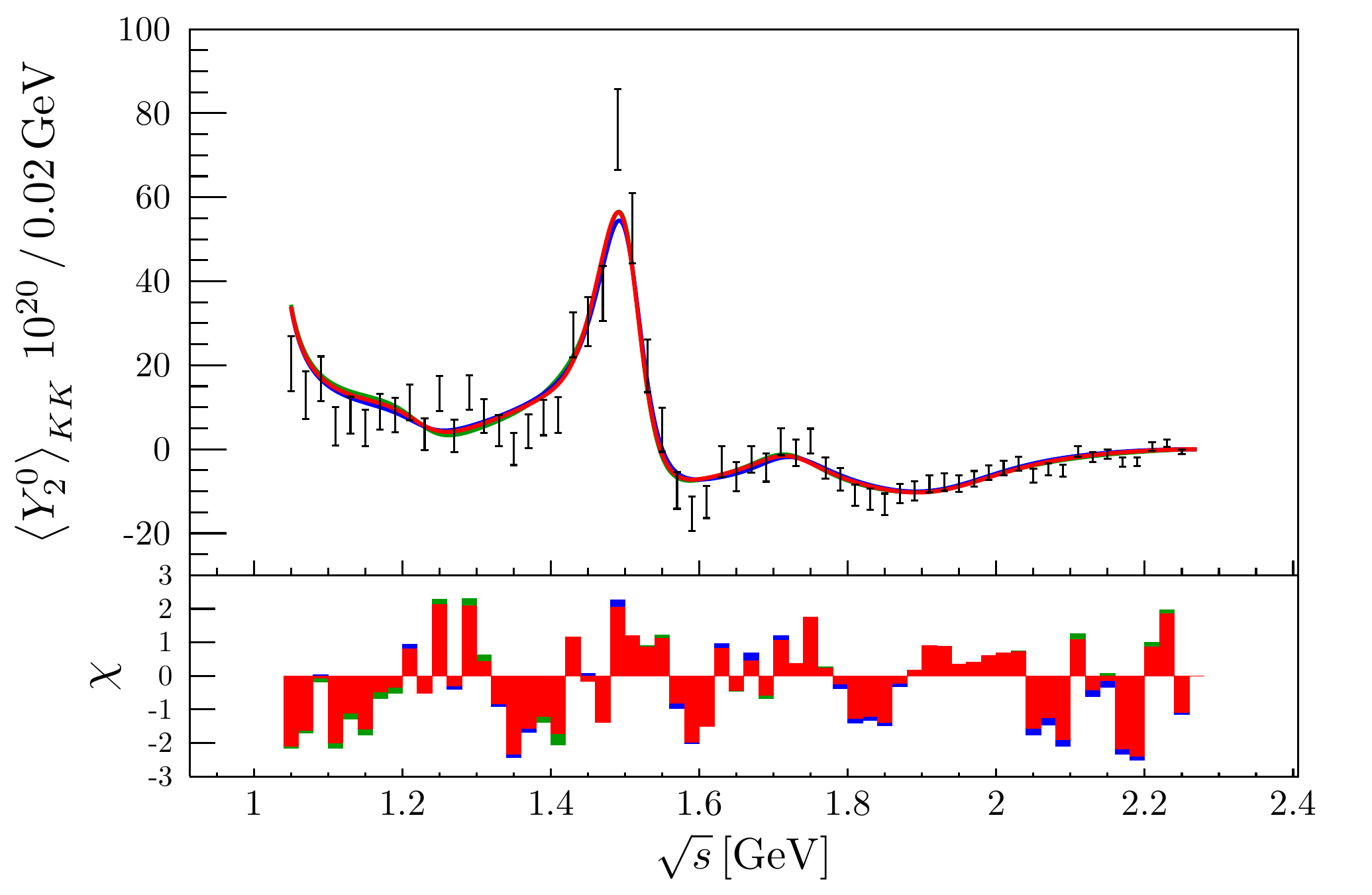}
\caption{Angular moments $\left<Y_0^0\right>$ and $\left<Y_2^0\right>$ for the decay $\bar{B}_s^0\rightarrow J/\psi\,\pi^+\pi^-$ (top two) and $\bar{B}_s^0\rightarrow J/\psi\,K^+K^-$ (bottom four) with an effective $\sigma\sigma$ channel. The picture shows Fit~1 in blue, Fit~2 in red, and Fit~3 in green. On the lower axis we show the fit residuals defined by $\chi=\left(\left<Y_L^0\right>_\mathrm{measured}-\left<Y_L^0\right>_\mathrm{fit}\right)/{\sigma_\mathrm{measured}}$.}
\label{Fig::Fit::AngularMoments_SigmaSigma}
\end{figure*}

\bsp
The LHCb collaboration extracted two additional $S$-wave resonances from their data~\cite{LHCb_PiPi}, 
namely $f_0(1500)$ and $f_0(1790)$. Since there is no $f_0(1790)$ in the listings of the Review of Particle
Physics by the PDG~\cite{PDG}, we use the name $f_0(2020)$ for the higher state,
in particular since the parameters we extract below are close to those reported for that resonance.
The first fit includes our parametrization with $N_R=2$ and $\gamma_i=0$ (Fit~1). 
To test the stability of this solution, we also include a fit with $N_R=2$ and $\gamma_i\neq 0$ (Fit~2) as well 
as a fit with $N_R=3$ and $\gamma_i=0$ (Fit~3). 
In order to obtain an estimate  of the systematic uncertainty, we repeat each fit with two different assumptions about the third channel, which we take to be dominated by either $\sigma\sigma$ or $\rho\rho$. 
The respective reduced $\chi^2$ of the best fit results are listed in Table~\ref{Tab::Fit::ReducedChi2}.
We show the corresponding angular moments in Figs.~\ref{Fig::Fit::AngularMoments_RhoRho} ($\rho\rho$) 
and~\ref{Fig::Fit::AngularMoments_SigmaSigma} ($\sigma\sigma$). 
In principle we could have also
investigated mixtures of $\sigma\sigma$ and $\rho\rho$ intermediate states or
 parametrizations representing the channels $\pi(1300)\pi$ or $a_1(1260)\pi$ reported
to be relevant for the $f_0(1500)$~\cite{PDG}, however, since with the given choices we already find
excellent fits to the data although the corresponding two-point function $\Sigma_{33}$ look vastly different
for the $\sigma\sigma$ and the $\rho\rho$ case (cf.\ the lower right panel of Fig.~\ref{Fig::Parametrization::SelfEnergy}), 
studying other possible decays will be postponed until data for further exclusive
final states become available.  
\esp

We note first of all that the $\rho\rho$ fits have a lower reduced $\chi^2$ compared to the $\sigma\sigma$ fits. 
Allowing for a linear term in the source further improves the data description,
as witnessed by the differences of Fits~1 and~2. 
The overall best reduced $\chi^2$ is obtained by including another, third, resonance.

For the $\rho\rho$ fit (see Fig.~\ref{Fig::Fit::AngularMoments_RhoRho}) we see that Fit~2 improves the description of $\left<Y_0^0\right>_{\pi\pi}$ in the energy region between $1.6$ and $2.0\,\GeV$. The biggest change between Fit~3 and the other ones is given by the better description of the high-energy 
tail in the decay $\bar{B}_s^0\rightarrow J/\psi K^+K^-$. 

For the $\sigma\sigma$ fit, Fig.~\ref{Fig::Fit::AngularMoments_SigmaSigma}, we observe a similar picture. 
Fit~2 provides a very slight overall improvement of Fit~1. However, here the main difference between Fit~3 and the rest resides 
in the better description of the $f_0(1500)$ especially for the decay $\bar{B}_s^0\rightarrow J/\psi \,\pi^+\pi^-$, while the high-energy tail of $\bar{B}_s^0\rightarrow J/\psi\,K^+K^-$ remains nearly untouched.
\begin{figure*}
	\includegraphics*[width=0.495\linewidth]{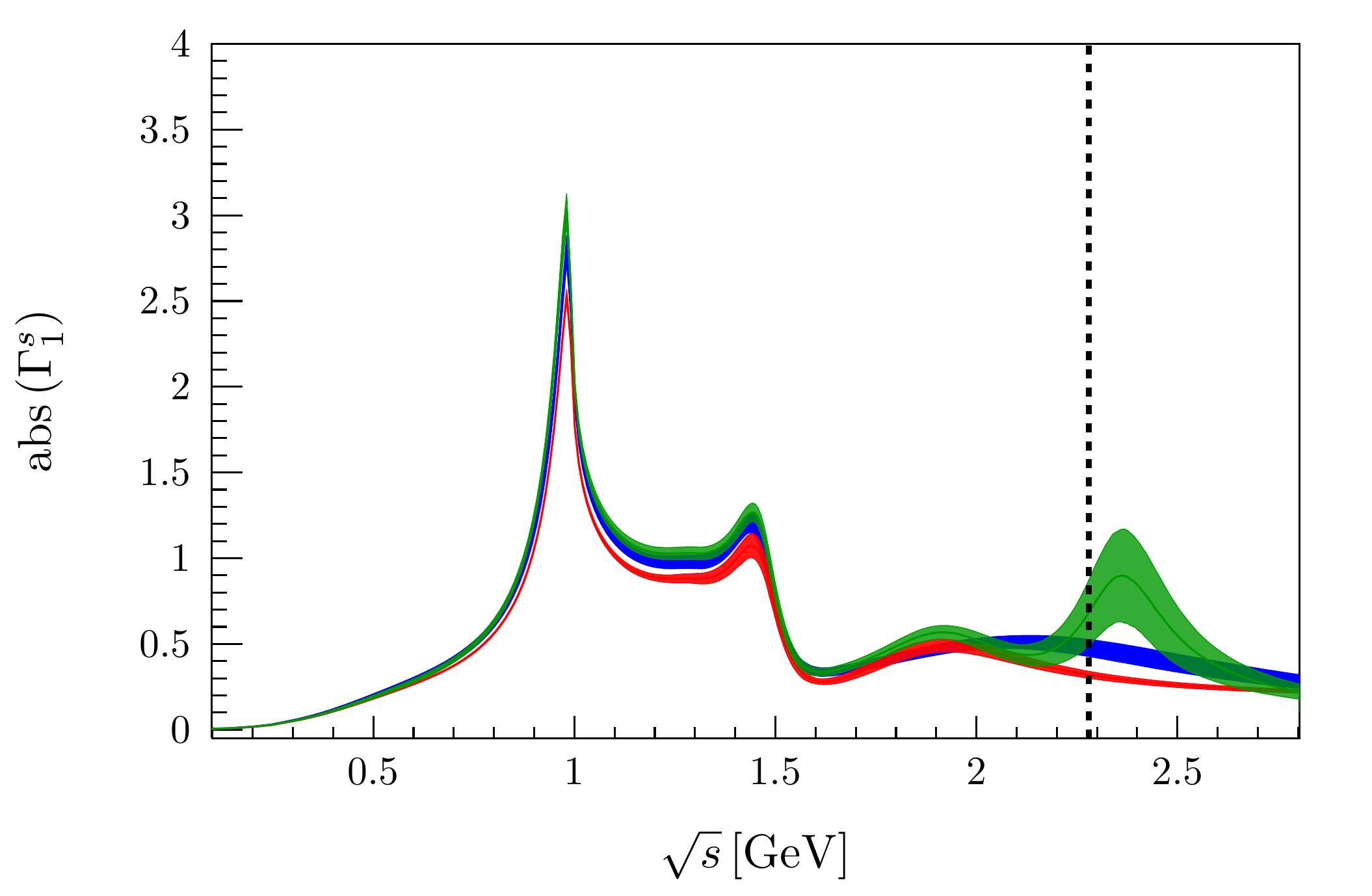} \hfill
	\includegraphics*[width=0.495\linewidth]{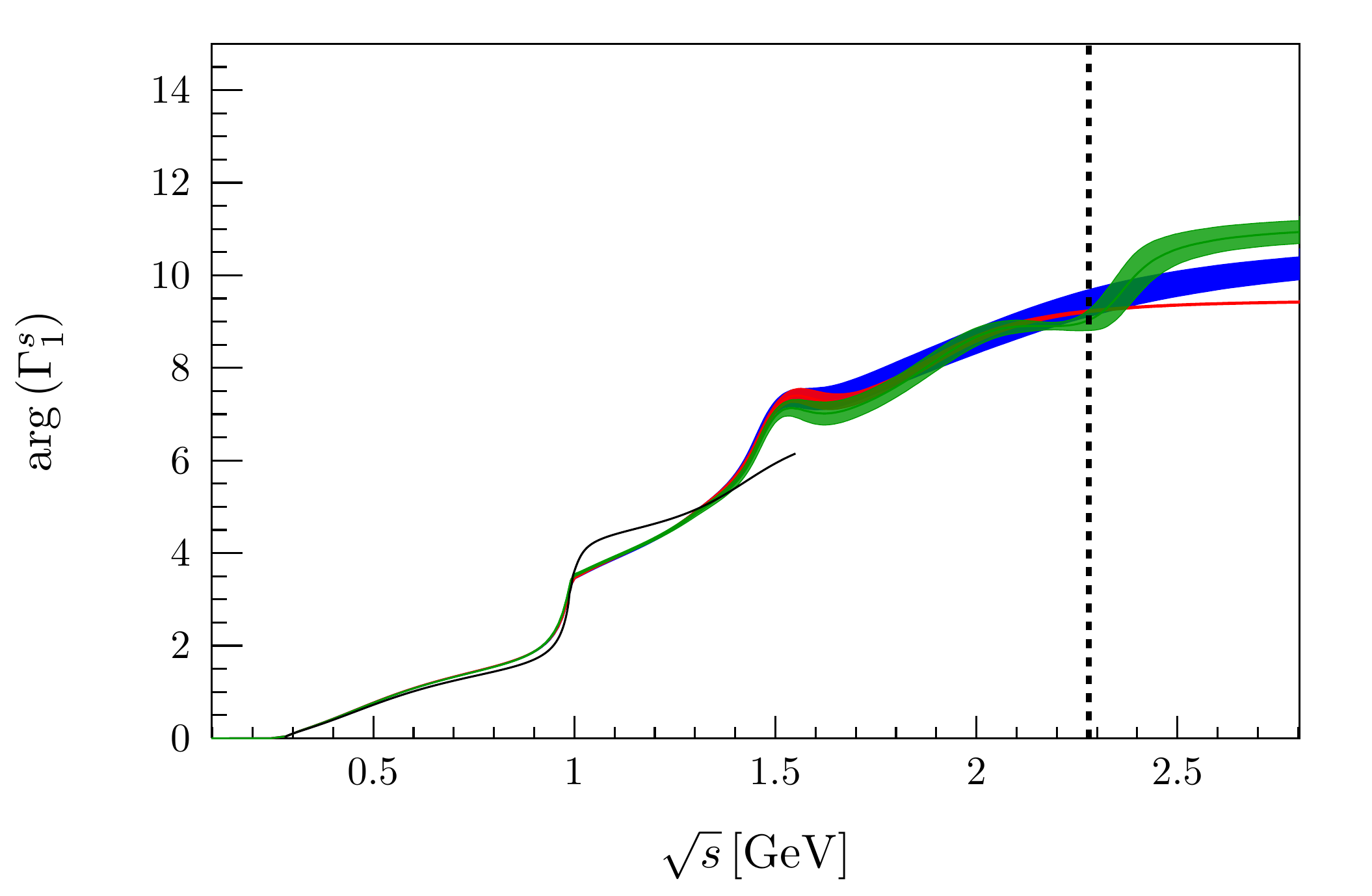} \\
	\includegraphics*[width=0.495\linewidth]{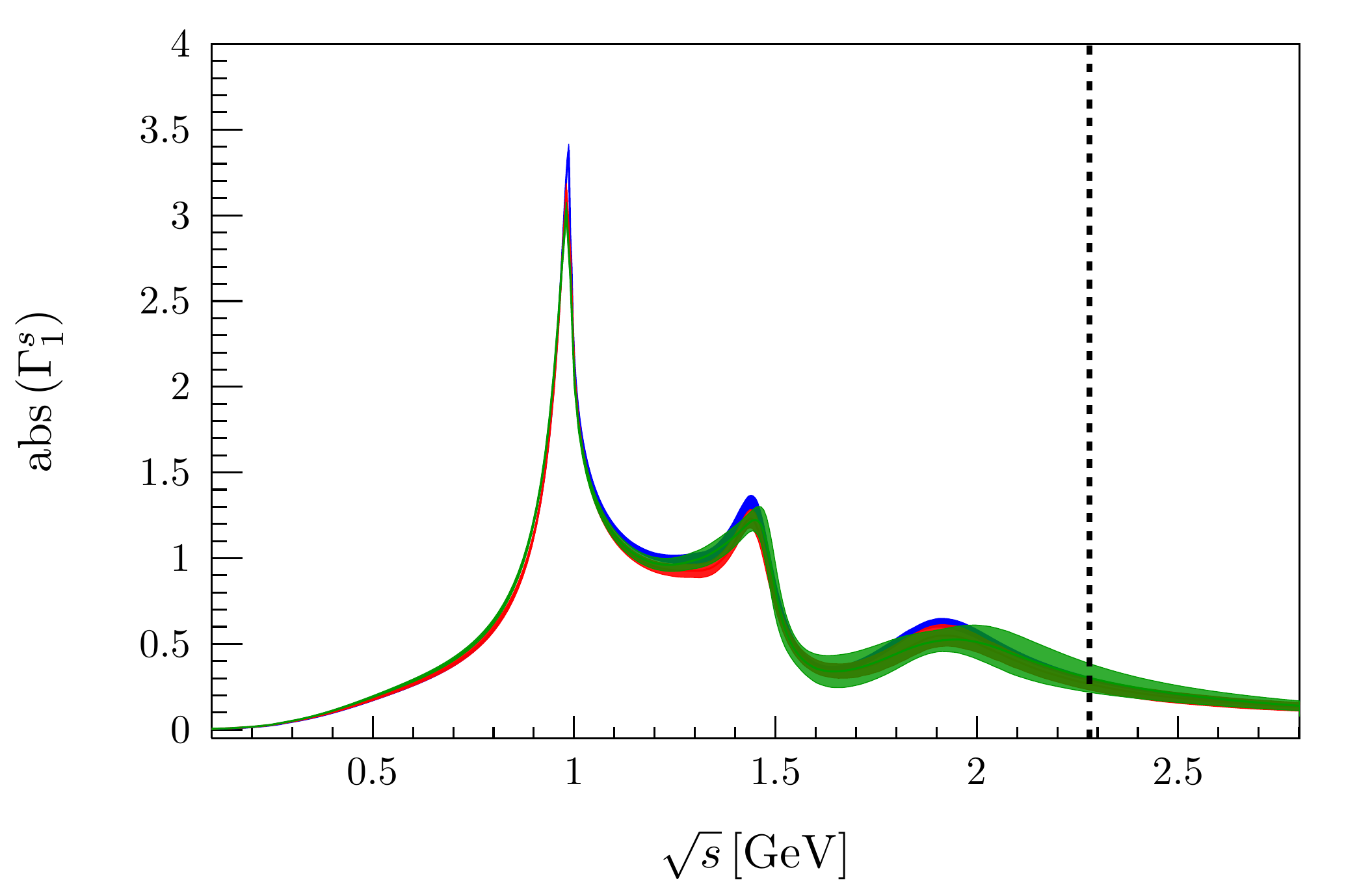} \hfill
	\includegraphics*[width=0.495\linewidth]{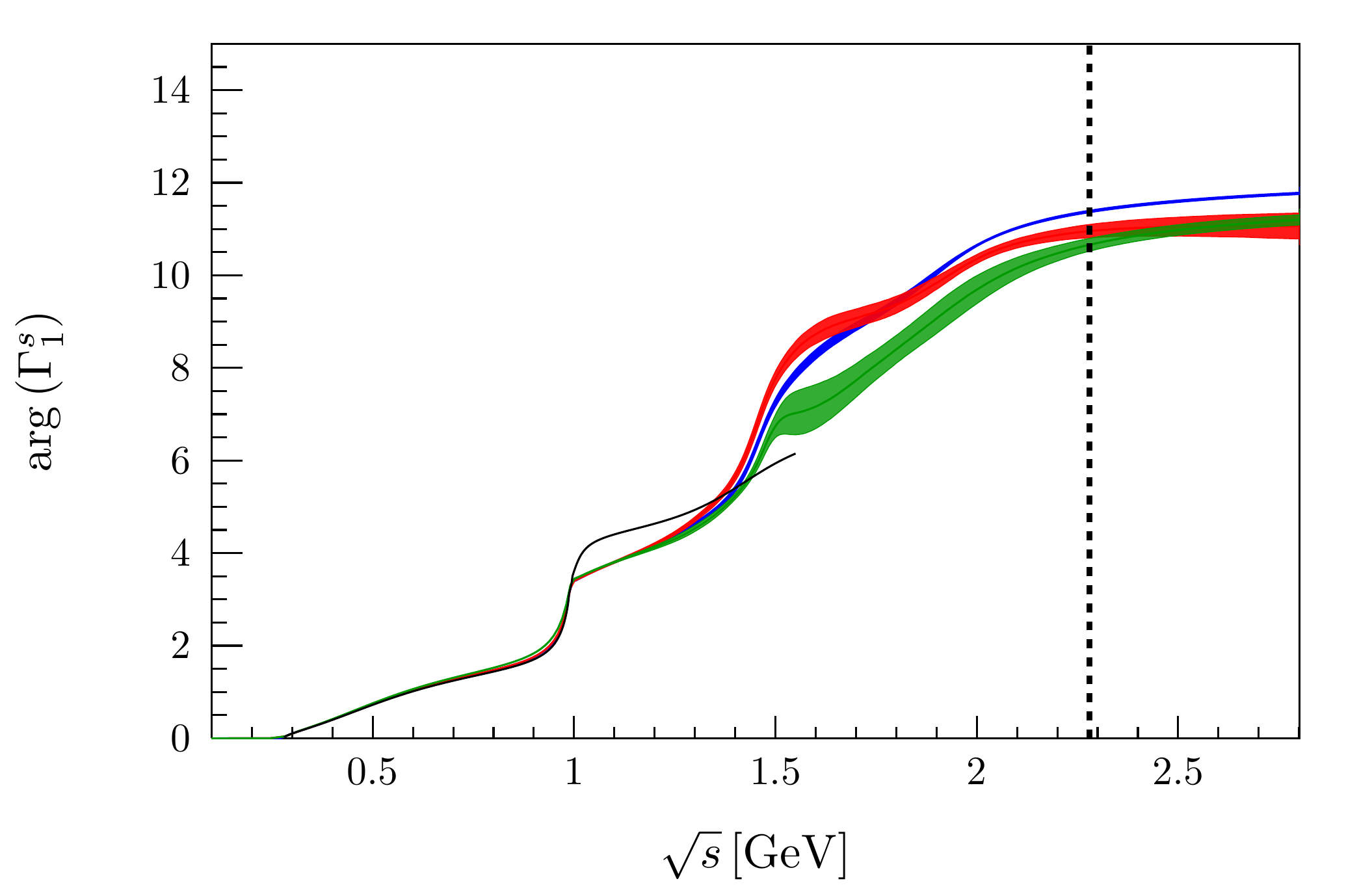}
\caption{Modulus (left) and phase (right) of the pion form factor $\Gamma_1^s$ for the fits with an additional $\rho\rho$ (top) and $\sigma\sigma$ (bottom) channel. The input scalar isoscalar scattering phase $\delta_0$ is depicted in black. Fit~1 is shown in blue, Fit~2 in red, and Fit~3 in green. The dotted vertical lines mark the kinematic upper limit for $\sqrt{s}$ in the $\bar B_s^0$ decay.}
\label{Fig::Formfactor::FF1}
\end{figure*}

\bsp
For a better comparison of the different fits we discuss the resulting form factors $\Gamma_i^s$ in some detail.
We begin by comparing the strange scalar pion form factor $\Gamma^s_1$ as shown in Fig.~\ref{Fig::Formfactor::FF1}. In all fits 
three resonances are clearly visible, namely the $f_0(980)$, $f_0(1500)$, and a broad structure around $2\,\GeV$ related to the $f_0(2020)$ resonance. Furthermore we also 
know that the input contains the broad $f_0(500)$ resonance. Fit~3 contains an additional resonance: 
in the case of the $\rho\rho$ fit, it has its pole around $2.4\,\GeV$ 
and is relatively narrow.  Notice that the maximum energy available for the $\pi\pi$ system  in the decay studied is $2.27\,\GeV$, thus this additional 
resonance in fact only contributes with its low-energy tail, giving small corrections for the high-energy parts of the angular moments. 
This is clearly visible in $\left<Y_0^0\right>_{KK}$ at high energies in Fig.~\ref{Fig::Fit::AngularMoments_RhoRho}, where Fit~3 can describe the last 
data points better than Fits~1 and~2. In comparison we see that the $\sigma\sigma$ fit lacks any such high-energy resonance. For this
fit the 
difference between Fit~3 and the rest is only visible in the argument of $\Gamma_1^s$, showing a shift in the range  $1.5\ldots2\,\GeV$. 
This improves the description of $\left<Y_2^0\right>_{\pi\pi}$ near the $f_0(1500)$ resonance. 
From this discussion it becomes clear that the data analyzed here do not allow us to extract information on any further resonance
beyond $f_0(500)$, $f_0(980)$, $f_0(1500)$, and $f_0(2020)$.
\esp

\begin{figure*}
	\includegraphics*[width=0.495\linewidth]{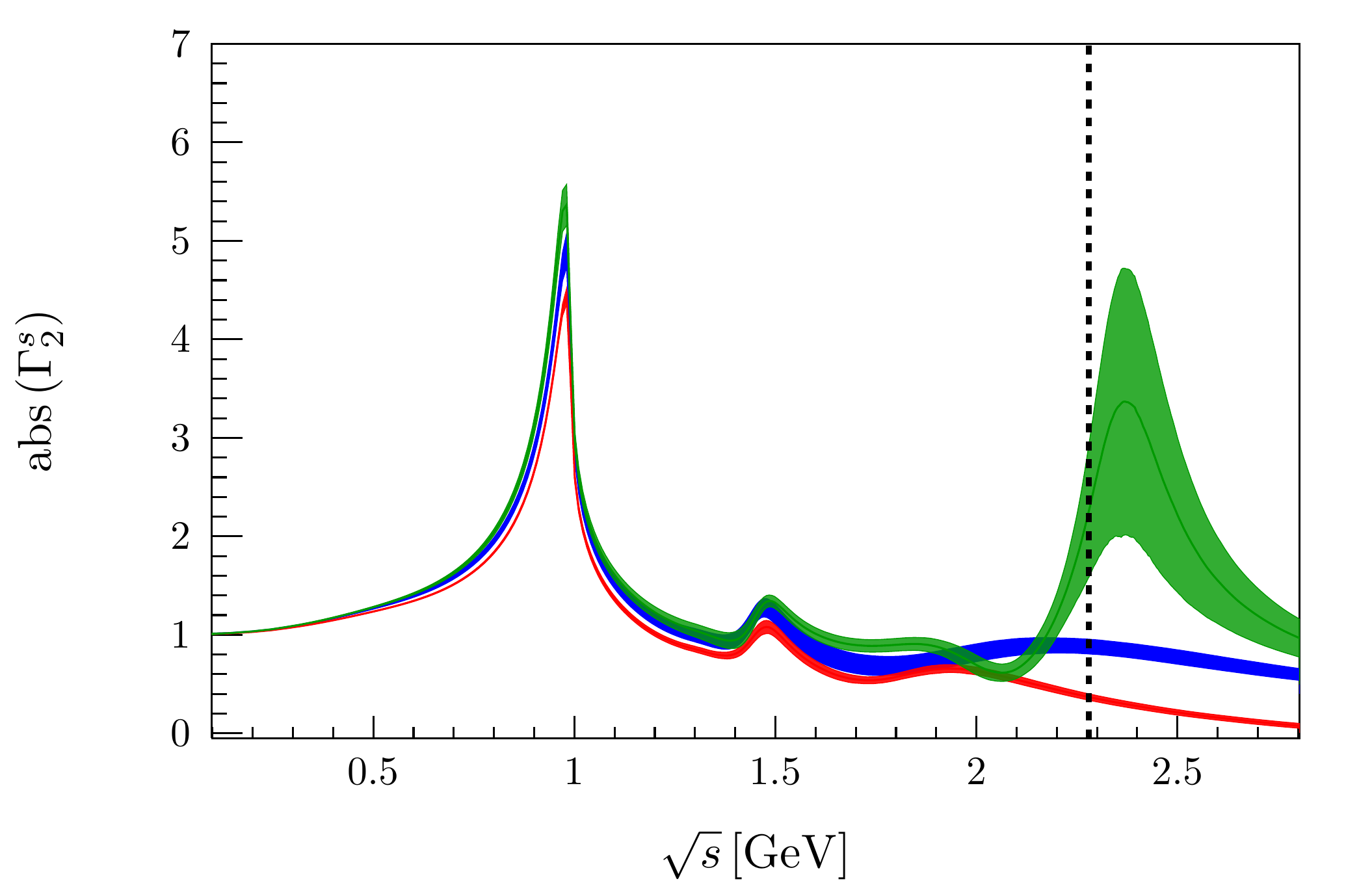} \hfill
	\includegraphics*[width=0.495\linewidth]{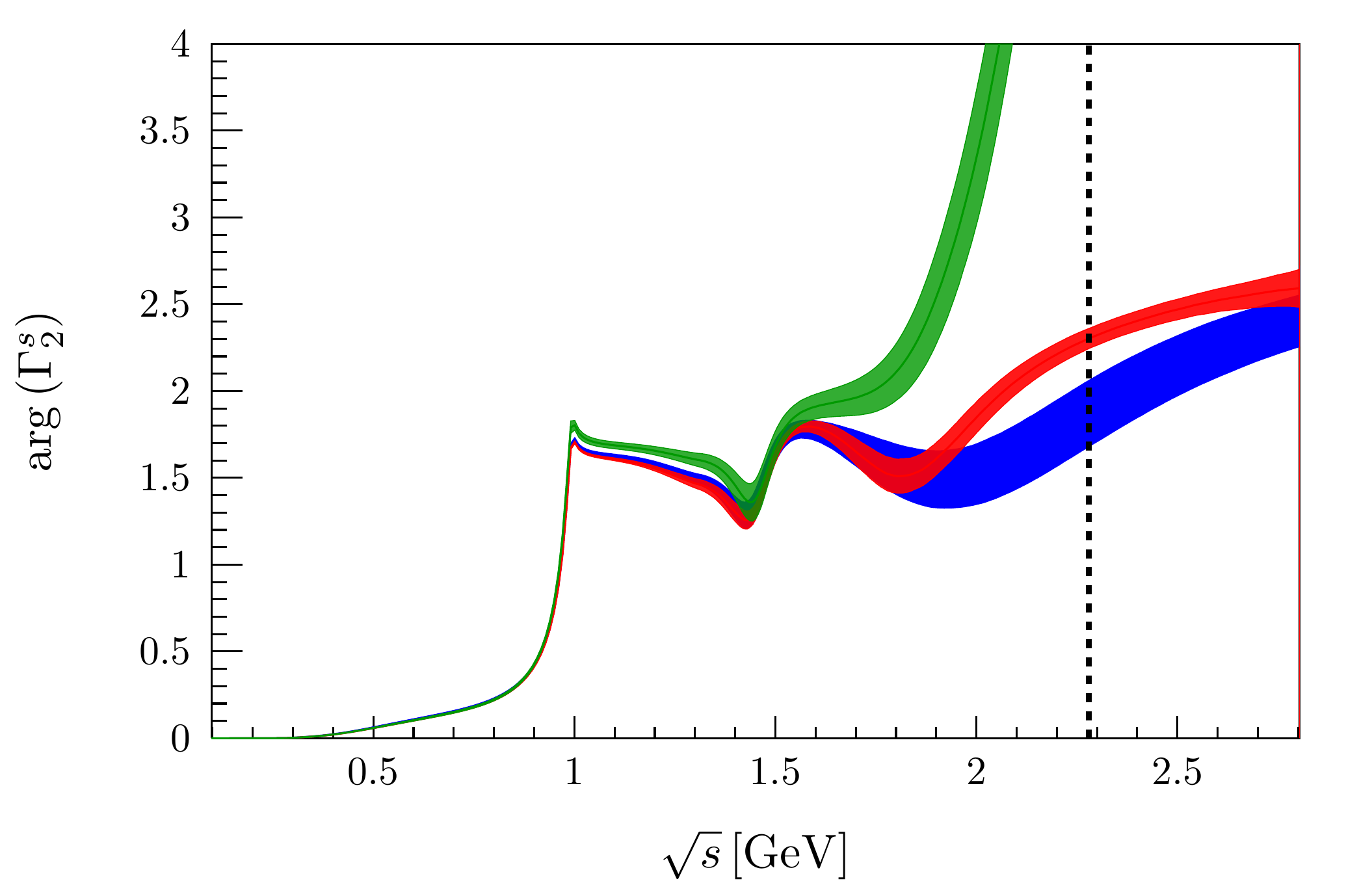} \\
	\includegraphics*[width=0.495\linewidth]{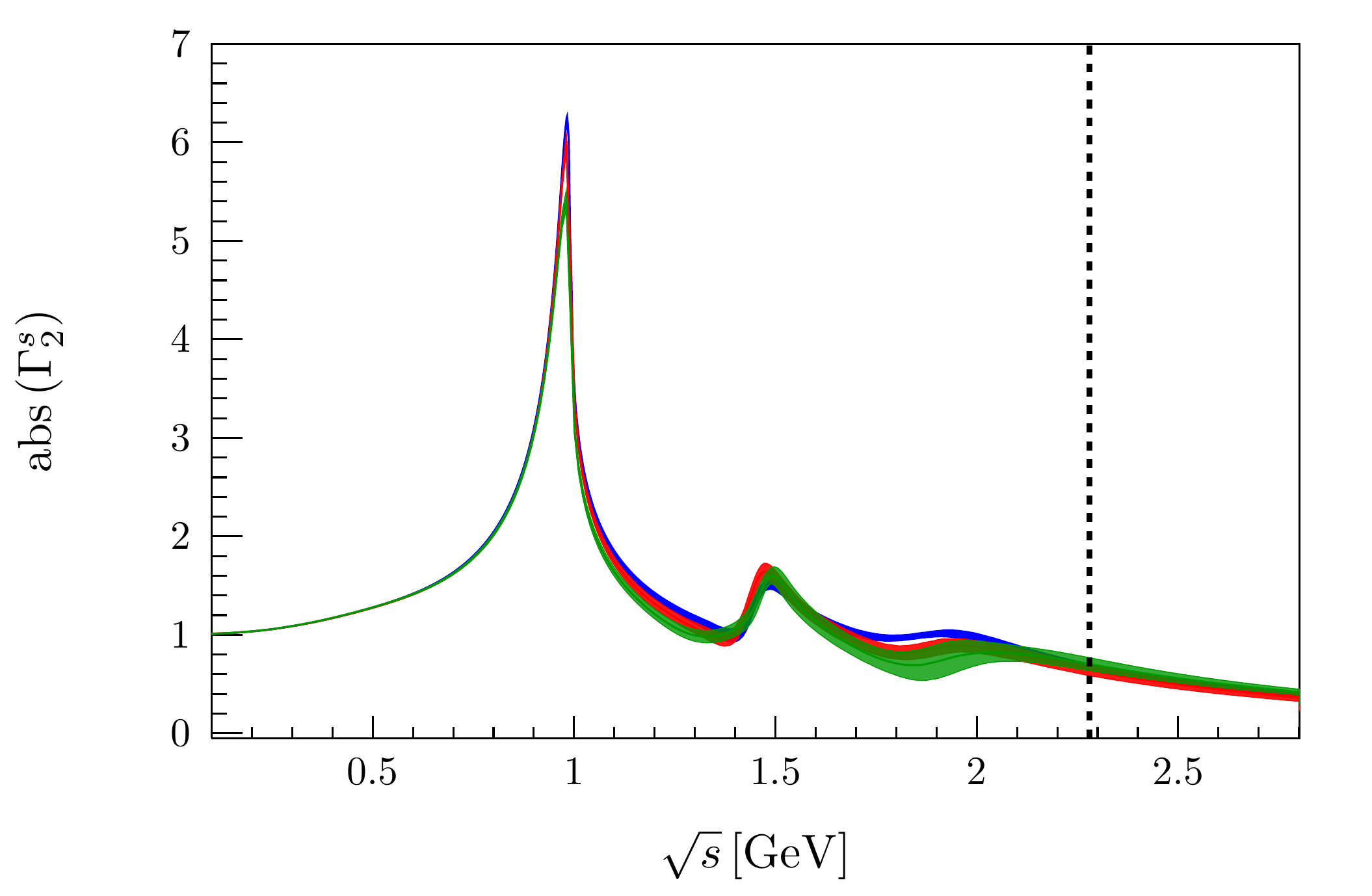} \hfill
	\includegraphics*[width=0.495\linewidth]{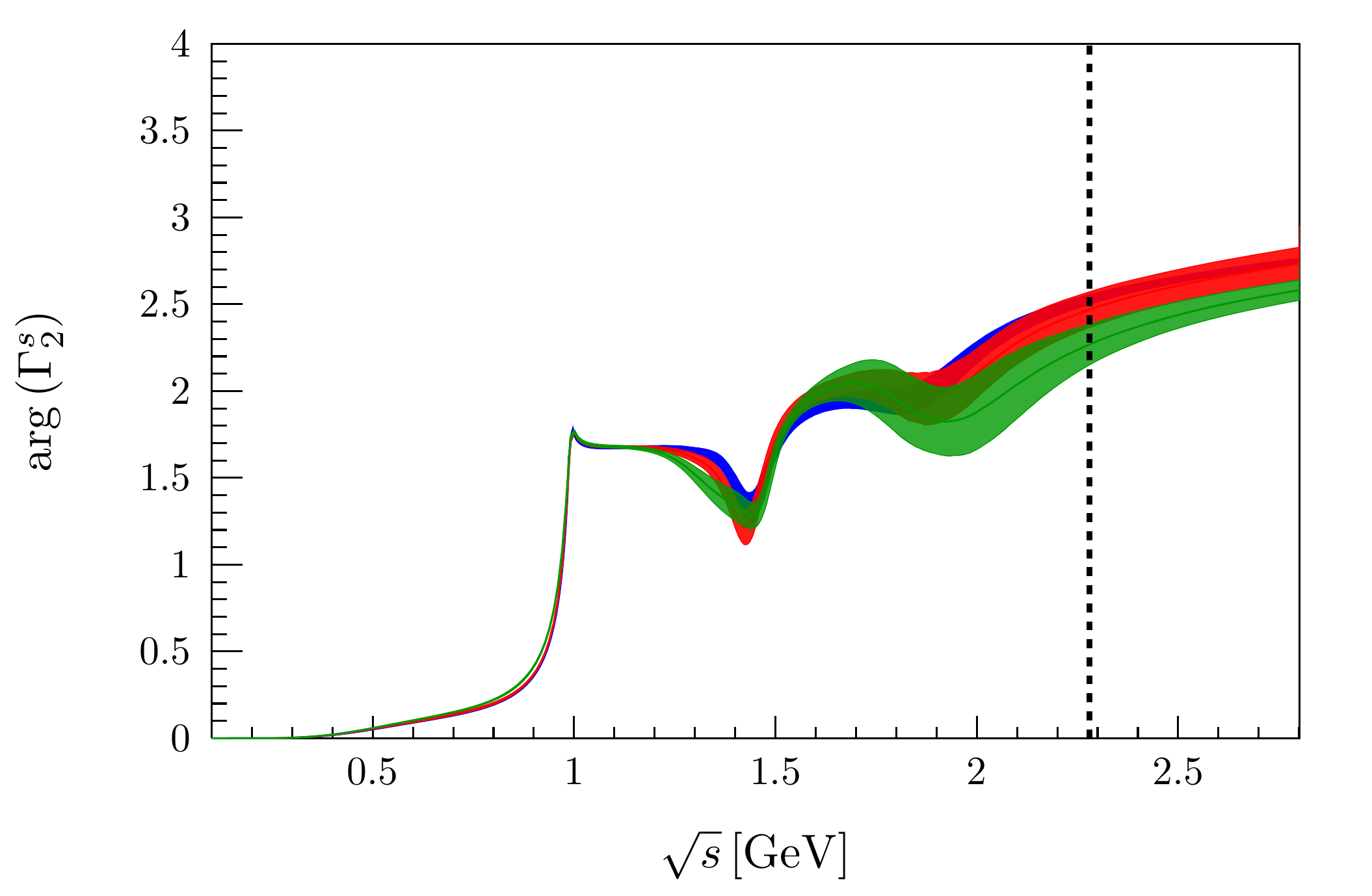}
\caption{Modulus (left) and phase (right) of the kaon form factor $\Gamma_2^s$ for the fits with an additional $\rho\rho$ (top) and $\sigma\sigma$ (bottom) channel. Fit~1 is shown in blue, Fit~2 in red, and Fit~3 in green. The dotted vertical lines mark the kinematic upper limit for $\sqrt{s}$ in the $\bar B_s^0$ decay.}
\label{Fig::Formfactor::FF2}
\end{figure*}

By comparing the extracted kaon form factors $\Gamma_2^s$ in Fig.~\ref{Fig::Formfactor::FF2} 
we see very similar  features as for the pion form factor.
However, the $f_0(1500)$ couples more weakly to the $K\bar{K}$ channel than to $\pi\pi$, which is in line with what is reported about this state by the PDG~\cite{PDG}. The impact of the additional resonance in Fit~3 that appears outside the accessible data range is even more pronounced.

\begin{figure*}
\centering
	\includegraphics[width=0.495\linewidth]{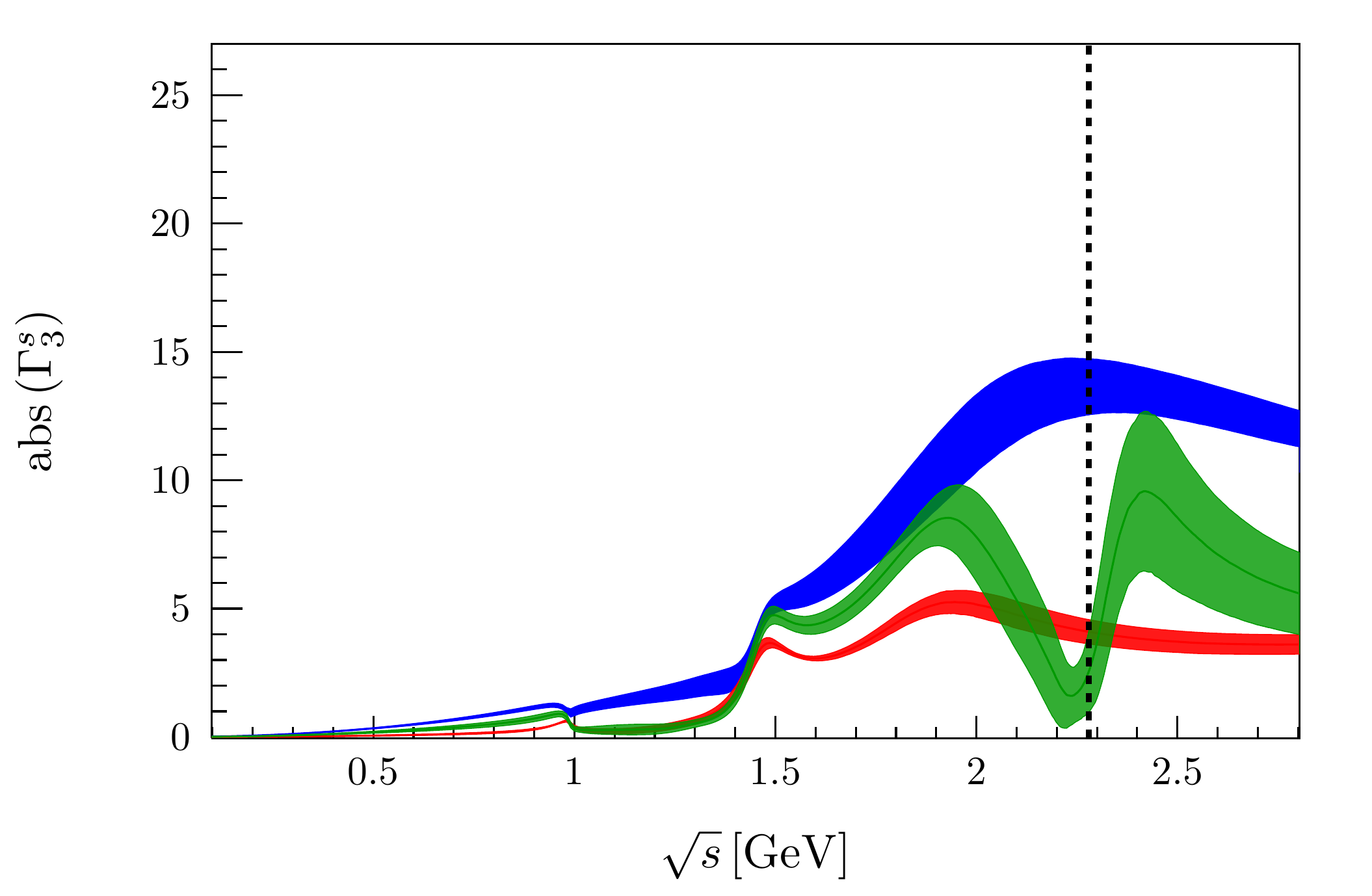} \hfill
	\includegraphics[width=0.495\linewidth]{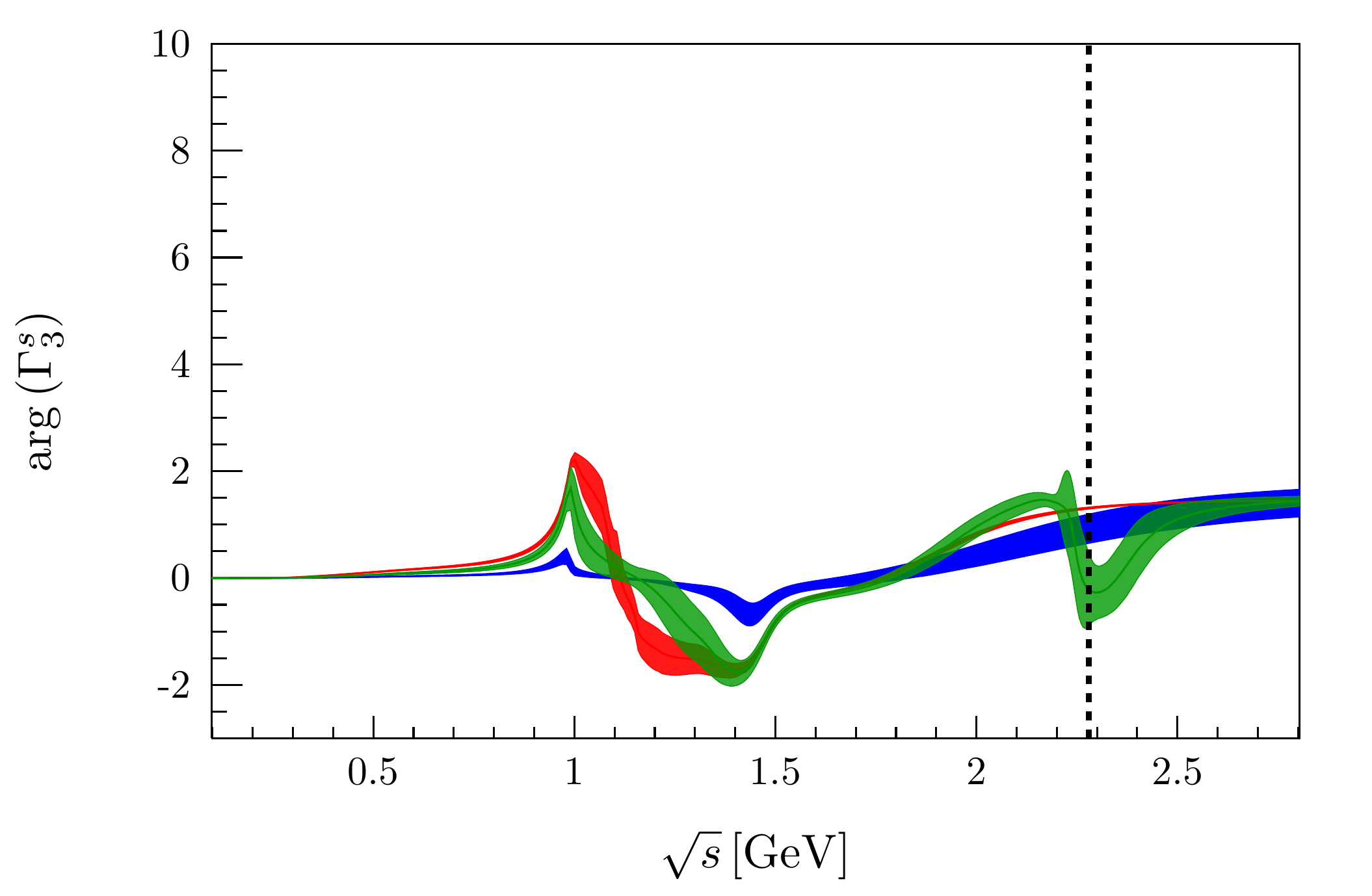} \\
	\includegraphics[width=0.495\linewidth]{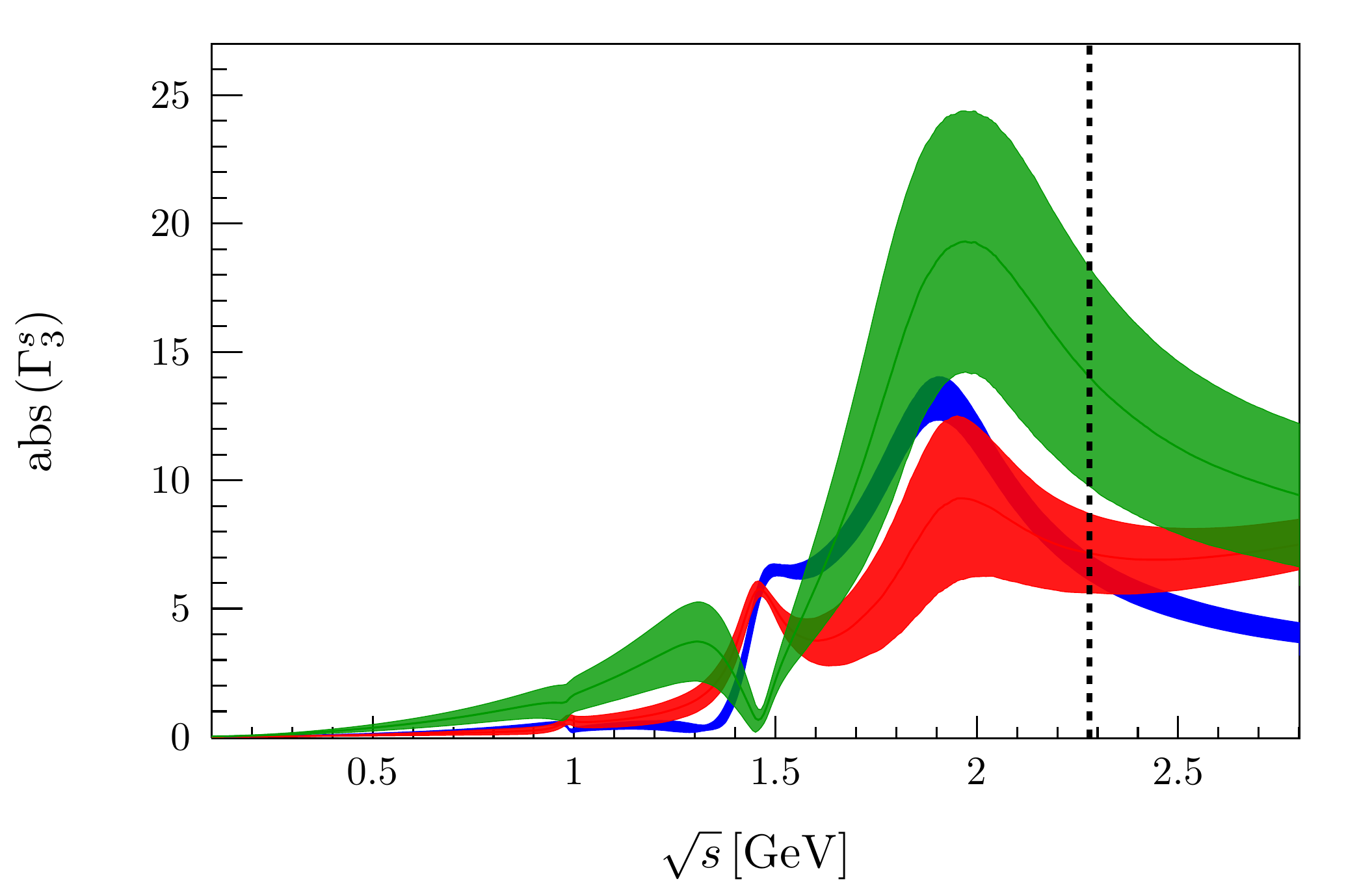} \hfill
	\includegraphics[width=0.495\linewidth]{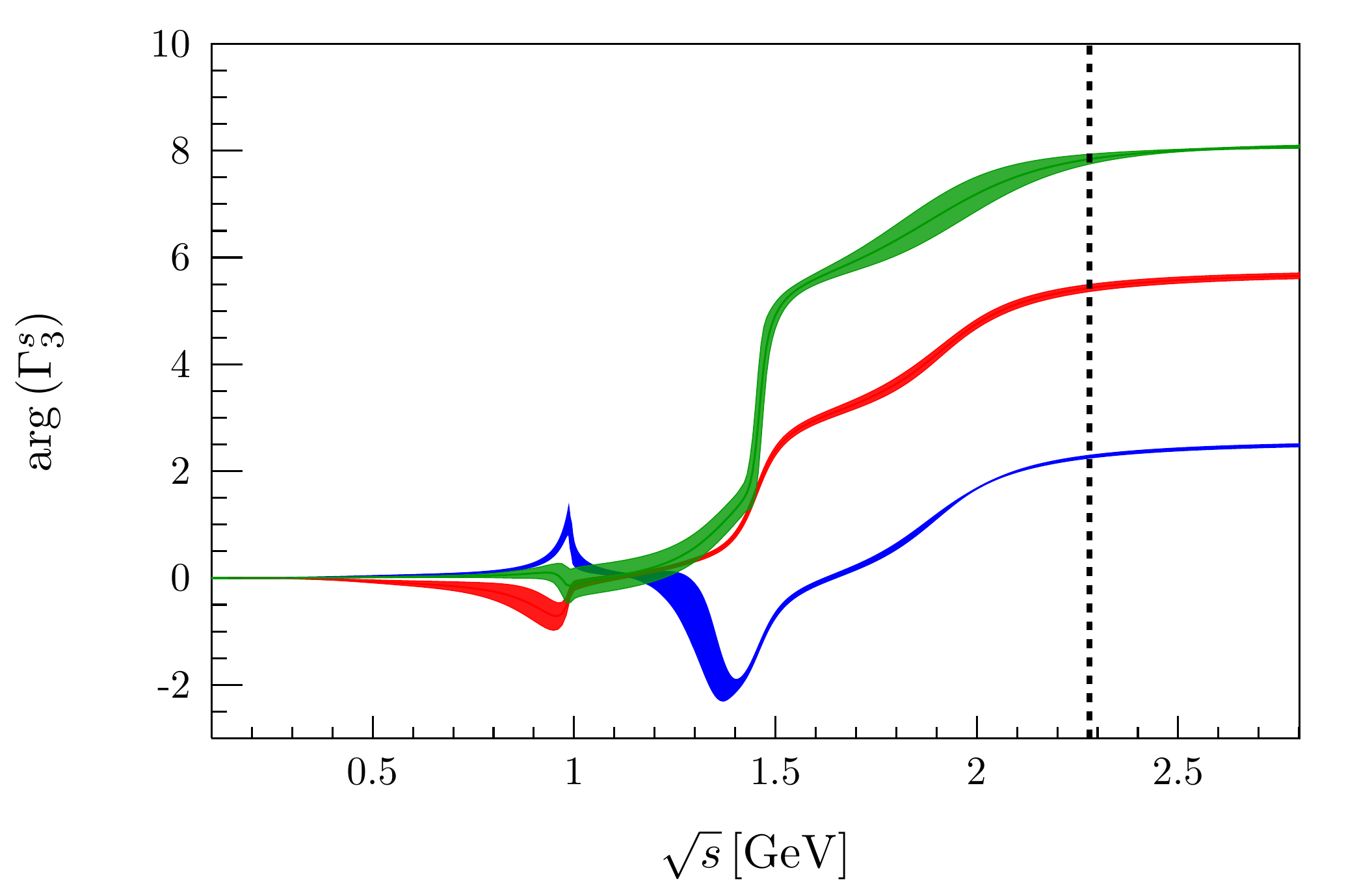}
\caption{Modulus (left, in arbitrary units) and phase (right) of the effective $4\pi$ form factor $\Gamma_3^s$ for the fits with an additional $\rho\rho$ (top) and $\sigma\sigma$ (bottom) channel. Fit~1 is shown in blue, Fit~2 in red, and Fit~3 in green. The dotted vertical lines mark the kinematic upper limit for $\sqrt{s}$ in the $\bar B_s^0$ decay.}
\label{Fig::Formfactor::FF3}
\end{figure*}

In Fig.~\ref{Fig::Formfactor::FF3} we compare the form factor of the additional, effective $4\pi$, channel $\Gamma_3^s$. 
We see that the results of the fits with the $4\pi$ channel parametrized as $\rho\rho$ 
differ significantly from the ones employing the $\sigma\sigma$ variant. Moreover, also Fits~1--3 
differ strongly from each other, even in the kinematic regime that can be reached in $\bar B_s^0$ decays.
To further constrain these amplitudes it is compulsory to include data on $\bar B_s^0\to J/\psi 4\pi$ in
the analysis, which is so far unavailable in partial-wave-decomposed form~\cite{Aaij:2013rja}.

Finally in Fig.~\ref{Fig::Poles::T11} we show the phases, $\delta$, and inelasticities, $\eta$, that result for $T_{11}$ in the different fits,
where we use the standard parametrization
\begin{equation}
T_{11}=\left(\eta e^{2i\delta}-1\right)/(2i\sigma_\pi) \,.
\end{equation}
In the figure we also show the two-channel input phase $\delta_0$ and inelasticity $\eta_0$ introduced in Eq.~\eqref{eq:inputdelandin} as
black solid lines.
The comparison of the different lines demonstrates that the high-energy extension maps smoothly onto the 
low-energy input, as it should. In the phases one clearly sees the effect of the $f_0(1500)$, which leads to a deviation
of the phase of $T_{11}$ from the input phase. In the inelasticity the full model starts to deviate from the input already 
at about $1.1\,\GeV$ as a consequence of the inclusion of the $4\pi$ channel. As in the phase the $f_0(1500)$ also leads
to a pronounced structure in the inelasticity. It is interesting to observe that neither in
the phase nor in the inelasticity there is a clear imprint of the $f_0(2020)$, which can be understood from its small coupling to 
the two-pion channel. 

In  Fig.~\ref{Fig::Poles::T11} we also show a comparison of our phases and inelasticities to those extracted in 
Ref.~\cite{Anisovich:2002ij} (plotted as purple dashed lines) and the preferred solution~\cite{Ochs:2013gi} 
of the CERN--Munich $\pi\pi$ experiment~\cite{Hyams:1975mc} (data points with error bars).
As one can see in the phase shifts,
all analyses agree up to about $1.5\,\GeV$. However, the effect of the $f_0(1500)$, present in all analyses, is very
different. Also for the inelasticity there is no agreement between our solution and those from the
two other sources, but here the deviation starts basically with the onset of the $\bar KK$ channel; for a
more detailed discussion of the current understanding of the inelasticity in the scalar isoscalar channel, we
refer to Ref.~\cite{GarciaMartin:2011cn}. 
Note that there is also no agreement between the amplitudes of Ref.~\cite{Ochs:2013gi} 
and Ref.~\cite{Anisovich:2002ij}. Thus, at this time one is to conclude that $T_{11}$ above $1.1\,\GeV$ 
is not yet known.

\begin{figure*}
	\centering
	\includegraphics[width=0.495\linewidth]{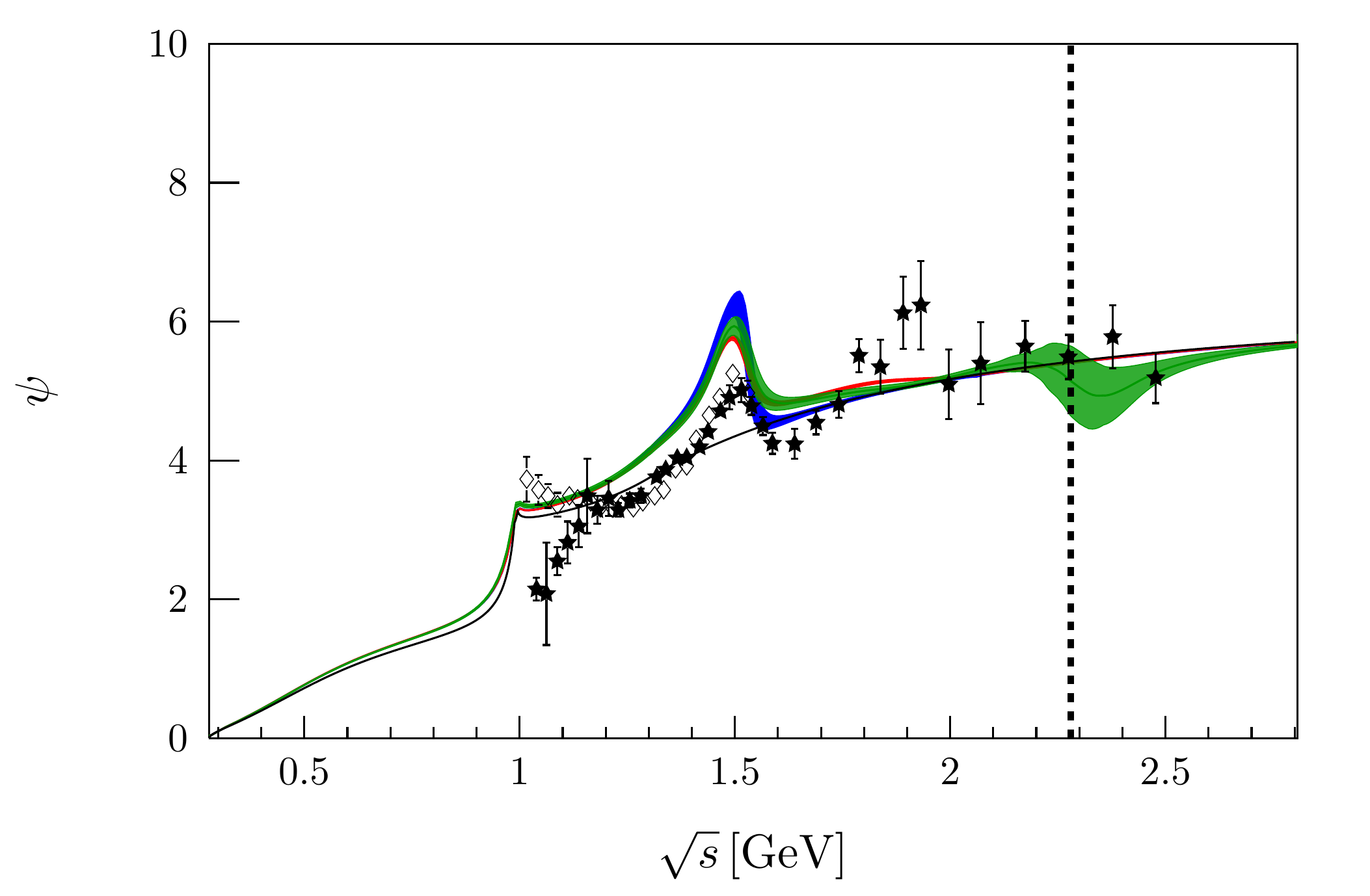} \hfill
	\includegraphics[width=0.495\linewidth]{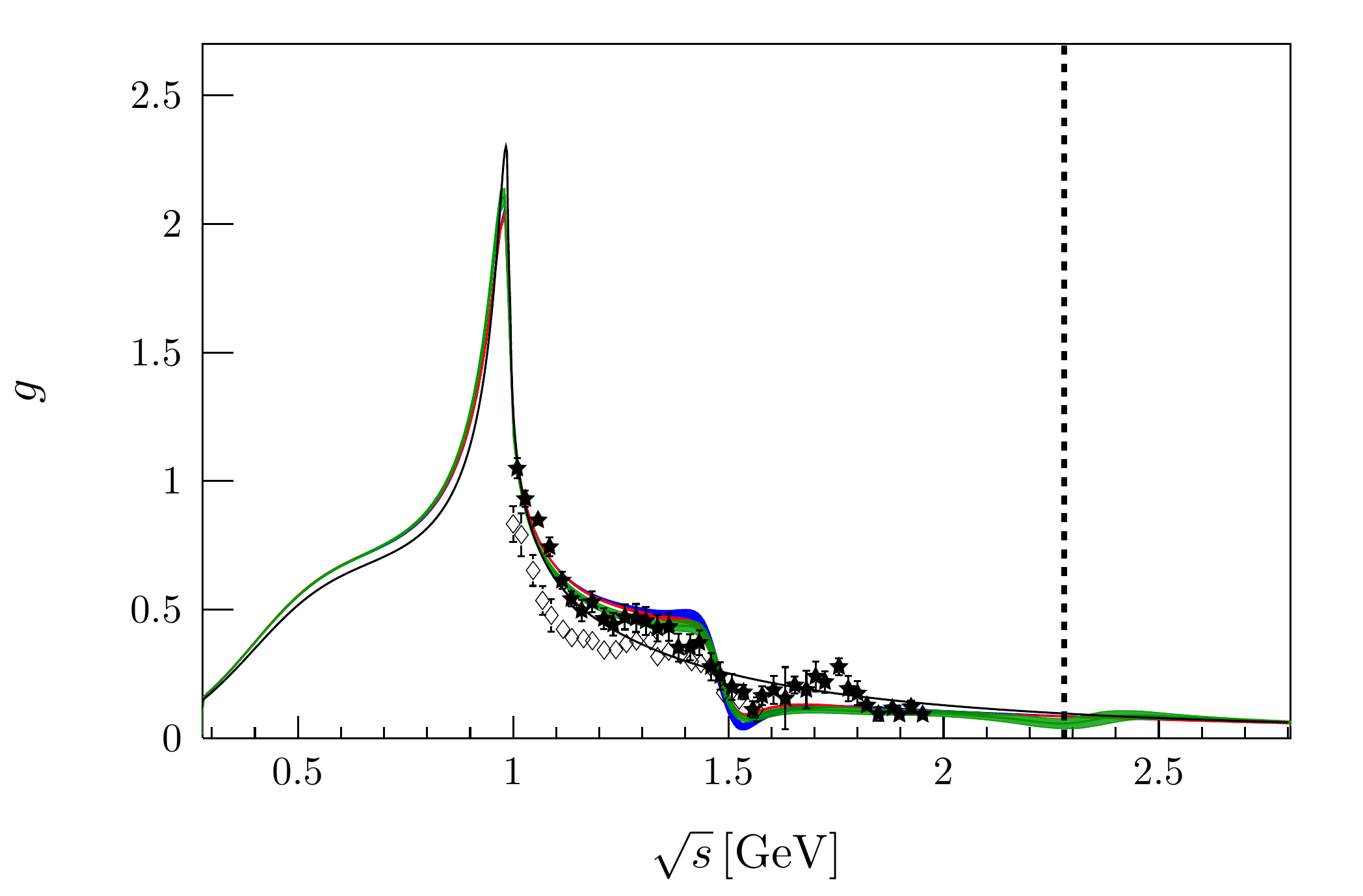} \\
	\includegraphics[width=0.495\linewidth]{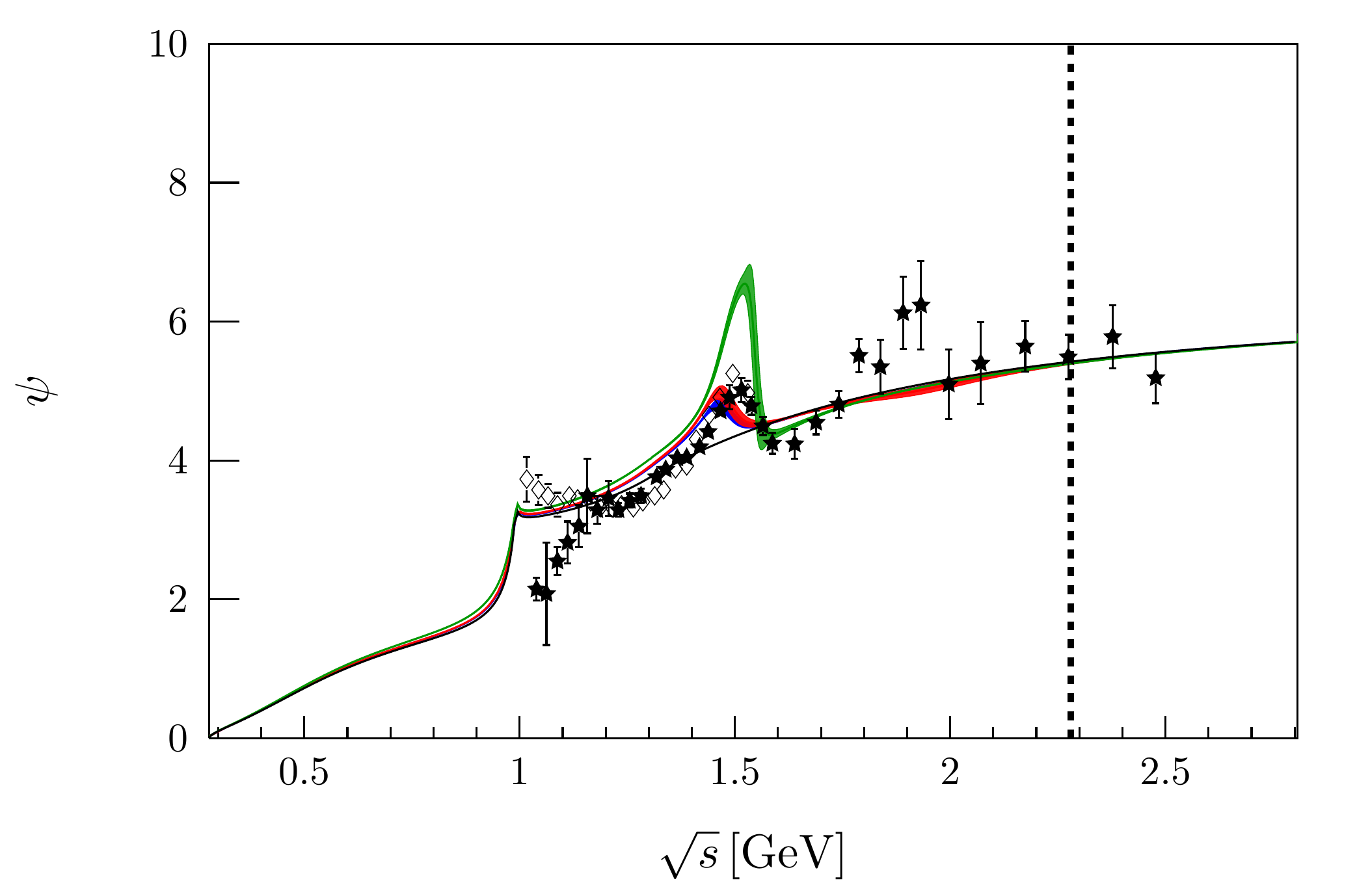} \hfill
	\includegraphics[width=0.495\linewidth]{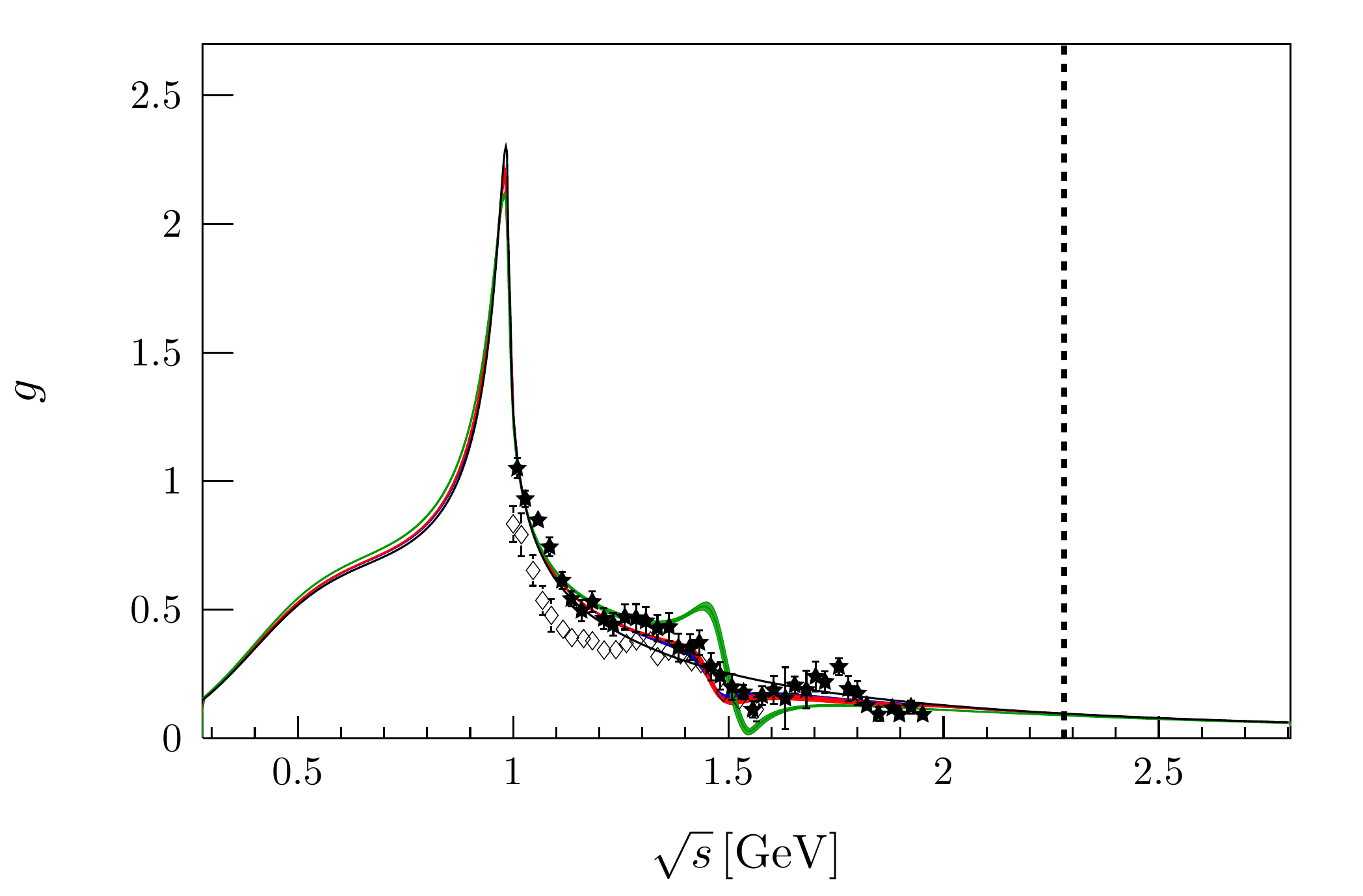}
\caption{
Scalar isoscalar $\pi\pi\to K\bar{K}$ scattering phase shift $\psi$ (left) and absolute value $g$ (right) defined by the $S$-wave amplitude $T_{12}=g\,e^{i\psi}$ for the fits with an additional $\rho\rho$ (top) and $\sigma\sigma$ (bottom) channel. Fit~1 is shown in blue, Fit~2 in red, Fit~3 in green, and the input $g_0$ and $\psi_0$~\cite{Dai:2014zta} in black. For comparison we show the amplitude analyses of Refs.~\cite{Cohen:1980cq} (open diamonds) and \cite{Etkin:1981sg} (filled stars).}\label{Fig::Poles::T12}
\end{figure*}

In a similar way, we can also compare the extracted $\pi\pi\to K\bar{K}$ scattering amplitude $T_{12}$ with its absolute value $g$ as well as its phase $\psi$, which are both shown in Fig.~\ref{Fig::Poles::T12}. While the resonance effects of the $f_0(1500)$ look qualitatively well-described by our high-energy extension, we see some differences to the actual data~\cite{Cohen:1980cq,Etkin:1981sg}. Note that the shown results are a prediction based solely on the $\bar B_s^0$ decay data and could be improved upon by explicitly taking the phase motion into account in the fit.

\section{Extraction of resonance poles}\label{sec::poles}
\bsp
In this section we present the extraction of resonance poles in the complex $s$-plane from the parametrizations discussed above. 
Traditionally those are given in terms of a mass $M$ and a width $\Gamma$, connected to the pole position $s_p$ via~\cite{PDG}
\begin{equation}
\sqrt{s_p} = M - i\frac{\Gamma}{2} \,.
\end{equation}
For narrow resonances located far from relevant thresholds, these parameters agree with the standard Breit--Wigner parameters. However,
for broad and/or overlapping states, significant deviations can occur between the parameters derived from the pole location and those from Breit--Wigner fits.
Since the analytic continuation to different Riemann sheets needs the on-shell scattering $T$-matrix as input, which, due to left-hand cuts
induced by crossing symmetry, has a complicated analytic structure
that cannot be deduced from the phase shifts straightforwardly,
we use the framework of Pad\'e approximants to search for the poles on the nearest unphysical sheets. 
For a thorough introduction into this topic, see e.g.\ Refs.~\cite{Masjuan:2013jha,Masjuan:2014psa,Caprini:2016uxy}.

\begin{figure*}
	\centering
	\includegraphics[width=0.495\linewidth]{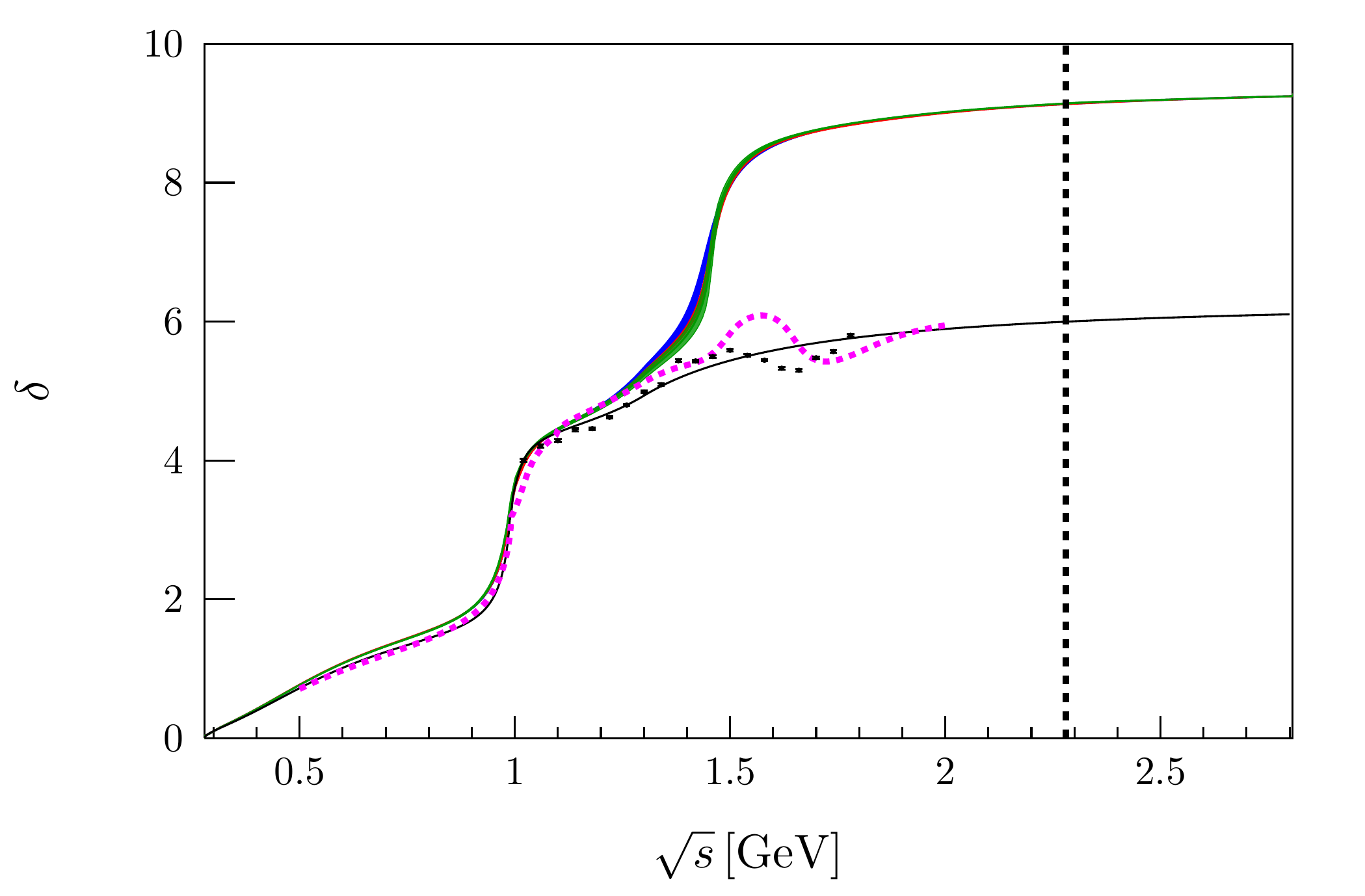} \hfill
	\includegraphics[width=0.495\linewidth]{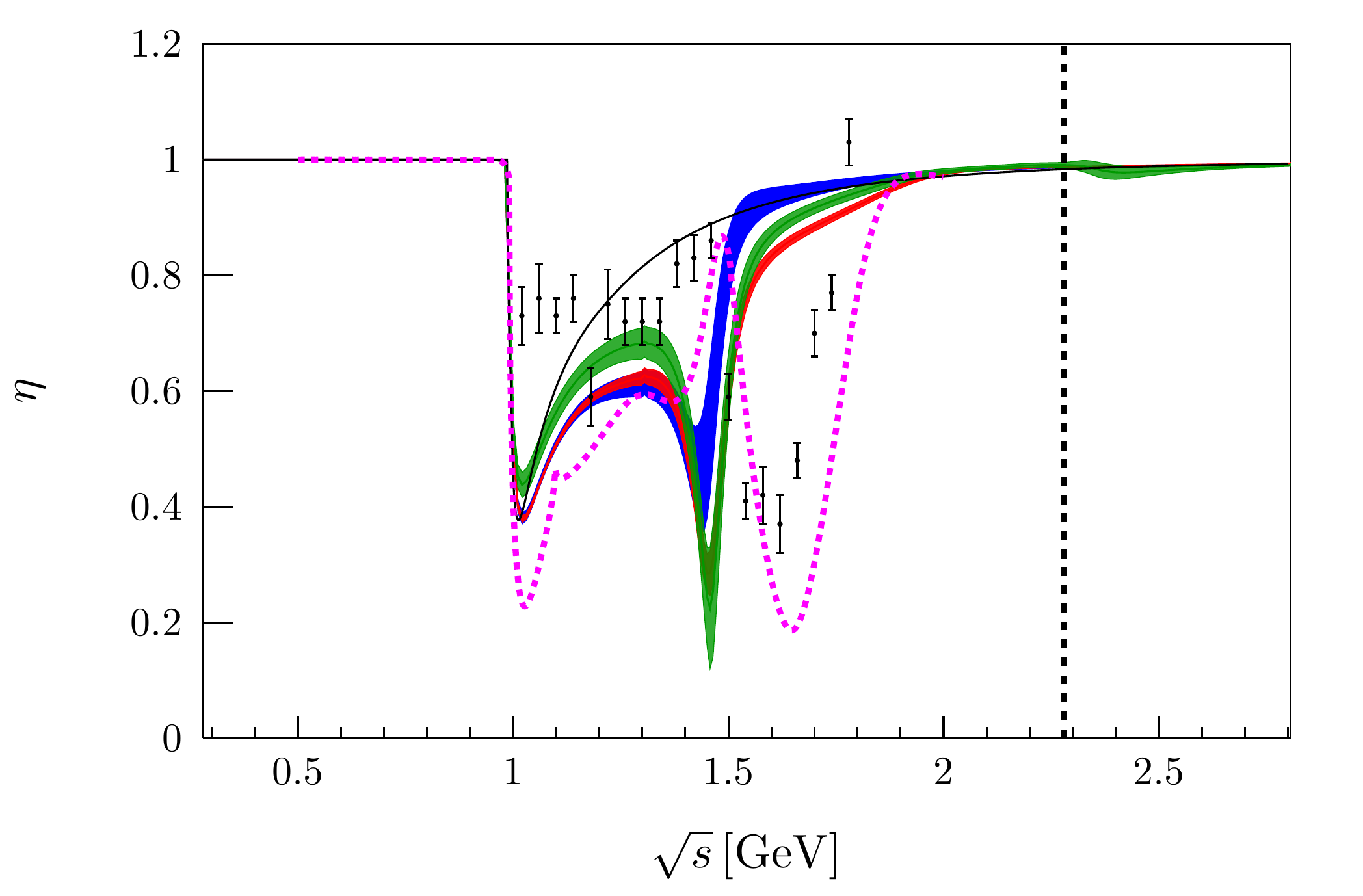} \\
	\includegraphics[width=0.495\linewidth]{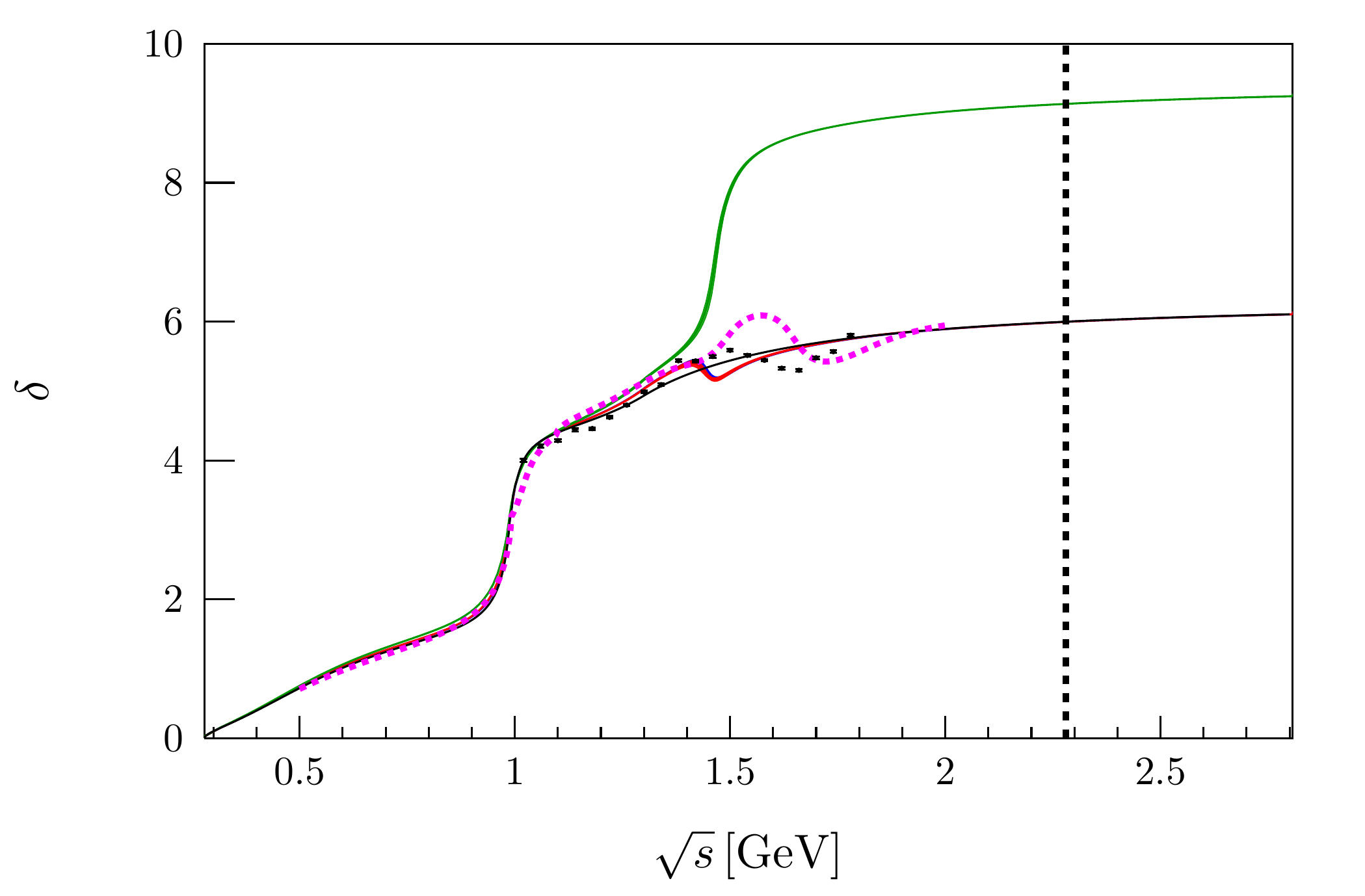} \hfill
	\includegraphics[width=0.495\linewidth]{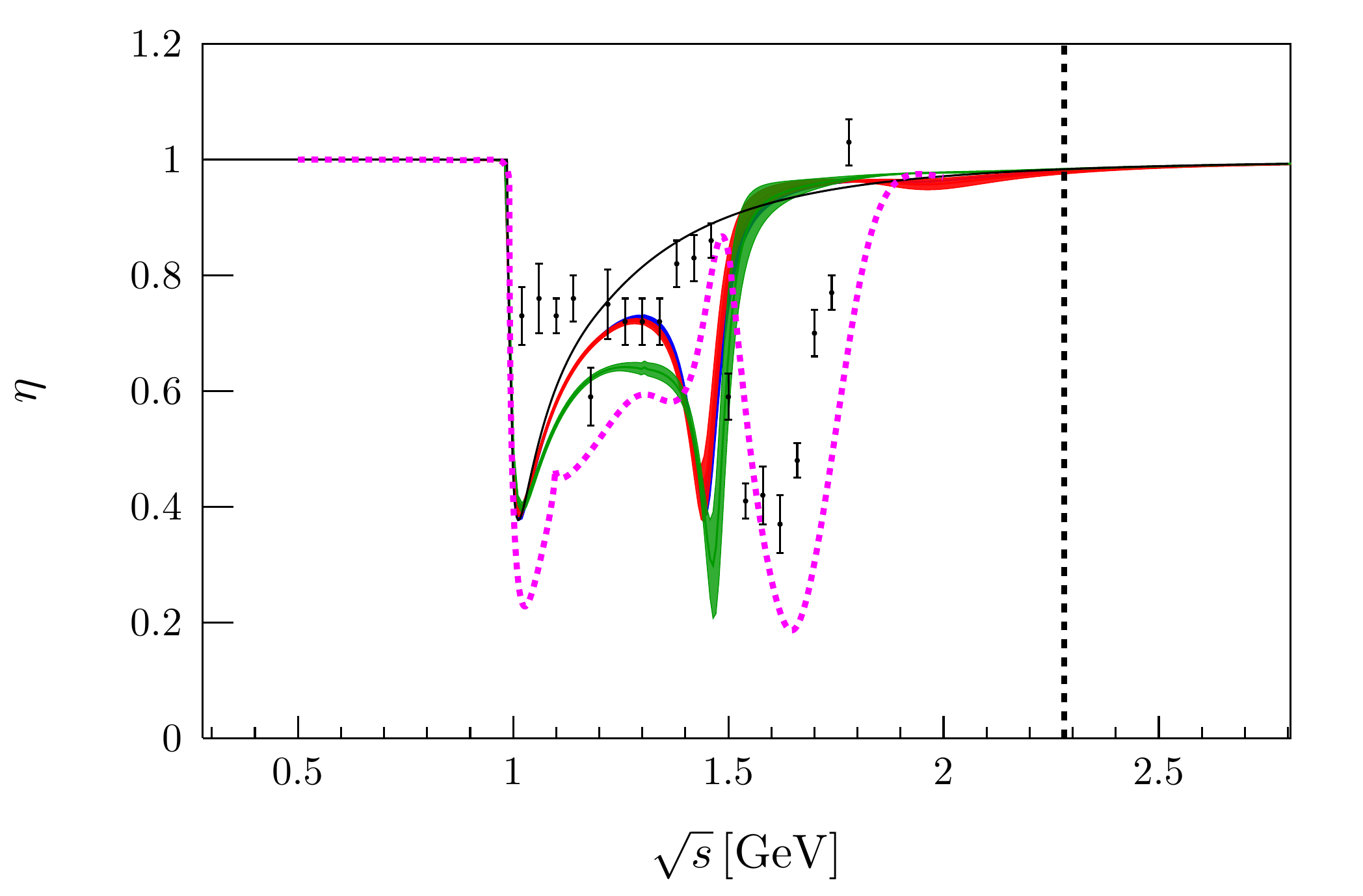}
\caption{Scalar isoscalar pion--pion scattering phase shift $\delta$ (left) and inelasticity $\eta$ (right) defined by the $\pi\pi$ $S$-wave amplitude $T_{11}=\left(\eta e^{2i\delta}-1\right)/(2i\sigma_\pi)$ for the fits with an additional $\rho\rho$ (top) and $\sigma\sigma$ (bottom) channel. Fit~1 is shown in blue, Fit~2 in red, Fit~3 in green, and the input $\delta_0$ and $\eta_0$~\cite{Dai:2014zta} in black. The purple dashed line shows the $K$-matrix solution of Ref.~\cite{Anisovich:2002ij}. In addition we plot the preferred phase shifts and inelasticities~\cite{Ochs:2013gi} of the CERN--Munich $\pi\pi$ experiment~\cite{Hyams:1975mc}, which are denoted by data points with error bars.}
\label{Fig::Poles::T11}
\end{figure*}

As the form factor $\Gamma_1^s(s)$ (Fig.~\ref{Fig::Formfactor::FF1}) as well as $T_{11}(s)$ (Fig.~\ref{Fig::Poles::T11}) are smooth functions
when moving from the upper complex $s$-plane of the first Riemann sheet to the lower complex $s$-plane of the neighboring unphysical sheet, we may expand both around
some properly chosen expansion point $s_0$ according to
\begin{align}
	P_M^N(s,s_0)=\frac{\sum_{n=0}^N a_n(s-s_0)^n}{1+\sum_{m=1}^M b_m (s-s_0)^m}\,.
\end{align}
The denominator allows for the inclusion of $M$ resonance poles lying on the unphysical Riemann sheet. In the following we set $M$ to $1$, allowing for the extraction of the resonance that lies closest to the expansion point $s_0$. The numerator ensures the convergence of the series to the form factor or the scattering matrix for $N\rightarrow \infty$. In order to obtain the complex parameters $a_n$ and $b_n$, we fit Pad\'e approximants to both the form factor and the scattering matrix simultaneously. As both $T_{11}$ and $\Gamma^{s}_1$ have the same poles, the parameters $b_n$ are the same for both, however, the $a_n$ are different. Note furthermore that the $a_0$ parameters 
are constrained by $\Gamma_1^s(s_0)$ or $T_{11}(s_0)$, respectively.

For near-threshold poles such as the $f_0(500)$ and $f_0(980)$, we perform the Pad\'e approximation not in $s$, but in the conformal variable
\begin{align}
	w(s)=\frac{\sqrt{s-4M_\pi^2}-\sqrt{4M_K^2-s}}{\sqrt{s-4M_\pi^2}+\sqrt{4M_K^2-s}}
\end{align}
instead~\cite{Masjuan:2013jha}. 
This variable transformation maps the upper half complex $s$-plane of the first Riemann sheet to the inner upper half of a unit circle in the complex $w$ plane, without introducing any unphysical discontinuities. The lower half of the second Riemann sheet is then mapped onto the lower half of the unit circle in the complex $w$-plane. 
This allows us to search for the two lowest poles
 within a circle around the expansion point $s_0$, without being limited by the proximity of the $\pi\pi$ and $K\bar{K}$ thresholds, 
which are automatically taken care of.

The statistical uncertainty is obtained through a bootstrap analysis of the fit results presented in Sect.~\ref{sec:fit}. The systematic uncertainty coming from the Pad\'e approximation on the other hand is estimated by~\cite{Masjuan:2014psa}
\begin{align}
	\Delta^N=\left|\sqrt{s_p^N}-\sqrt{s_p^{N-1}}\right|\,,
\end{align}
where $s_p^N$ denotes the pole extracted by employing $P_1^N(s,s_0)$.

As in principle the results still depend on the expansion point $s_0$, we proceed as follows. We first calculate Pad\'e approximants for a varying $s_0$; near the true pole position, the extracted Pad\'e pole stabilizes. Finally we choose the $s_0$ that minimizes $\Delta^N$ for the maximum order of $N$ employed.

Corresponding residues of the poles are then described by the coupling strength $g_{R\pi\pi}$ of the resonance $R$ to $\pi\pi$ and the coupling $g_{Rss}$ of the $\bar{s}s$ source to the resonance $R$. They are defined by the near-pole expansions~\cite{Moussallam:2011zg,GarciaMartin:2011jx}
\begin{align}
	\lim_{s\rightarrow s_p}T_{11}(s) &=\frac{r_T}{s_p-s}=\frac{g_{R\pi\pi}^2}{32\pi (s_p-s)} \,,\nl
	\lim_{s\rightarrow s_p}\Gamma_1^s(s) &=\frac{r_\Gamma}{s_p-s}=-\frac{g_{R\pi\pi}g_{Rss}}{\sqrt{3}(s_p-s)}\,.
\end{align}
The extracted poles and residues for the resonances are shown in Table~\ref{tab::poles::lowerpoles3}. 

\begin{table*}
	\centering
	\renewcommand{\arraystretch}{1.4}
	\resizebox{\textwidth}{!}{
	\begin{tabular}{cccccccccc}
		\toprule
		& & Fit & $\frac{\sqrt{s_0}}{\GeV}$ & $\Re\sqrt{s_p}/\MeV$ & $-2\times\Im\sqrt{s_p}/\MeV$ & $|r_T|/\GeV^2$ & $\arg(r_T)$& $|r_\Gamma|/\GeV^2$ & $\arg(r_\Gamma)$\\
		\midrule
		$f_0(500)$ & $\rho\rho$ & $1$ & $0.481$ & $441\pm1$  & $504\pm2$  & $0.204\pm0.002$  & $-145\pm1$  & $0.0309\pm0.0028$  & $-160\pm3$  \\
		$f_0(500)$ & $\sigma\sigma$ &  $1$ & $0.466$ & $440\pm1$ & $ 521\pm1$ & $0.205\pm0.001$ & $-149\pm1$ & $0.0254\pm0.0010$ & $-169\pm2$ \\
		$f_0(500)$ & $\rho\rho$ & $2$ & $0.483$ & $441\pm1$  & $503\pm1$  & $0.204\pm0.001$  & $-145\pm1$  & $0.0275\pm0.0010$  & $-159\pm2$ \\
		$f_0(500)$ & $\sigma\sigma$ & $2$ & $0.486$ & $443\pm1$ & $521\pm2$ & $0.205\pm0.002$ & $-147\pm1$ & $0.0279\pm0.0032$ & $-161\pm4$ \\
		$f_0(500)$ & $\rho\rho$ & $3$ & $0.481$  & $441\pm2$  & $505\pm3$  & $0.202\pm0.002$  & $-145\pm2$  & $0.0279\pm0.0039$  & $-159\pm4$  \\
		$f_0(500)$ & $\sigma\sigma$ & $3$ & $0.485$ & $442\pm1$ & $510\pm1$ & $ 0.203\pm0.001$ & $-146\pm1$ & $0.0284\pm0.0023$ & $-161\pm3$\\
		\bottomrule
		$f_0(980)$ & $\rho\rho$ & $1$ & $0.941$  & $998\pm2$  & $65\pm3$  & $0.099\pm0.006$  & $-164\pm3$  & $0.258\pm0.016$  & $107\pm4$   \\
		$f_0(980)$ & $\sigma\sigma$ & $1$ & $0.941$ & $998\pm1$ & $48\pm2$ & $0.082\pm0.007$ & $-164\pm5$ & $0.258\pm0.019$ & $109\pm5$  \\
		$f_0(980)$ & $\rho\rho$ & $2$ & $0.941$  & $1001\pm2$  & $65\pm3$  & $0.114\pm0.011$  & $-160\pm6$  & $0.270\pm0.020$  & $109\pm5$   \\
		$f_0(980)$ & $\sigma\sigma$ & $2$ & $0.941$ & $998\pm1$ & $50\pm2$ & $0.082\pm0.006$ & $-166\pm5$ & $0.249\pm0.014$ & $108\pm4$ \\
		$f_0(980)$ & $\rho\rho$ & $3$ & $0.941$  & $993\pm3$ & $65\pm3$  & $0.094\pm0.005$ & $-168\pm3$  & $0.261\pm0.012$  & $103\pm3$  \\
		$f_0(980)$ & $\sigma\sigma$ & $3$ & $0.941$ & $998\pm2$ & $60\pm2$ & $0.099\pm0.007$ & $-163\pm5$ & $0.281\pm0.016$ & $109\pm4$ \\
		\midrule
		$f_0(1500)$ & $\rho\rho$ & $1$ & $1.459$  & $1460\pm6$  & $109\pm7$  & $0.131\pm0.017$  & $-82\pm3$  & $0.18\pm0.03$  & $-53\pm5$  \\
		$f_0(1500)$ & $\sigma\sigma$ & $1$ & $1.449$ & $1456\pm4$ & $107\pm8$ & $0.047\pm0.005$ & $-86\pm3$ & $0.23\pm0.02$ & $-74\pm4$\\
		$f_0(1500)$ & $\rho\rho$ & $2$ & $1.517$  & $1465\pm4$  & $116\pm4$  & $0.115\pm0.007$  & $-86\pm 2$  & $0.18\pm0.02$  & $-50\pm2$  \\
		$f_0(1500)$ & $\sigma\sigma$ & $2$ & $1.449$ & $1452\pm5$ & $103\pm8$ & $0.045\pm0.005$ & $-82\pm6$ & $0.23\pm0.02$ & $-54\pm6$ \\
		$f_0(1500)$ & $\rho\rho$ & $3$ & $1.466$  & $1465\pm5$  & $105\pm7$  & $0.097\pm0.018$  & $-87\pm3$  & $0.18\pm0.03$  & $-57\pm4$  \\
		$f_0(1500)$ & $\sigma\sigma$ & $3$ & $1.476$ & $1477\pm6$ & $90\pm9$ & $0.097\pm0.010$ & $-86\pm7$ & $0.12\pm0.04$ & $-51\pm16$ \\
		\midrule
		$f_0(2020)$ & $\rho\rho$ & $1$ & $2.145$  & $1996\pm67$  & $998\pm163$  & $0.215\pm0.407$  & $4\pm82$  & $2.23\pm0.62$  & $18\pm15$\\
		$f_0(2020)$ & $\sigma\sigma$ & $1$ & $1.900$ & $1888\pm9$ & $344\pm12$ & $0.005\pm0.002$ & $-104\pm24$ & $0.48\pm0.04$ & $106\pm4$ \\
		$f_0(2020)$ & $\rho\rho$ & $2$  & $1.949$  & $1869\pm9$  & $461\pm15$  & $0.026\pm0.013$  & $31\pm33$  & $0.51\pm0.06$  & $-10\pm11$ \\
		$f_0(2020)$ & $\sigma\sigma$ & $2$ & $1.900$ & $1908\pm10$ & $344\pm19$ & $0.008\pm0.006$ & $-101\pm64$ & $0.41\pm0.10$ & $103\pm13$ \\
		$f_0(2020)$ & $\rho\rho$ & $3$ & $1.949$  & $1919\pm23$  & $366\pm47$  & $0.011\pm0.006$  & $77\pm51$  & $0.45\pm0.11$  & $32\pm15$ \\
		$f_0(2020)$ & $\sigma\sigma$ & $3$ & $1.900$ & $1910\pm50$ & $414\pm42$ & $0.014\pm0.016$ & $82\pm69$ & $0.72\pm0.34$ & $66\pm34$ \\
		\bottomrule
	\end{tabular}
	}
	\caption{Pad\'e poles for $f_0(500)$, $f_0(980)$, and $f_0(1500)$ for $N=5$, as well as $f_0(2020)$ for $N=6$. The error is the uncorrelated sum of statistical and systematic uncertainty.}
	\label{tab::poles::lowerpoles3}
	\renewcommand{\arraystretch}{1}
\end{table*}

As we did not include any variation of the input phases, we see that the statistical uncertainty coming from the fit parameters of the higher-mass resonances has only a small impact on the poles of $f_0(500)$ and $f_0(980)$. In fact the uncertainty is dominated by the systematic error coming from the Pad\'e expansion. At higher energies the statistical uncertainty becomes more significant. 

\begin{figure*}
	\centering
	\includegraphics[width=0.495\linewidth]{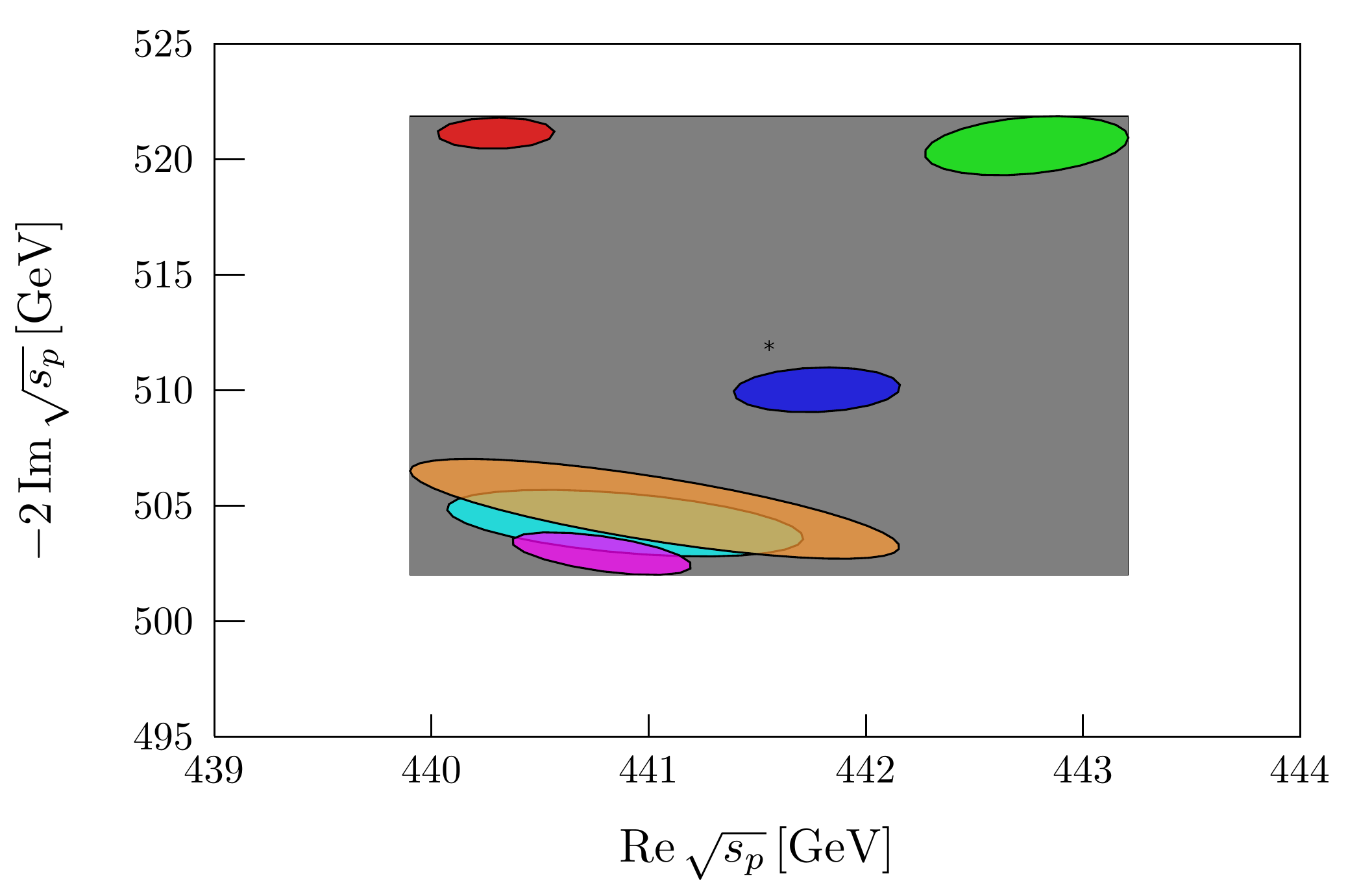} \hfill
	\includegraphics[width=0.495\linewidth]{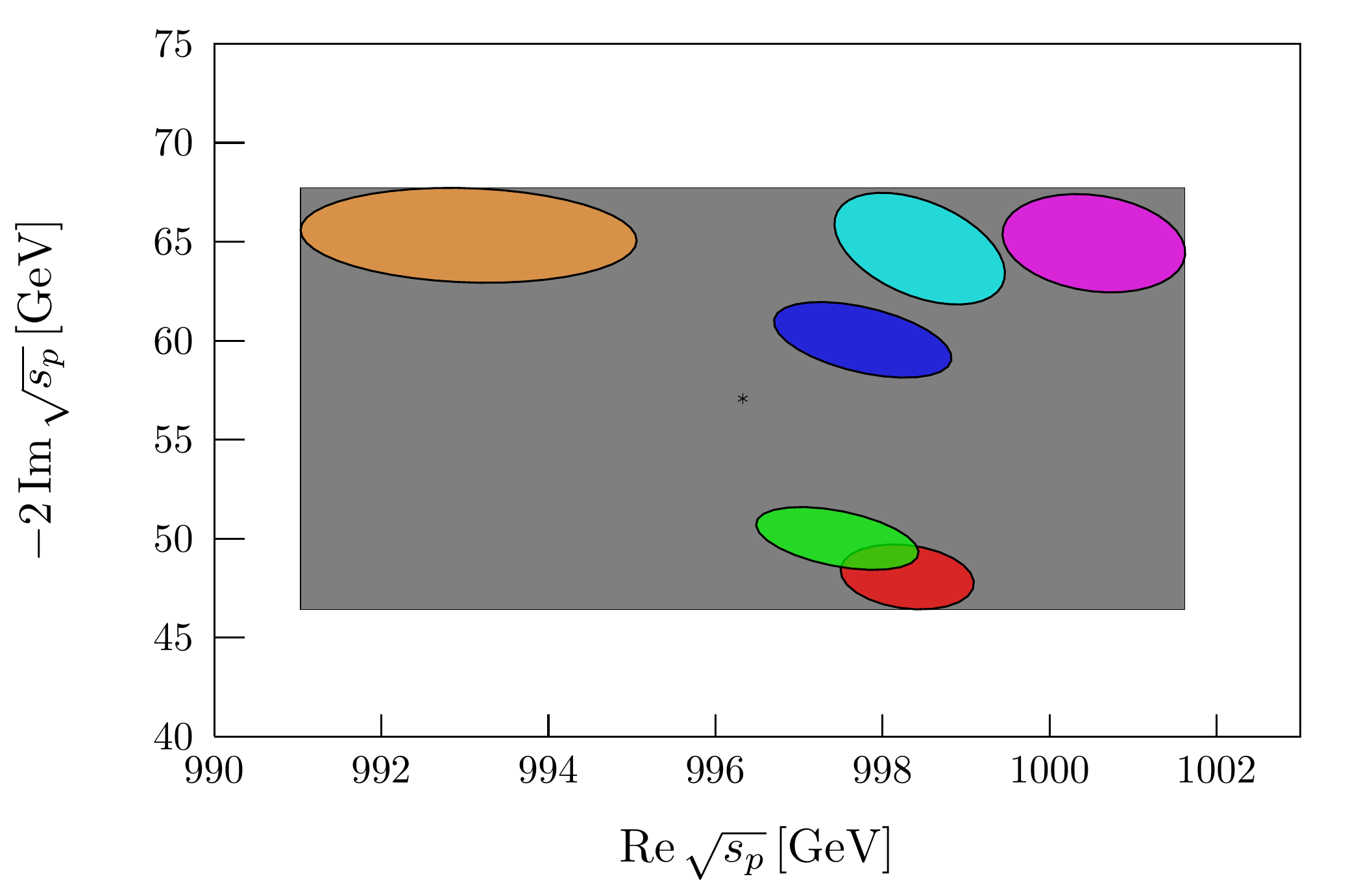} \\
	\includegraphics[width=0.495\linewidth]{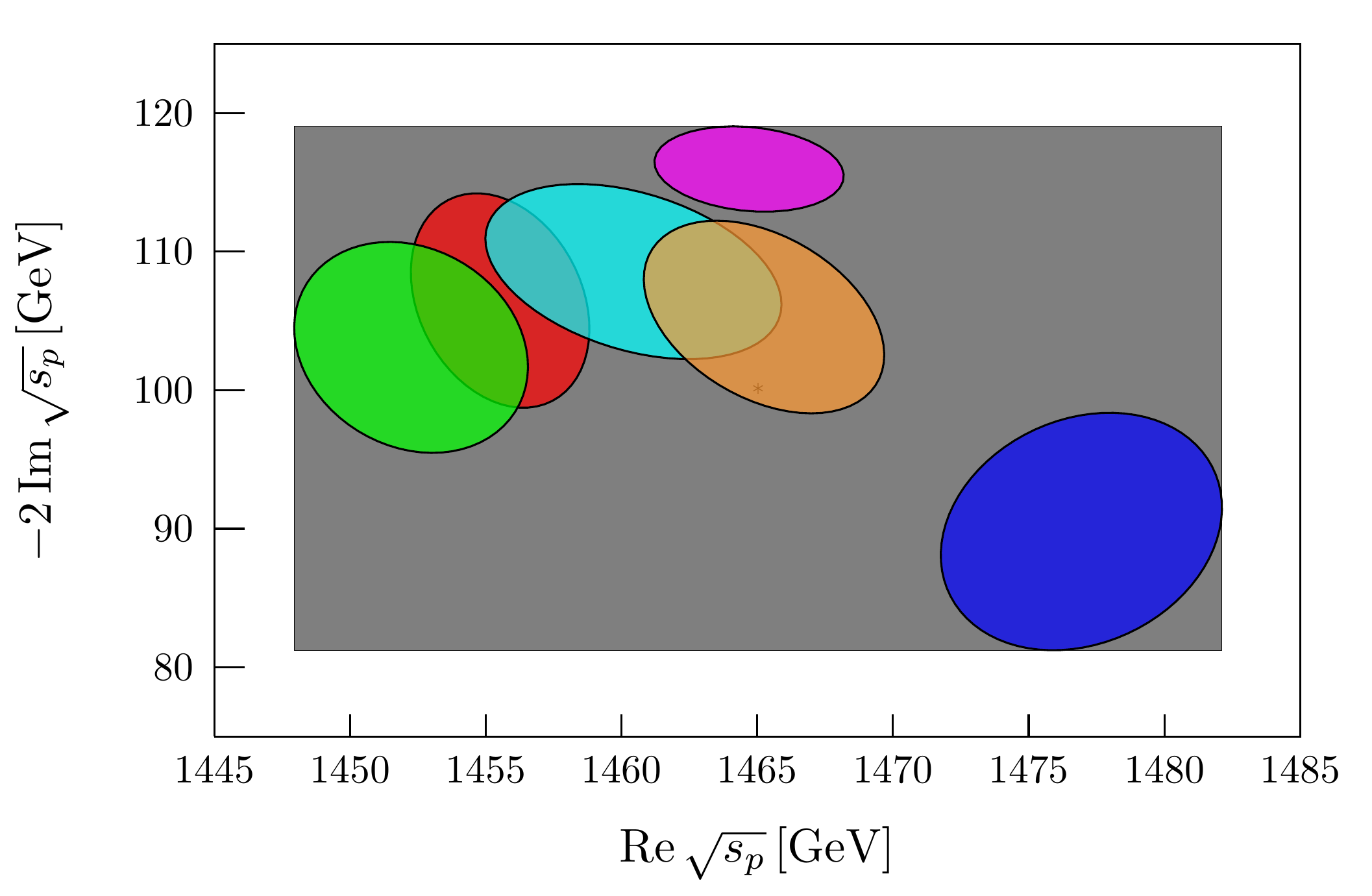} \hfill
	\includegraphics[width=0.495\linewidth]{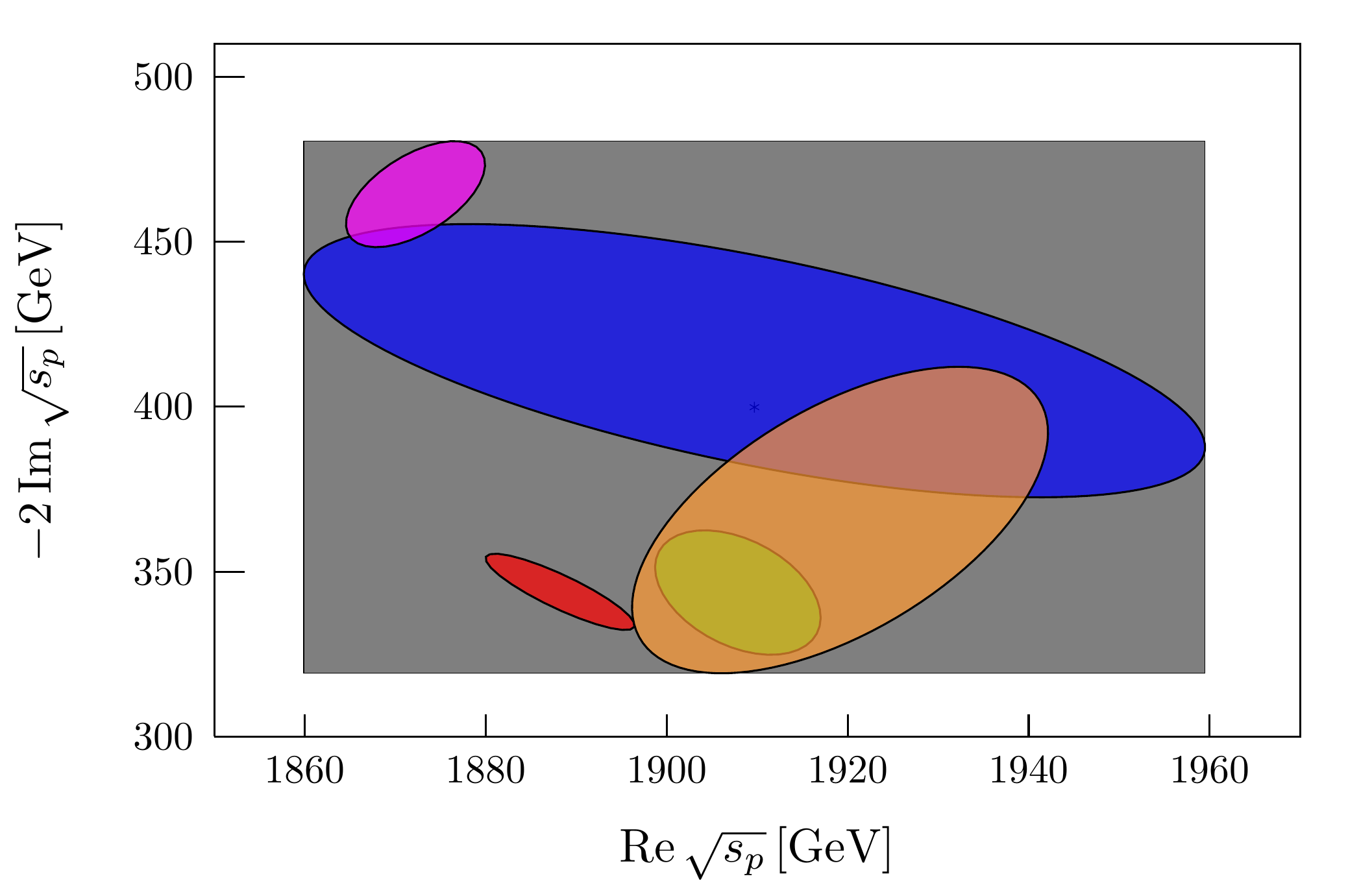}
	\caption{Poles for the $f_0(500)$ (left top), $f_0(980)$ (right top), $f_0(1500)$ (left bottom), and $f_0(2020)$ (right bottom). We show the three fits with a $\sigma\sigma$ channel, namely Fit~1 (red), Fit~2 (green), and Fit~3 (blue), as well as the fits with the $\rho\rho$ channel with Fit~1 (cyan), Fit~2 (magenta), and Fit~3 (orange). The mean values are shown in black.}
	\label{Fig::Poles::Mean}
\end{figure*}

However, overall we have strong systematic effects due to the assumptions on the parametrization such as the number of additional resonances and the linear terms in the polynomials. As we do not have a criterion that allows us to decide which fits we should prefer, 
we keep them all and perform a conservative estimate of the uncertainty: we choose a range for the resonance parameters such
that all poles with their corresponding errors are included.
The quoted mean is the middle of the resulting box as illustrated in Fig.~\ref{Fig::Poles::Mean}.

In order to see whether the pole extraction leads to sensible results, we first compare our findings for the $f_0(500)$ and $f_0(980)$ to the literature~\cite{Caprini:2005zr,GarciaMartin:2011jx,Moussallam:2011zg,Dai:2014zta}. In our parametrization the $f_0(500)$ has a mass of $\left(442\pm2\right)\MeV$ with a width of $\left(512\pm 10\right)\MeV$. For the $f_0(980)$ we find a mass of $\left(996\pm6\right)\MeV$ and a width of $\left(57\pm11\right)\MeV$. As Ref.~\cite{Dai:2014zta} serves as our input below $1\,\GeV$, their pole positions are taken as a benchmark, which lie at $\left(441-i\,544/2\right)\MeV$ and $\left(998-i\,42/2\right)\MeV$, respectively.
While the real parts are therefore perfectly consistent, we see that our parametrization slightly shifts the imaginary parts of the poles with respect to the input. 

Furthermore we can compare the coupling strengths $g_{R\pi\pi}$ and $g_{Rss}$ to the ones found in Ref.~\cite{Moussallam:2011zg}, which we adjust for the fact that the latter are quoted for the complex conjugate poles. 
For the $f_0(500)$, we obtain
\begin{align}
\left|g_{f_0(500)\pi\pi}\right| &=\left(4.53\pm0.03\right)\GeV \,, \nl
\arg\left(g_{f_0(500)\pi\pi}\right) &= \left(-73\pm2\right)^\circ \,, \nl
\left|g_{f_0(500)ss}\right| &=\left(11\pm2\right)\MeV \,,\nl
\arg\left(g_{f_0(500)ss}\right) &= \left(90\pm7\right)^\circ \,.
\end{align}
This is to be compared to
$\left|g_{f_0(500)\pi\pi}\right|=4.76\,\GeV$ and 
$\arg\left(g_{f_0(500)\pi\pi}\right)= -76.4^\circ$ as well as
$\left|g_{f_0(500)ss}\right|=\left(17\pm5^{+1}_{-7}\right)\MeV$ and
$\arg\left(g_{f_0(500)ss}\right) = 80.2^\circ$~\cite{Moussallam:2011zg}. 
With the exception of $|g_{f_0(500)\pi\pi}|$, which appears to be shifted by about 5\%, 
these numbers are consistent with our findings. 
For the $f_0(980)$ pole, we find
\begin{align}
\left|g_{f_0(980)\pi\pi}\right| &=\left(3.1\pm0.5\right)\GeV \,, \nl
\arg\left(g_{f_0(980)\pi\pi}\right) &= \left(-81\pm5\right)^\circ \,, \nl
\left|g_{f_0(980)ss}\right| &=\left(147\pm14\right)\MeV \,,\nl
\arg\left(g_{f_0(980)ss}\right) &= \left(9\pm4\right)^\circ \,,
\end{align}
in comparison to the
reference values 
$\left|g_{f_0(980)\pi\pi}\right|=2.80\,\GeV$,
$\arg\left(g_{f_0(980)\pi\pi}\right) = -85.3^\circ$, 
$\left|g_{f_0(980)ss}\right|=\left(146\pm44^{+14}_{-7}\right)\MeV$, and
$\arg\left(g_{f_0(980)ss}\right) = 14.2^\circ$~\cite{Moussallam:2011zg}.
In this case therefore all parameters are consistent within uncertainties, with a small tension for the argument of $g_{f_0(980)ss}$. 
In particular, we reproduce the well-known hierarchy in the couplings to the $\bar ss$
current: the $f_0(980)$ couples to the strange scalar current an order of magnitude more
strongly than the $f_0(500)$ does.
Overall we find good agreement of our pole parameters for $f_0(500)$ and $f_0(980)$ 
with the literature. 
We see that, a posteriori, the subtraction of 
the additional term in the scattering amplitude that introduces the 
explicit resonances, cf.\ Eq.~\eqref{eq::Formalism::subtractedPotential}, 
suppresses its influence on the lower-mass poles sufficiently.
The agreement between the reference parameters and ours gives us confidence
for an extraction of the higher poles via Pad\'e approximants.

As a reference for the higher resonance poles, we compare to the Breit--Wigner parameters of LHCb~\cite{LHCb:2012ae}. For the $f_0(1500)$, 
the collaboration quotes a resonance with mass $\left(1465.9\pm3.1\right)\MeV$ and width $\left(115\pm7\right)\MeV$.
The pole we extract corresponds to a mass of $\left(1465\pm18\right)\MeV$ and a width of $\left(100\pm19\right)\MeV$, which lies within the previously quoted uncertainties of LHCb.
The uncertainties we find are significantly larger: this is most likely due to the more flexible range of resonance parametrizations we employ; the masses and widths extracted using Breit--Wigner functions only are probably too optimistic.
In addition we can extract the corresponding residues, which are given by
\begin{align}
\left|g_{f_0(1500)\pi\pi}\right| &=\left(2.9\pm1.0\right)\GeV \,, \nl
\arg\left(g_{f_0(1500)\pi\pi}\right) &= \left(-42\pm4\right)^\circ \,, \nl
\left|g_{f_0(1500)ss}\right| &=\left(125\pm76\right)\MeV \,,\nl
\arg\left(g_{f_0(1500)ss}\right) &= \left(167\pm21\right)^\circ \,.
\end{align}
The main uncertainties stem from the assumptions made on the parametrization of the form factor, such as the number of resonances and the additional channels. Nevertheless, we note that, despite a large uncertainty, the central value for $\left|g_{f_0(1500)ss}\right|$ seems to be comparable to $\left|g_{f_0(980)ss}\right|$.  For further comparison, according to Refs.~\cite{Maltman:1999jn,Moussallam:2011zg} the $a_0(1450)$ couples to an isovector scalar $\bar{u}d$ current with $\left|g_{a_0(1450)ud}\right|=\left(284\pm54\right)\MeV$, which is of the same order as our extracted value for $g_{f_0(1500)ss}$. The precise relation between the two couplings might be used to elucidate the structure of a scalar nonet around $1.5\,\GeV$, which is however beyond the scope of the present study.

For broad, overlapping resonances a definition of branching ratios is not straightforward. Here we follow a
prescription originally proposed to define the width of $f_0(500)\to \gamma\gamma$~\cite{Morgan:1990kw}
by using the narrow-width formula of the form
\begin{align}
\BR_{R\rightarrow\pi\pi}=\frac{\Gamma_{R\rightarrow \pi\pi}}{\Gamma_R}=\frac{\left|g_{R\pi\pi}\right|^2}{32\pi m_R \Gamma_R}\sqrt{1-\frac{4M_\pi^2}{m_R^2}} \,,\label{eq::Pade::Narrowwidth}
\end{align}
with the residues as coupling constants. With this 
we can deduce a branching ratio $\BR_{f_0(1500)\rightarrow\pi\pi}=(58\pm31)\%$, where the main uncertainty stems from the difference between Fits~1 and~2 with an additional $\sigma\sigma$ channel compared to the rest of the fits. This is compatible with the (much more precise) branching ratio quoted by the PDG, $\BR_{f_0(1500)\rightarrow\pi\pi}=\left(34.9\pm2.3\right)\%$~\cite{PDG}.

The last resonance identified by LHCb as the $f_0(1790)$ has a mass of $\left(1809\pm22\right)\MeV$ with a width of $\left(263\pm30\right)\MeV$. As we do not impose a Breit--Wigner line shape, our fits seem to prefer a significantly heavier and much broader resonance with mass $\left(1910\pm50\right)\MeV$ and a width of $\left(398\pm79\right)\MeV$. Note that for the average we neglected the pole extracted from Fit~1 with the $\rho\rho$ parametrization, since this fit describes the prominent resonance structure in the $\pi\pi$ spectrum less accurately than the rest of the fits. As the pole position of the higher pole extracted in our analysis is in better agreement with the $f_0(2020)$ of the PDG (which quotes a mass of $\left(1992\pm16\right)\MeV$ and a width of $\left(442\pm60\right)\MeV$~\cite{PDG}), we will refer to it as such in the following. Furthermore we see that this pole allows for a stronger variance in the different fits. As its line shape does not only depend on the interference with other resonances, but also on further inelasticities, additional information about these channels would be appreciable.

Finally, we can also constrain the coupling strengths of this resonance to $\pi\pi$ and $\bar{s}s$, which are given as 
\begin{align}
\left|g_{f_0(2020)\pi\pi}\right| &=\left(1.2\pm0.9\right)\GeV \,, \nl
\arg\left(g_{f_0(2020)\pi\pi}\right) &= \left(2\pm89\right)^\circ \,, \nl
\left|g_{f_0(2020)ss}\right| &=\left(1019\pm786\right)\MeV \,,\nl
\arg\left(g_{f_0(2020)ss}\right) &= \left(-72\pm149\right)^\circ \,.
\end{align}
As we can see the coupling strength to the $\pi\pi$-channel is consistent with $0$ within $1.5\sigma$. The big uncertainty also strongly influences the extraction of $g_{f_0(2020)ss}$, which in addition is affected by a strong systematic uncertainty coming from the parametrization and can hardly be constrained in a meaningful manner. Using the narrow-width formula of Eq.~\eqref{eq::Pade::Narrowwidth}, the branching ratio into $\pi\pi$ is  $\BR_{f_0(2020)\rightarrow\pi\pi}=(1.3\pm 1.8)\%$, which is obviously also consistent with zero. No meaningful branching ratios are quoted by the PDG in this case.

Since the bare resonance coupling strengths $g_i^r$ as well as the bare resonance masses $m_r$ are source-independent, we can use the same parameters for any decay with $\pi\pi$ $S$-wave final-state interactions and negligible left-hand cuts. Therefore a simultaneous study of $\bar{B}_s^0\rightarrow J/\psi \pi\pi$ and $\bar{B}_d^0\rightarrow J/\psi\pi\pi$~\cite{Aaij:2013zpt} should be useful to constrain the resonances in the scalar isoscalar channel further. 
\esp

\section{Summary and outlook}\label{sec:summary}
\bsp
In this article,
we have shown that the parametrization of Ref.~\cite{H_1} for the pion vector form factor can be adapted to the scalar form factors of pions and kaons, 
marrying the advantages of a rigorous dispersive description at low energies with the phenomenological success of a unitary and analytic isobar
model beyond.
For the scalar isoscalar channel, the low-energy part must already be 
provided in terms of a dispersively constructed coupled-channel Omn\`es matrix. 
We rely on the conjecture that
the resulting strange scalar form factors can be tested
in a simultaneous study of the $S$-waves in the helicity amplitudes for the
decays $\bar{B}_s^0\rightarrow J/\psi \pi\pi$ and $\bar{B}_s^0\rightarrow J/\psi K\bar{K}$,
whose leading angular moments we can describe successfully.
In this way, we have in fact determined the corresponding strange scalar form factors up to $\sqrt{s} \approx 2\,\GeV$, 
in particular for the pion with rather good accuracy.
To quantify the uncertainties of the method, we compared fits based on different assumptions, such as different numbers of resonances 
as well as different final-state channels. Although they describe the data almost equally well, we see a significant systematic uncertainty at higher energies,
which should be reduced significantly, however, once further information about the inelastic channels becomes available. 
For now, we only included an effective $4\pi$ channel modeled either by $\rho\rho$ or $\sigma\sigma$ intermediate states; 
for a more detailed description of the branching ratios of the heavier scalar isoscalar resonances,
we might need to include further inelastic channels such as $a_1\pi$, $\eta\eta$, or $\eta\eta^\prime$.

As the parametrization developed is fully unitary and analytic, we extracted resonance parameters as pole positions and residues in the complex energy plane, employing Pad\'e approximants.
In particular, we determined resonance poles as well as coupling constants for 
 $f_0(1500)$ and $f_0(2020)$. While the pole location for the $f_0(1500)$ is consistent with the one derived from the LHCb Breit--Wigner extraction, we find a significantly shifted pole for the $f_0(2020)$. 
This shift ought to be tested experimentally in other processes with prominent $S$-wave pion--pion final-state interactions.
Alternatively---or in addition---we might also include scattering data at higher energies in the fits explicitly~\cite{Bugg:1996ki,Anisovich:2002ij}.
\esp

\begin{acknowledgements}
\bsp
We thank T.~Isken, B.~Moussallam, J.~Niecknig, W.~Ochs, J.~Ruiz de Elvira, and A.~Sarantsev for useful discussions. 
Financial support by DFG and NSFC through funds provided to
the Sino--German CRC~110 ``Symmetries and the Emergence of
Structure in QCD'' (DFG Grant No.~TRR110 and NSFC Grant
No.~11621131001) is gratefully acknowledged.
\esp
\end{acknowledgements}


\end{document}